\documentclass[longauth]{aa}  

\usepackage[utf8]{inputenc} 
\usepackage[T1]{fontenc}
\usepackage{textgreek}
\usepackage{gensymb}
\usepackage{graphicx}
\usepackage{subfig}
\usepackage{txfonts}
\usepackage{hyperref}

\usepackage{color}

\begin{document} 

   \title{Dynamical masses of M-dwarf binaries in young moving groups}
   
   \subtitle{I - The case of TWA~22 and GJ~2060}

   \author{L. Rodet\inst{1} \and M. Bonnefoy\inst{1} \and S. Durkan\inst{2,3} \and H. Beust\inst{1} \and A-M Lagrange\inst{1} \and  J. E. Schlieder\inst{4,5} \and M. Janson\inst{3} \and A. Grandjean\inst{1} \and G. Chauvin\inst{1,6}  \and S. Messina\inst{7} \and A.-L. Maire\inst{5} \and W. Brandner\inst{5} \and J. Girard\inst{8,1} \and P. Delorme\inst{1} \and B. Biller\inst{9} \and C. Bergfors\inst{5,10} \and S. Lacour\inst{11} \and M. Feldt\inst{5} \and T. Henning\inst{5} \and A. Boccaletti\inst{11} \and J.-B. Le Bouquin\inst{1} \and J.-P. Berger\inst{1} \and J.-L. Monin\inst{1} \and S. Udry\inst{12} \and S. Peretti\inst{12} \and D. Segransan\inst{12} \and F. Allard\inst{13} \and D. Homeier\inst{14} \and A. Vigan\inst{15} \and M. Langlois\inst{13,15} \and J. Hagelberg\inst{1} \and F. Menard\inst{1} \and A. Bazzon\inst{18} \and J.-L. Beuzit\inst{1} \and A. Delboulb\'e\inst{1} \and S. Desidera\inst{7} \and R. Gratton\inst{7} \and J. Lannier\inst{1} \and R. Ligi\inst{15}  \and  D. Maurel\inst{1} \and D. Mesa\inst{7} \and M. Meyer\inst{18} \and A. Pavlov\inst{5} \and J. Ramos\inst{5}  \and R. Rigal\inst{17} \and R. Roelfsema\inst{16}  \and G. Salter\inst{15} \and M. Samland\inst{5} \and T. Schmidt\inst{11} \and E. Stadler\inst{1} \and L. Weber\inst{12}}
   
      \institute{Univ. Grenoble Alpes, CNRS, IPAG, F-38000 Grenoble, France\\ 
         	\email{laetitia.rodet@univ-grenoble-alpes.fr} \and 
         	Astrophysics Research Centre, Queens University Belfast, Belfast, Northern Ireland, UK \and 
         	Department of Astronomy, Stockholm University, Stockholm, Sweden \and 
         	Exoplanet and Stellar Astrophysics Laboratory, Code 667, NASA Goddard Space Flight Center, Greenbelt, MD 20771, USA \and 
         	Max Planck Institute for Astronomy, K\"onigstuhl 17, D-69117 Heidelberg, Germany \and 
         	Unidad Mixta Internacional Franco-Chilena de Astronom\`ia, CNRS/INSU UMI 2286 and Departemento de Astronom\`ia, Universidad de Chile, Casilla 36-D, Santiago, Chile \and 
         	INAF - Osservatorio Astronomico di Padova, Vicolo della Osservatorio 5, 35122, Padova, Italy \and 
         	European Southern Observatory (ESO), Alonso de C\`ordova 3107, Vitacura, 19001 Casilla, Santiago, Chile \and 
         	Institute for Astronomy, University of Edinburgh, Blackford Hill, Edinburgh EH9 3HJ, UK \and 
         	University College London, London, UK \and 
         	LESIA, Observatoire de Paris, PSL Research University, CNRS, Sorbonne Universit\'es, UPMC, Univ. Paris 06, Univ. Paris Diderot, Sorbonne Paris Cit\'e, 5 place Jules Janssen, 92195 Meudon, France \and 
          	Geneva Observatory, University of Geneva, Chemin des Mailettes 51, 1290 Versoix, Switzerland \and 
          	CRAL, UMR 5574, CNRS, Universit\'e de Lyon, Ecole Normale Sup\'erieure de Lyon, 46 All\'ee d'Italie, F-69364 Lyon Cedex 07, France \and 
         	Zentrum für Astronomie der Universität Heidelberg, Landessternwarte, Königstuhl 12, 69117 Heidelberg, Germany \and 
         	Aix Marseille Universit\'e, CNRS, LAM (Laboratoire d'Astrophysique de Marseille) UMR 7326, 13388 Marseille, France \and 
         	NOVA Optical Infrared Instrumentation Group, Oude Hoogeveensedijk 4, 7991 PD Dwingeloo, The Netherlands \and 
         	Anton Pannekoek Institute for Astronomy, Science Park 904, NL-1098 XH Amsterdam, The Netherlands \and 
         	Institute for Particle Physics and Astrophysics, ETH Zurich, Wolfgang-Pauli-Strasse 27, 8093 Zurich, Switzerland} 

   \date{Received March 1, 2018; accepted June 13, 2018}
 
  \abstract
{Evolutionary models are widely used to infer the mass of stars, brown dwarfs, and giant planets. Their predictions are thought to be less reliable at young ages ($<$ 200 Myr) and in the low-mass regime ($\mathrm{<1~M_{\odot}}$). GJ~2060~AB and TWA~22~AB are two rare astrometric M-dwarf binaries respectively members of the AB Doradus (AB Dor) and Beta Pictoris (\textbeta~Pic) moving groups. As their dynamical mass can be measured within a few years, they can be used to calibrate the evolutionary tracks and set new constraints on the age of young moving groups.}
   {We aim to provide the first dynamical mass measurement of GJ~2060 and a refined measurement of the total mass of TWA~22. We also aim to characterize the atmospheric properties of the individual components of GJ~2060 that can be used as inputs to the evolutionary models.}
   {We used NaCo and SPHERE observations at VLT and archival Keck/NIRC2 data to complement the  astrometric monitoring of the binaries. We combined that astrometry with new HARPS radial velocities (RVs) and FEROS RVs of GJ~2060. We used a Markov Chain Monte-Carlo (MCMC) module to estimate posteriors on the orbital parameters and dynamical masses of GJ~2060~AB and TWA~22~AB from the astrometry and RVs.  Complementary data obtained with the integral field spectrograph VLT/SINFONI were gathered to extract the individual near-infrared (1.1-2.5 $\mu$m) medium-resolution (R$\sim$1500-2000) spectra of GJ~2060~A and B. We compared the spectra to those of known objects and to grids of BT-SETTL model spectra to infer the spectral type, bolometric luminosities, and temperatures of those objects.}
   {We find a total mass of $\mathrm{0.18\pm 0.02~M_\odot}$ for TWA~22. That mass is in good agreement with model predictions at the age of the \textbeta Pic moving group. We obtain a total mass of $\mathrm{1.09 \pm 0.10~M_{\odot}}$ for GJ~2060. We estimate a spectral type of M$1\pm0.5$, $\mathrm{L/L_{\odot}=-1.20\pm0.05}$ dex, and $\mathrm{T_{eff}=3700\pm100}$ K for GJ~2060~A. The B component is a  M$3\pm0.5$ dwarf with $\mathrm{L/L_{\odot}=-1.63\pm0.05}$ dex and $\mathrm{T_{eff}=3400\pm100}$ K. The dynamical mass of GJ~2060~AB is inconsistent with the most recent models predictions (BCAH15, PARSEC) for an ABDor age in the range 50-150 Myr. It is 10 to 20\% (1-2\textsigma, depending on the assumed age) above the models predictions, corresponding to an underestimation of $0.10$ to $0.20~\mathrm{M_\odot}$. Coevality suggests a young age for the system ($\sim$ 50 Myr) according to most evolutionary models.}
{TWA~22 validates the predictions of recent evolutionary tracks at $\sim$20 Myr. On the other hand, we evidence a 1-2~\textsigma~ mismatch between the predicted and observed mass of GJ~2060~AB. This slight departure may indicate that one of the star hosts a tight companion. Alternatively, this would confirm the models tendency to underestimate the mass of young low-mass stars.}


  \maketitle

\section{Introduction}

Our understanding of stellar evolution has made a lot of progress since the introduction of the Hertzsprung-Russell diagram (HRD) a hundred years ago. The beginning of a star life, before it reaches the zero age main sequence, has been in particular deeply investigated through the development of evolutionary models. The latter rely on equations of state describing the stellar interior structure, and can make use of atmospheric models to define boundary conditions and predict emergent spectra. Different families of models exist \citep{dantona1997,siess2000,tognelli2012,bressan2012,feiden2015,baraffe2015}, and their physical and chemical ingredients (e.g., nuclear rates, opacity, atmospheric parameters) have been updated in the recent years \citep[e.g.,][]{baraffe2015}. The models can predict the age and mass of stellar and substellar objects from the measured broad band photometry, surface gravity, radius, luminosity, and effective temperature. The mass is the fundamental parameter which allows to comprehend the object nature and formation pathways.

The models predictions remain to be calibrated in various mass and age regimes \citep[e.g.,][]{hillenbrand2004,mathieu2007}. Uncertainties related to the object formation process (formation mechanism, early accretion history, etc.) exist in the pre-main sequence (PMS) regime \citep[e.g.,][]{baraffe2002,baraffe2010}. Further uncertainties may be added for low-mass stars, which have strong convection, rotation and magnetic activity \citep{mathieu2007}. About 50 low-mass (below 1 $\mathrm{M_\odot}$) pre-main sequence stars had their mass determined thus far \citep[e.g.,][]{simon2000,gennaro2012,stassun2014,mizuki2018}. Most of these systems have been studied through their disk kinematics, and are thus younger than 10 million years  \citep[e.g.,][]{guilloteau2014,simon2017}. Moreover, this method only allows to determine the total mass of the system, disk included. The disk mass can be a non negligible fraction of the total mass \citep[e.g.,][Fig. 9]{andrews2013}, so that uncertainties remain on the stellar mass. A dozen of the stars with dynamical mass are SB2 eclipsing binaries, for which the orbital inclination can be strongly constrained and the mass determined from the orbit. However, eclipsing binaries are very tight stellar pairs (orbital periods 1-10 d) so that each star strongly influences the other one (tides, high rotation speed, convection inhibition). Thus, their evolution may not be representative of typical stars \citep{chabrier2007,kraus2011,stassun2014}. Consequently, evolution models remain poorly constrained for low-mass stars for most of the pre-main sequence stellar evolution. This can induce systematic offsets and disparate mass predictions \citep{hillenbrand2004,mathieu2007}. 

Some rare young (age$<$200 Myr) and nearby (d$<$100 pc) binaries resolved with high resolution imaging techniques (adaptive optics, speckle interferometry, lucky imaging, sparse aperture masking) have orbital periods shorter than a decade. Combined with a precise parallax, astrometric follow-up of the two components relative orbit gives the total dynamical mass of the system. Knowledge about the individual masses can then be gained from additional radial velocity measurements. These systems offer a good prospect for calibrating the PMS tracks and the underlying models physics. To date and to our knowledge, only 9 such systems in the intermediate PMS regime (10-100 Myr) have dynamical mass estimates below 1 $\mathrm{M_\odot}$, with various model agreements: HD 98800 B \citep{boden2005}, TWA 22 \citep{bonnefoy2009},  HD 160934 \citep{azulay2014}, ABDor \citep{azulay2015,close2007}, GJ 3305 \citep{montet2015}, V343 Nor A \citep{nielsen2016}, NLTT 33370 \citep{dupuy2016}, GJ 2060 (this work) and  GJ 1108 \citep{mizuki2018}. We will here provide a refined dynamical mass for TWA 22 and a first determination for GJ 2060.

The calibration is nonetheless often limited by uncertainties on the age and distance of these benchmarking systems. These uncertainties are mitigated for systems belonging to known young nearby associations and moving groups (YMG). The age of the YMG can be inferred from several approaches (lithium depletion boundary, kinematics,...) and parallaxes can be measured for individual members (Gaia, \citeauthor{arenou2017} \citeyear{arenou2017}, Hipparcos, \citeauthor{vanleeuwen2007} \citeyear{vanleeuwen2007}). Moreover, those systems share the same age (8-150 Myr) as the substellar companions resolved during direct imaging surveys \citep[planets and brown dwarfs; e.g.,][]{chauvin2004,lagrange2010,marois2008,marois2010,rameau2013a,rameau2013b} and whose mass determination also depends on PMS evolutionary models.

 TWA~22 and GJ~2060 are two precious astrometric M-dwarf binaries with orbital period of a few years. They are proposed members of the young \textbeta Pic and AB Dor moving groups, respectively. Both systems have a well-measured parallax. We initiated their follow-up in 2004 with various ground-based facilities in order to measure their dynamical masses and characterize their components. This paper presents an in-depth study of these two systems using published and additional observations and discusses the agreement between their orbit, their atmospheric properties, the age of their moving group and the PMS evolutionary models. We first review the observations and membership studies previously performed (section \ref{section:presentation}), and then present new imaging and spectroscopic data (section \ref{section:Observations}). We analyze the spectroscopic properties of GJ 2060 (section \ref{section:Spectrophotometry}). We derive in section \ref{section:MCMC} the dynamical masses from orbital fits, and use them to probe the evolutionary models (section \ref{section:Models}). The agreement between models and data is finally discussed in section \ref{section:Discussion}.  

\section{Age and Membership of TWA~22 and GJ~2060}
\label{section:presentation}

\subsection{TWA~22}

\textbf{TWA~22} (2MASS J10172689-5354265), located at $d=17.5 \pm 0.2$ pc \citep{teixeira2009}, was originally proposed as a member of the $\sim$10 Myr old \citep{bell2015} TW Hydrae association (TWA) by \cite{song2003}. This classification was based on its strong Li 6708 \AA~absorption and H$\alpha$ emission lines and sky position near other TWA members. A subsequent kinematic analysis of all TWA members proposed at the time by \cite{mamajek2005} revealed that the available kinematics of TWA~22 were largely inconsistent with the bulk of other TWA members and provided a low probability of membership. Possible membership in either TWA or the older \textbeta~Pictoris moving group \citep[$\sim$ 25 Myr,][]{bell2015} was then proposed by \cite{song2006}. 

TWA~22 was included in 2003 as a target in an adaptive optics (AO) imaging survey to search for low-mass companions \citep{chauvin2010}. It was resolved into a $\sim$100 mas, equal luminosity binary and was as a potential benchmark target for dynamical mass measurements and model calibration. For this purpose, \cite{teixeira2009} measured the parallax, provided revised proper motion and radial velocity measurements, and performed a detailed kinematic analysis of TWA~22 and found further evidence for membership in the \textbeta~Pic group, but were unable to fully rule out TWA membership. Then, \cite{bonnefoy2009} presented resolved spectra of the components, measured spectral types (later refined by \citeauthor{bonnefoy2014a} \citeyear{bonnefoy2014a} to M5 $\pm$ 1 for TWA 22 A and M5.5 $\pm$ 1 for TWA 22 A), and performed an astrometric orbit fit to the available observations. This revealed that the total mass of the system was incompatible with model predictions, considering an age range consistent with the age of TWA. The authors however noted that the models may simply be underpredicting the system mass at such a young age.

TWA~22 has now been adopted as a bona fide member of the \textbeta~Pic group on the basis of Bayesian methods to determine membership to kinematic moving groups (BANYAN I,\citeauthor{malo2013} 2013; BANYAN II, \citeauthor{gagne2014} 2014). TWA~22’s kinematics were used to develop the \textbeta~Pic group kinematic model implemented in the BANYAN Bayesian estimator (it has >99\% probability of membership). The amount of lithium observed in TWA 22 is consistent with the age of the TWA association, but we know now that it is also compatible with its membership to the \textbeta~Pic group, as Li may still subsist in the components at the age of \textbeta~Pic. The age of the \textbeta~Pic group has been revised multiple times in recent years using isochronal methods that rely on all group members \citep{malo2014,bell2015}, the lithium depletion boundary of the group \citep{binks2014,malo2014,messina2016,shkolnik2017}, the rotation distribution of known members \citep{messina2016}, and model comparisons to dynamical masses of binaries in the group \citep{montet2015,nielsen2016}. This large variety of age determination methods converge toward a group age of $\sim$25 Myr (see Table \ref{table:BetaPicMG}).

\begin{table}[h]
	\centering
	\caption{Age estimates of the Beta Pictoris moving group.}\label{table:BetaPicMG}	
	\small
	\begin{tabular}{lcc}
		\hline\\[-0.8em]
		Paper & Age & Method\\
		 & (Myr) & \\
		\hline
		\hline\\[-0.8em]
		\cite{malo2014} & 15-28  & Isochronal methods\\
		& $26 \pm 3$ & Lithium depletion boundary\\
		\cite{bell2015} & $24 \pm 3$  & Isochronal methods\\
		\cite{binks2014} & $21 \pm 4$ & Lithium depletion boundary\\
		\cite{shkolnik2017} & $22 \pm 6$ & Lithium depletion boundary\\
		\cite{messina2016} & $25 \pm 3$ & Rotation distribution and\\ 
		 & & Lithium depletion boundary\\ 
		\cite{montet2015} & $37 \pm 9$ & Dynamical mass of binaries\\
		\cite{nielsen2016} & $26 \pm 3$ & Dynamical mass of binaries\\
		\hline
	\end{tabular}
\end{table}  

In this work we adopt the \textbeta~Pic group age for TWA~22, provide new astrometric measurements of the binary components, combine these data with previous data to perform an updated orbital fit and measure the system mass, and compare the derived mass to estimates from the latest stellar evolution models. The binary period is relatively short ($\sim\!\!5$ yr) and TWA~22 has been regularly observed from 2004 to 2007 and later in 2013 and 2015, enabling a very good characterization through the orbital fit. The two components are the least massive stars in the \textbeta~Pic group for which a dynamical mass has been computed. They complete the mass sampling between the giant planet \textbeta~Pictoris b \citep{lagrange2010} and the higher mass binaries GJ~3305 \citep[total mass 1.1 $\mathrm{M_\odot}$;][]{montet2015} and V343~Nor \citep[total mass 1.4 $\mathrm{M_\odot}$;][]{nielsen2016}. TWA 22 is thus an essential benchmark to test the predictions of the evolutionary models in the young group in the 0.1 $ \mathrm{M_\odot}$ mass range.

\subsection{GJ~2060}

\textbf{GJ~2060} (2MASS J07285137-3014490) is an early M-dwarf at $d = 15.69 \pm 0.45$ pc \citep{vanleeuwen2007} that was first identified as a small separation binary by the Hipparcos satellite \citep{dommanget2000}. The star was subsequently identified as a nearby young star in the paper presenting the discovery of the AB Doradus moving group \citep{zuckerman2004}. This work presented GJ~2060 and $\sim$30 other stars as having both Galactic kinematics consistent with the well studied young system AB Dor and independent indicators of youth (X-ray and H-alpha emission, large v$\sin i$, etc.). Along with AB Dor itself and six other nearby stars within a $\sim$5 pc radius, GJ~2060 is a member of the AB Dor moving group nucleus. The system has since been verified as a bona fide member of the AB Dor moving group using revised group kinematic distributions and Bayesian methods with an estimated membership probability of >99\% \citep{malo2013,gagne2014}. GJ~2060 was first resolved into an 0.175" multiple system by \cite{daemgen2007} using adaptative optics imaging. The system has been observed multiple times since with high resolution imaging and exhibited rapid orbital motion \citep[see ][]{janson2014}.
      The age of the AB Dor moving group, and thereby GJ~2060, was first proposed to be $\sim$50 Myr by \cite{zuckerman2004}. Yet, the age of the group remains relatively poorly constrained and ages ranging from the original $\sim$50 Myr to $\sim$150 Myr have been proposed over the last decade \citep[e.g., ][]{close2005,nielsen2005,luhman2005,lopez-santiago2006,ortega2007,torres2008,barenfeld2013,bell2015}. The system components of the groups’ namesake quadruple system AB Dor have been studied in detail \citep{close2005,nielsen2005,guirado2011,azulay2015} and comparisons to stellar evolution models indicate discrepancies between the measured masses of the components and point toward ages <100 Myr. This is in conflict with group ages estimated both from the individual components of AB Dor and from the ensemble of stars using HR diagram placement \citep{luhman2005,bell2015}, rotation periods \citep{messina2010}, and  Li depletion \citep{barenfeld2013}. These works find that individual members and the ensemble of proposed AB Dor members have properties consistent with the Pleiades open cluster and likely have a comparable age \citep[$\sim$120 Myr; ][]{stauffer1998,barrado2004,dahm2015}. Here we use new astrometric and radial velocity measurements of the GJ~2060 system to derive component masses and perform similar comparisons to stellar evolution models. No orbital fit had been performed on this system yet, so that its orbital elements and dynamical mass are first determined in the present article.

\begin{table}[h]
	\centering
	\caption{Age estimates of the AB Doradus moving group.}\label{table:ABDorMG}
	\small
	\begin{tabular}{lcc}
		\hline\\[-0.8em]
		Paper & Age & Method\\
		 & (Myr) & \\
		\hline
		\hline\\[-0.8em]
		\cite{zuckerman2004} & $50 \pm 10$  & Isochronal methods\\
		\cite{luhman2005} & 100-125  & Isochronal methods\\
		\cite{lopez-santiago2006} & 30-50 & Isochronal methods\\
		\cite{bell2015} & $149 ^{+51}_{-19}$ & Isochronal methods\\
		\cite{ortega2007} & $119 \pm 20$ & Stellar dynamics\\  
		\cite{messina2010} & $\sim$ 70 & Rotation periods\\
		\cite{barenfeld2013} & > 110 & Kine-chemical analysis\\
		\cite{nielsen2005} &50-100 & AB Dor C\\
		\cite{boccaletti2008} & $75 \pm 25$ & AB Dor C\\ 
		\cite{guirado2011} & 40-50 & AB Dor A \\
		\cite{azulay2015} & 40-50 & AB Dor B \\
		\hline
	\end{tabular}
\end{table} 

\section{Observation and data processing}
\label{section:Observations}

A summary of the new observations of TWA~22 and GJ~2060 is given in Table \ref{table:observations}. We describe the datasets and related reduction processes in more details below. 

\begin{table*}[t]
\caption{\label{table:observations}Observing log. The field rotation $\theta$  is given when the observations are performed in pupil-tracking mode.}
\centering
\begin{tabular}{cccccccccc}
\hline\\[-0.8em]
UT date& Target	&	Instrument	& Mode &  $\mathrm{DIT \times NDIT \times N_{EXPO}}$ 	& $\theta$ & $\mathrm{\langle Seeing \rangle}$ \tablefootmark{a} & $\langle \tau_{0} \rangle$ & Airmass \\
& & & &   & (deg) & (")	&	(ms)	& \\
\hline \hline\\[-0.8em]
11/02/2013	&	TWA~22AB	& NaCo	&	H-S13	&	$0.345s\times30\times32$	&	 n.a. & 1.0 & 5.7 & 1.15 \\
11/02/2013	& GSC08612-01565	&	NaCo	&	H-S13	&	$0.345s\times30\times32$	& n.a.&	1.0 & 6.0 & 1.22 \\
12/02/2013	&	$\Theta$ Ori C	&	NaCo	&	H-S13	&	$3s\times3\times25$ & n.a. & 0.9 & 6.1 & 1.06 \\
03/02/2015  & TWA~22AB	& SPHERE	&	IRDIS-K12	&	$4s\times16\times15$ & n.a. & 2.5 & 1.4 & 1.15  \\
03/02/2015  & TWA~22AB	& SPHERE	&	IFS-YH	&	$32s\times2\times17$ & n.a. & 2.5  & 1.4 & 1.16  \\
\hline
21/11/2012	&	$\Theta$ Ori C	&	NaCo	&	H-S13	&	$3s\times5\times26$ & n.a. & 0.6 & 3.7 & 1.08 \\
25/11/2012	&	GJ~2060~AB	&	NaCo	&	H-S13	&	$0.15s\times100\times7$ & n.a. & 1.0	&	1.7	&	1.01 \\
25/11/2012	&	GJ~3305AB	&	NaCo	& H-S13	&		$0.12s\times200\times4$  & n.a.	&	0.8 &	2.1 & 1.08 \\
22/11/2013	&	 GJ~2060~AB	&	SINFONI	&	J 	&	$1s\times4\times11$ &  n.a.& 0.8 & 1.9	&	0.93 \\
22/11/2013	&	 GJ~2060~AB	&	SINFONI	&	H+K 	&	$0.83s\times4\times11$ & n.a. & 0.8 & 2.1	&	0.92 \\
22/11/2013	&	 HIP~036092	&	SINFONI	&	J	&	$6s\times2\times1$ & n.a. & 0.9 & 2.2	&	1.01 \\
22/11/2013	&	 HIP~036092	&	SINFONI	&	H+K	&	$5s\times2\times1$ & n.a. & 0.8 & 2.4	&	1.01 \\
05/02/2015	&	 GJ~2060~AB	&	SPHERE	&	IRDIS-K12	&	$2s\times32\times10$ & n.a. & 2.0 & 2.6	&	1.14 \\
05/02/2015	&	 GJ~2060~AB	&	SPHERE	&	IFS-YH	&	$32s\times2\times11$ & n.a. & 1.8 & 2.6	&	1.14 \\
16/03/2015   &   GJ~2060~AB   &  AstraLux   &       z'     &  $0.015s\times20000\times1$  &  n.a.   &  n.a.   &  n.a.  & 1.48  \\
16/03/2015   &   GJ~2060~AB   &  AstraLux   &       i'      &  $0.015s\times20000\times1$  &  n.a.   &  n.a.   &  n.a.  & 1.54  \\
01/10/2015	&	 GJ~2060~AB	&	NIRC2	&	$\mathrm{K_{cont}}$	&	$0.2s\times50\times6$ & 0.67 & n.a.	& n.a. &	1.81 \\
18/11/2015	&	 GJ~2060~AB	&	NIRC2	&	$\mathrm{K_{cont}}$	&	$0.2s\times50\times9$ & 1.71 & n.a.	& n.a. &	1.56 \\
29/11/2015	&	GJ~2060~AB &	 SPHERE & IRDIS-H23 & $4s\times40\times4$ & n.a.& 1.12 & 3.4 & 1.06 \\
29/11/2015 &	 GJ~2060~AB & SPHERE & IFS-YJ &	$16s\times10\times4$ & n.a. &  1.12 & 3.4 & 1.06\\
25/12/2015   &   GJ~2060~AB   &  AstraLux   &       z'     &  $0.015s\times10000\times1$  &  n.a.   &  n.a.   &  n.a.  & 1.01  \\
26/12/2015 & GJ~2060~AB & SPHERE & IRDIS-H23 & $16s\times14\times16$ &  2.3  &  0.8 & 3.5 & 1.26 \\
26/12/2015 & GJ~2060~AB & SPHERE & IFS-YJ &	$ 8s\times7\times16$ &  2.3 &  0.8 & 3.5 & 1.26 \\
28/12/2015   &   GJ~2060~AB   &  AstraLux   &       z'     &  $0.015s\times10000\times1$  &  n.a.   &  n.a.   &  n.a.  & 1.15  \\
27/03/2016 & GJ~2060~AB & SPHERE 	& IRDIS-H23 &	$ 2s\times40\times16 $ &  n.a. &  0.5 & 2.6 & 1.08 \\
 27/03/2016 	& GJ~2060~AB & SPHERE & IFS-YJ 	&	$ 16s\times20\times5 $ &  n.a. &  0.5 & 2.6 & 1.08 \\
07/02/2017	&	 GJ~2060~AB	&	SPHERE	&	IRDIS-K12	&	$4s\times8\times16$ & 1.87 & 0.6 & 15.4	&	1.11 \\
07/02/2017	&	 GJ~2060~AB	&	SPHERE	& IFS-YH	&	$16s\times2\times16$ & 1.65 & 0.6 & 15.2	&	1.11 \\
\hline
\end{tabular}
\tablefoot{\tablefoottext{a}{DIMM for the VLT}}
\end{table*}

	\begin{table*}[h]
		\centering
		\caption{Summary of TWA~22 astrometry}\label{table:astromTWA22}
		\begin{tabular}{cccccc}
			\hline\\[-0.8em]
			UT Date & Band & $\Delta$Ra & $\Delta$Dec & Instrument & Reference \\
			 & & (mas) & (mas) & &\\
			\hline
			\hline\\[-0.8em]
			05/03/2004 & NB2.17 & 99 $\pm$ 3 & -17 $\pm$ 3 &	NaCo & \cite{bonnefoy2009}\\
			27/04/2004 & NB1.75 & 98 $\pm$ 6 & -36 $\pm$ 6 &	NaCo & \cite{bonnefoy2009}\\
			06/05/2005 & H-ND & 15 $\pm$ 3 & -89 $\pm$ 3 & NaCo & \cite{bonnefoy2009}\\
			08/01/2006 & H & -68 $\pm$ 2 & -49 $\pm$ 2 &	NaCo & \cite{bonnefoy2009}\\
			26/02/2006 & H & -74 $\pm$ 3 & -30 $\pm$ 3 &	NaCo & \cite{bonnefoy2009}\\
			06/03/2007 & H & -57 $\pm$ 4 & 80 $\pm$ 2 &	NaCo & \cite{bonnefoy2009}\\
			04/12/2007 & H & 19 $\pm$ 3 & 98 $\pm$ 3 &	NaCo & \cite{bonnefoy2009}\\
			26/12/2007 & H & 26 $\pm$ 3 & 97 $\pm$ 3 &	NaCo & \cite{bonnefoy2009}\\
			11/02/2013 & H & 2 $\pm$ 1 & 100 $\pm$ 1 &	NaCo & This work\\
			05/02/2015 & IFS-YH & -43 $\pm$ 1 & 93 $\pm$ 1 &	SPHERE & This work\\
			\hline
		\end{tabular}
	\end{table*}

\subsection{TWA~22}

	\subsubsection{NaCo observations}
	
	TWA~22~AB was observed in field-tracking mode on February 11, 2013 with the NAOS-CONICA (NaCo) adaptive-optics instrument mounted on the VLT/UT4  \citep{2003SPIE.4841..944L, 2003SPIE.4839..140R} as part of a program dedicated to the orbit monitoring of young binaries (PI Bonnefoy; program ID 090.C-0819).  The S13 camera  was associated to the H-band filter ($\lambda_{c}=1.66~\mu m, \Delta \lambda=0.33~\mu  m$), yielding a square field of view of 13.5 arcsec. The wavefront sensing was achieved in the near-infrared on the pair (seen as a whole). We acquired 32 frames (NEXPO)  of the binaries consisting of $0.345~s \times 30$ ($DIT \times NDIT$) averaged exposures each.  Small ($\pm3"$) dithers were applied  every four frames to allow for an efficient sky and bias subtraction at the data processing step. 
	We observed immediately after TWA~22~AB the M6 star GSC08612-01565 to calibrate the point-spread function (PSF) of the instrument using the same adaptive-optics setup and the same DIT, NDIT, and NEXPO as for TWA~22~AB. 
	We observed the following night the crowded field of stars around $\Theta$ Ori C to calibrate the platescale and field orientation with the same filter and camera and the visible wavefront sensor. That astrometric field was already used in \cite{bonnefoy2009} for the previous observations of TWA~22.

	All the data were reduced with the \texttt{eclipse} sofware \citep{1997Msngr..87...19D}. The  \texttt{eclipse} routines carried out the basics cosmetic steps: bad pixel flagging and interpolation, flat field calibration, sky subtraction, and cross-correlation and shift of the dithered frames. We extracted the position of the  $\Theta$ Ori stars  and compared them to those reported in \cite{1994AJ....108.1382M} to infer a platescale of $13.19 \pm 0.08$ mas/pixel and a True North of $-0.90 \pm 0.15^{\circ}$ for those  observations. 
	We used a deconvolution algorithm dedicated to the stellar field blurred by the adaptive-optics corrected point spread functions  to deblend the overlaping point-spread-functions of TWA22~A and B in the final NaCo image \citep{1998SPIE.3353..426V} and measure the position and the photometry of each components. The same tool was used in \cite{bonnefoy2009}.  The algorithm is based on the minimization in the Fourier domain of a regularized least square objective function using the Levenberg-Marquardt method. It is well suited to our data which are Nyquist-Sampled. We cross-checked our results using the IDL \texttt{Starfinder} PSF fitting package \citep{2000SPIE.4007..879D} which  implements a custom version of the CLEAN algorithm to build a flux distribution model of the binary but does not perform any spatial deconvolution.  We find a contrast $\Delta H=0.52\pm0.05$ mag consistent with the values derived at previous epochs \citep{bonnefoy2009}. The binary is found at a $PA=1.15\pm0.15^{\circ}$ and separation $\rho=100\pm3$ mas.

	\subsubsection{SPHERE observations}
	\label{sec:SPHEREdatTWA22}
	
	The binary was observed on February 3, 2015 as part of the SHINE (SpHere INfrared survey for Exoplanets) survey \citep{chauvin2017} with the high-contrast instrument SPHERE at UT3/VLT \citep{2008SPIE.7014E..18B}.  The observations were scheduled as part of a sub-program (filler) of SHINE devoted to the astrometric monitoring of tight binaries. 
	
 SPHERE was operated in field-tracking mode. No coronagraph  was inserted into the light path. The IRDIFS\_EXT mode enabled for simultaneous observations with the dual-band imaging sub-instrument IRDIS \citep{2008SPIE.7014E..3LD, 2010MNRAS.407...71V} in the K1 ($\lambda_{c}=2.110$ \textmu m; $\Delta \lambda=0.102$ \textmu m) and K2 ($\lambda_{c}=2.251$ \textmu m; $\Delta \lambda=0.109$ \textmu m) filters in parallel with the lenslet-based integral field spectrograph \citep[IFS,][]{2008SPIE.7014E..3EC, 2015A&A...576A.121M} in the Y to H band ($0.96-1.64$ \textmu m). Only the IRDIS data were exploited because the low-resolution (R$\sim$30) IFS observations are superseeded by the SINFONI spectra (R$\sim1500-2000$) of the binary exploited in \cite{bonnefoy2009} and \cite{bonnefoy2014a}.  
 
 We acquired 240$\times$4s IRDIS frames of the binary. The IRDIS dataset was reduced at the SPHERE Data Center\footnote{http://sphere.osug.fr} (DC) using the SPHERE Data Reduction and Handling (DRH) automated pipeline \citep{2008SPIE.7019E..39P,delorme2017}. The DC carried out the basic corrections for bad pixels, dark current, and flat field.  It also includes correction for the instrument distortion \citep{2016A&A...587A..56M}.
 
 The wavefront-sensing of the adaptive optics system SAXO \citep{2006OExpr..14.7515F, 2014SPIE.9148E..0OP} could operate on the target in spite of its faintness at optical wavelengths \citep[V=13.8 mag; ][]{2005yCat.1297....0Z} and of the adverse observing conditions (Table \ref{table:observations}). But the tip-tilt mirror occasionally produced strong undesired offset of TWA~22 in the field of view and part of the sequence was affected by low Strehl ratio. We then selected by eye 71  frames with the best angular resolution. We measured the relative position of the binary in the remaining frames using a custom cross-correlation $\mathrm{IDL}$ script. The frames were then re-aligned using sub-pixel shifts with a tanh interpolation kernel. The registered frames were averaged to produce a final frame using the \texttt{Specal} pipeline (Galicher et al., in prep). 

\begin{figure}[h]
	\centering
	\includegraphics[width=\linewidth]{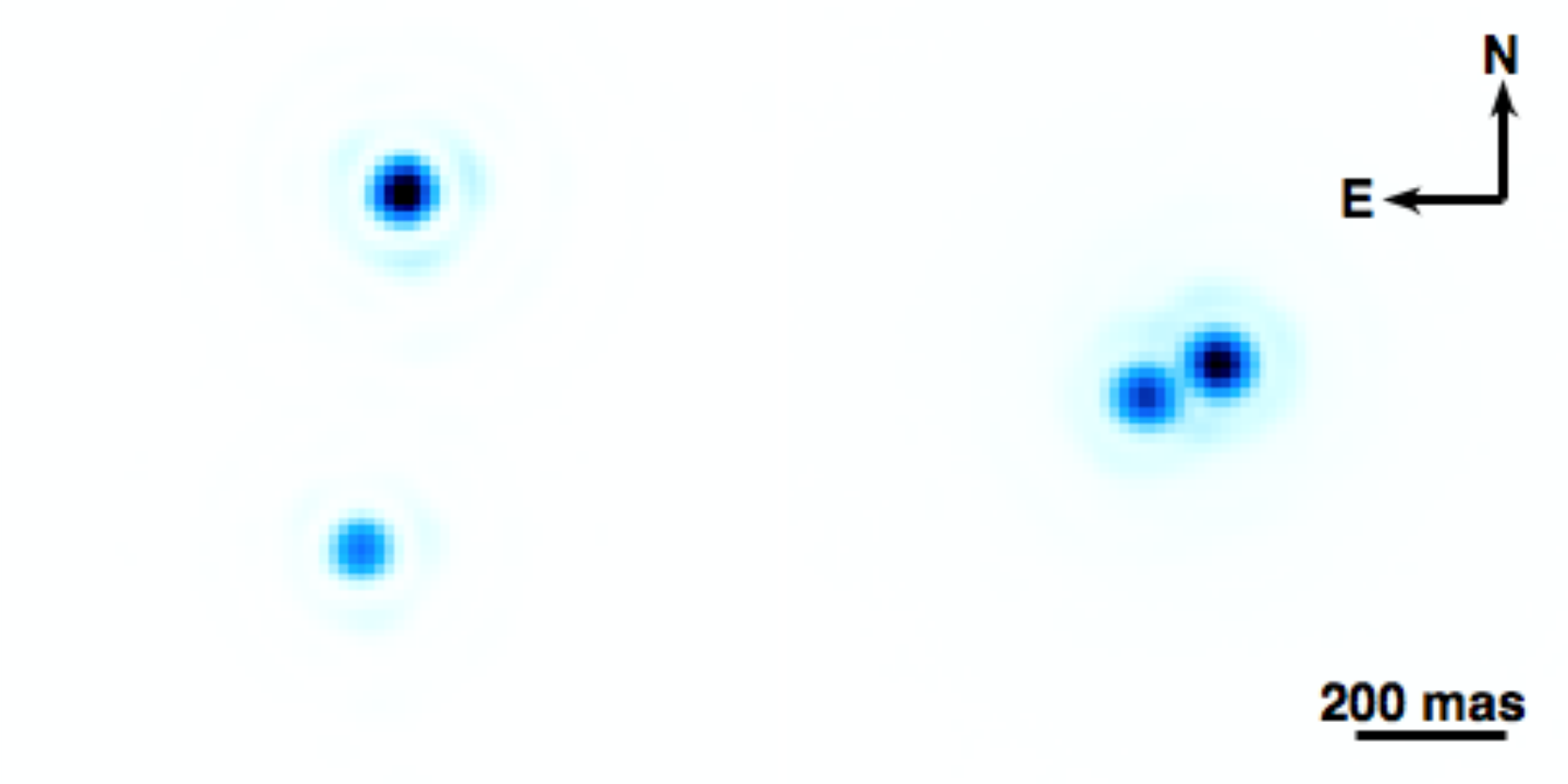}
    \caption{SPHERE/IRDIS K1 (\textlambda~= 2.11 nm) observations of GJ 2060 AB (left) and TWA 22 AB (right). They were taken respectively in February 2017 and February 2015. }
    \label{fig:Images}
\end{figure}

TWA~22 A and B are well resolved into the final K1 and K2 images (see Fig. \ref{fig:Images}). We did not observe any reference star to calibrate the point-spread-function so that we could not use deconvolution algorithm for that epoch. But the high Strehls of the SPHERE observations mitigate the cross-contamination of the binary components. We measured their position in the K1 image (offering the best angular resolution) fitting a Moffat function within an aperture mask (4 pixel radii) centered on the guessed position of the stars. We varied the aperture size ($\pm$1 pixels in radius) and considered alternative fitting function (Gaussian, Lorentzian) to estimate an error on the astrometry. We used a True North value of $1.72 \pm 0.06^{\circ}$ and a platescale of $12.267 \pm 0.009$ mas/pixel derived from the observations of $\Theta$ Orionis C as part of the long term analysis of the SHINE astrometric calibration \citep[same field as the one observed with NaCo; ][]{2016A&A...587A..56M, 2016SPIE.9908E..34M}. This leads to a position angle $PA=114.90\pm0.10^{\circ}$ and a separation $\rho=103\pm1$ mas between the two components of TWA~22. 

\subsection{GJ~2060}

	\subsubsection{NaCo data}
	
	We observed GJ~2060 with NaCo (Program 090.C-0698; PI Delorme) in the H-band in the course of a direct imaging survey of M-dwarfs \citep{2012A&A...539A..72D,lannier2016}.  The observations were performed in field tracking mode with the detector cube mode enabling for short integration time (0.15s).  We also observe the astrometric calibrator $\Theta$ Ori C with the same setup. 
	The data were all reduced with the  \texttt{eclipse} tool.  We find a platescale of $13.19\pm0.06$ mas/pixel and a True North value of $-0.60\pm0.33^{\circ}$ for those observations. GJ~2060~AB  was tight (69 mas) in the images. This required the use of a deconvolution algorithm to deblend the binary components. We reduced for that purpose the data of GJ~3305 observed the same night (Table~\ref{table:observations}). GJ~3305 is itself a tight pair of M-dwarfs and is a member of the Beta Pictoris moving group \citep{2001ApJ...562L..87Z}.  The separation of GJ~3305 in November 2012  (290 mas)  and the Strehl ratio were sufficient to mitigate the self-contamination of the binary component. We could then extract a subfield centered on GJ~3305~A that could serve as a reference PSF. We nonetheless used in addition three isolated bright stars from the $\theta$ Orionis field observed four nights before GJ~2060 at a close airmass as  to evaluate the dependency of the results related to the PSF choice. 
	The deconvolution algorithm of \cite{1998SPIE.3353..426V} yields a $PA=232.2\pm2.3^{\circ}$ and $\rho=69\pm2$ mas for GJ~2060~AB. This measurement is confirmed by the \texttt{Starfinder} tool.  
	
	\subsubsection{NIRC2 archival data}
	
	We collected and reduced two sets of archival data obtained in pupil-tracking mode with the Keck/NIRC2 adaptive optics instrument \citep{2004ApOpt..43.5458V} on October 1, 2015 (program N101N2; PI Mann) and November 18, 2015 (program H269N2; PI Gaidos). They were both obtained with the $K_{cont}$ filter ($\lambda_{c}=2.2706$ \textmu m, $\Delta \lambda=0.0296$ \textmu m).
	
	Both sequences contain two sets of frames which correspond each to a position of the source on the detector. We averaged each set of three frames to produce two resulting frames. Those resulting frames were then used to subtract the sky and bias contributions into the 6 original frames. We registered the frames on a common origin,  applied a rotation to re-align them to the North, and averaged them to produce the final frames. The last step enabled to filter out part of the bad pixels. 
		
	We fitted a Moffat function on each star flux distribution to retrieve their relative position. For both epochs, we  considered the platescale ($9.971 \pm 0.005$ mas/pixel) and  the absolute orientation on the sky ($0.262 \pm 0.022^{\circ}$) reported in \cite{service2016}. The estimated contrasts and astrometry are reported in Tables \ref{table:photGJ2060} and \ref{table:astromGJ2060}, respectively. 
	
	\begin{table}[h]
		\small
		\caption{Contrasts and apparent magnitude of GJ~2060~A and B}\label{table:photGJ2060}
		\centering
		\begin{tabular}{ccccc}
\hline\\[-0.8em]
Date	&	Band	 & Contrast 	&	GJ~2060~A	&	GJ~2060~B	\\
		&			&		(mag)	&	(mag)	&	(mag)	\\
		\hline \hline\\[-0.8em]
25/11/2012	&	H	&	$0.80\pm0.20$	& & \\
01/10/2015	&	$\mathrm{K_{cont}}$	& $0.95\pm0.10$	&	&	\\
18/11/2015	&	$\mathrm{K_{cont}}$	& $0.87\pm0.10$	&	&	\\
22/11/2013	&	$\mathrm{J_{synth}}$	&	$0.94\pm0.06$	&	&	\\
22/11/2013	&	$\mathrm{H_{synth}}$	&	$0.98\pm0.06$	&	$6.34\pm0.06$ & $7.32\pm0.09$	\\
22/11/2013	&	$\mathrm{Ks_{synth}}$	&	$0.90\pm0.06$	& $6.09\pm0.06$	&	$7.04\pm0.07$ \\
22/11/2013	&	$\mathrm{K1_{synth}}$	&	$0.90\pm0.06$	&	&	\\
22/11/2013	&	$\mathrm{K2_{synth}}$	&	$0.87\pm0.06$	&	&	\\
05/02/2015	&	K1	&	$1.00\pm0.04$	& $6.12\pm0.04$	&	$7.08\pm0.07$ \\
05/02/2015	&	K2	&	$0.90\pm0.04$	& $6.07\pm0.04$	&	$6.95\pm0.06$ \\
05/02/2015	&	$\mathrm{J_{synth}}$	&	$0.97\pm0.01$	&	$6.99\pm0.03$ & $7.96\pm0.04$	\\
29/11/2015	&	H2	&	$0.99\pm0.05$	&	 & 	\\
29/11/2015	&	H3	&	$0.99\pm0.05$	&	 & 	\\
26/12/2015 &	H2	&	$1.00\pm0.01$	&  $6.33\pm0.04$	&	$7.33\pm0.05$ \\
26/12/2015	&	H3	&	$0.99\pm0.02$	&	 & 	\\
27/03/2016	&	H2	&	$1.02\pm0.02$	&	 & 	\\
27/03/2016	&	H3	&	$1.00\pm0.01$	&  $6.33\pm0.04$	&	$7.33\pm0.05$ \\
07/02/2017	&	K1	&	$0.96\pm0.06$	& 	&	\\
07/02/2017	&	K2	&	$0.88\pm0.05$	&	&	\\
07/02/2017	&	$\mathrm{J_{synth}}$	&	$0.97\pm0.03$	&	&	\\
\hline
		\end{tabular}
	\end{table}

\begin{table*}[h]
		\small
		\caption{Summary of GJ 2060 astrometry.}\label{table:astromGJ2060}
		\centering
		\begin{tabular}{cccccccc}
			\hline\\[-0.8em]
			UT Date & Band & Position Angle & Separation & $\Delta$RA & $\Delta$Dec & Instrument & Reference \\
			 & & (deg) & (mas) & (mas) & (mas) & & \\
			\hline
			\hline\\[-0.8em]
			28/12/2002 & Kp & 180.3 $\pm$ 0.2 & 425 $\pm$ 4 & -2 $\pm$ 2 & -425 $\pm$ 4 &	Keck\_ NIRC2 & \cite{janson2014}\\
			30/11/2005 & H & 143.7 $\pm$ 1.5 & 175 $\pm$ 11 & 104 $\pm$ 11 & -141 $\pm$ 12 &	Gemini\_ NIRI & \cite{daemgen2007}\\
			12/11/2008 & z’ & 169.7 $\pm$ 0.3 & 479 $\pm$ 5 & 86 $\pm$ 4 & -471 $\pm$ 6 & Astralux & \cite{bergfors2010}\\
			31/01/2010 & z’ & 176.2 $\pm$ 0.3 & 458 $\pm$ 5 & 30 $\pm$ 3 & -457 $\pm$ 6 & Astralux & \cite{janson2012}\\
			25/10/2010 & z’ & 181.1 $\pm$ 0.3 & 423 $\pm$ 4 & -8 $\pm$ 3 & -423 $\pm$ 4 & Astralux & \cite{janson2014}\\
			06/01/2012 & z’ & 191.6 $\pm$ 0.3 & 294 $\pm$ 3 & -59 $\pm$ 3 & -288 $\pm$ 4 & Astralux & \cite{janson2014}\\
			25/11/2012 & H & 232.3 $\pm$ 3.0 & 69 $\pm$ 5 & -55 $\pm$ 4 & -42 $\pm$ 4 & NaCo & \cite{janson2014}\\
			05/02/2015 & IRD\_ EXT & 161.7 $\pm$ 0.2 & 393 $\pm$ 1 & 123 $\pm$ 2 & -373 $\pm$ 2 & SPHERE & This work\\
			16/03/2015 & z' & 162.3 $\pm$ 0.5 & 399 $\pm$ 4 & 121 $\pm$ 5 & -380 $\pm$ 5 & Astralux & This work\\
			01/10/2015 & Kc & 166.3 $\pm$ 0.2 & 439 $\pm$ 4 & 105 $\pm$ 3 & -426 $\pm$ 5 & Keck\_ NIRC2 & This work\\
			18/11/2015 & Kc & 167.1 $\pm$ 0.2 & 447 $\pm$ 4 & 101 $\pm$ 3 & -436 $\pm$ 5 & Keck\_ NIRC2 & This work\\
			29/11/2015 & IRD\_ EXT & 66.8 $\pm$ 0.1 & 449 $\pm$ 1 & 103 $\pm$ 1 & -437 $\pm$ 1 & SPHERE & This work\\
			25/12/2015 & z' & 167.5 $\pm$ 0.2 & 453 $\pm$ 2 & 98 $\pm$ 2 & -442 $\pm$ 3 & Astralux & This work\\
			26/12/2015 & IRD\_ EXT & 167.0 $\pm$ 0.1 & 453 $\pm$ 1 & 102 $\pm$ 1 & -441 $\pm$ 1 & SPHERE & This work\\
			28/12/2015 & z' & 167.5 $\pm$ 0.2 & 454 $\pm$ 2 & 98 $\pm$ 2 & -443 $\pm$ 3 & Astralux & This work\\
			27/03/2016 & IRD\_ EXT & 168.6 $\pm$ 0.1 & 463 $\pm$ 1 & 92 $\pm$ 1 & -454 $\pm$ 1 & SPHERE & This work\\
			07/02/2017 & IRD\_ EXT & 173.1 $\pm$ 0.2 & 478 $\pm$ 1 & 57 $\pm$ 2 & -474 $\pm$ 2 & SPHERE & This work\\
			\hline
		\end{tabular}
	\end{table*}

	\subsubsection{SPHERE observations}
	
	The binary was observed as part of the SHINE survey on February 2015 and February 2017 in field and pupil tracking mode, respectively. For both nights, the IRDIFS\_EXT mode was used. 
	
	The reduction of the IFS data was performed following the procedure described in \cite{2015A&A...576A.121M}  and \cite{2014A&A...572A..85Z}. The calibrated spectral datacubes are made of 39 narrow band images. We rotated the datacubes corresponding to each exposures to align them to the North and averaged them. We extracted from the resulting cube the flux ratio between each component of the binary for both epochs (74 mas wide circular aperture).
		
		We  made use of the IRDIS data for the astrometric monitoring. We followed the procedure  described in Section \ref{sec:SPHEREdatTWA22} to reduce those data.  We used the True North and platescale values reported in Section \ref{sec:SPHEREdatTWA22} for the 2015 observations. We adopted a true North on sky of $1.702\pm0.058$\degree~and a platescale of $12.250 \pm 0.009$ mas/pix from the observations of NGC3603 obtained on February 7, 2017 as part of the long-term astrometric calibration of the instrument \citep{2016SPIE.9908E..34M}. The binary position in the final images was measured with a Moffat function and is reported in Table \ref{table:astromGJ2060}. Fig. \ref{fig:Images} displays the 2017 epoch.
		
	\subsubsection{AstraLux observations}
	
	Three of the AstraLux data points presented here are previously unpublished. These were obtained as a continuation of the AstraLux orbital monitoring campaign for young M-dwarf binaries, with a particular focus on young moving group members \citep{janson2014,janson2017}. The new data were acquired in March and December of 2015 with the lucky imaging camera \texttt{AstraLux Sur} \citep{hippler2009} at the ESO NTT telescope (programs 094.D-0609(A) and 096.C-0243(B)). They were reduced in an identical way as previously in the survey \cite[e.g.,][]{janson2014}. For the March run, the cluster NGC 3603 was used as astrometric calibrator, giving a pixel scale of 15.23 mas/pixel and a true North angle of 2.9\degree. In the December run, the Trapezium cluster was used for astrometric calibration, yielding a pixel scale of 15.20 mas/pixel, and a true North angle of 2.4\degree.
		
	\subsubsection{SINFONI integral field spectroscopy} 
	
	GJ~2060~AB was finally observed on November 22, 2013 with the SINFONI instrument mounted on the VLT/UT4  as part of our dedicated program for the orbital characterization of dynamical calibrators (PI Bonnefoy; program ID 090.C-0819). SINFONI (Spectrograph for INtegral Field Observations in the Near Infrared) couples a modified version of the adaptive optics module MACAO \citep{2003SPIE.4839..329B} to the integral field spectrograph SPIFFI  \citep{2003SPIE.4841.1548E} operating in the near-infrared (1.10-2.45 \textmu m). SPIFFI slices the field of view into 32 horizontal slitlets that sample the horizontal spatial direction and rearranges them to form a pseudo long slit. That pseudo-slit is dispersed by the grating on the 2048 $\times$ 2048 SPIFFI detector.  GJ~2060~A was bright enough at R band to allow for an efficient adaptive optics correction. We used the pre-optics providing 12.5 mas $\times$ 25 mas rectangular spaxels on sky and a square field of view of 0.8" side.  The target was observed during two consecutive sequences with the J and H+K gratings, covering the $1.10-1.40$ and $1.45-2.45$ \textmu m wavelength range at  R$\sim$2000 and 1500 resolving powers, respectively.  We obtained 11 frames with the binary in the field of view. In-between each frame, the binary was dithered to increase the final field of view and filter out residual non-linear and hot pixels. We also obtained an exposure on the sky at the end of each sequence to efficiently subtract the sky emission lines, detector bias, and residual detector defects. The observatory obtained observations of HIP~036092 immediately after GJ~2060. HIP~036092 is a B8V star that was used to evaluate and remove the telluric absorption lines.  
	
	We used the version 3.0.0 of the ESO data handling pipeline \citep{2006NewAR..50..398A} through the workflow engine \texttt{Reflex} \citep{2013A&A...559A..96F} which allowed for an end-to-end automatized reduction. \texttt{Reflex} performed the usual cosmetic steps on the bi-dimensional raw frames (flat field removal, bad-pixel flagging and interpolation). Those steps rely on calibration frames taken the days following our observations. The distortion and wavelength scale were calibrated on the entire detector. The positions of the slitlets on the detector were measured and used to build the datacubes containing the spatial (X,Y) and spectral dimensions (Z). In the final step, the cubes corresponding to individual exposures were merged into a master cube. 
	
 	GJ~2060 is well resolved into the J and H+K mastercubes but the sources contaminate each other. We applied the \texttt{CLEAN3D} tool described in \cite{2017A&A...597A..91B} to deblend the sources at each wavelength. The PSF at each wavelength is built from the duplication of the profile of GJ~2060~A following a PA=0$^{\circ}$. The tool produced two datacubes where one of the two components of the system is removed.  We extracted the J and H+K band spectra of each component integrating the flux within circular apertures of radius 147 and 110 mas at each wavelength in the datacubes, respectively. We extracted the telluric standard star spectrum using the same aperture sizes and corrected its continuum with a 12120 K black body  \citep{1991Ap&SS.183...91T}. The hydrogen and helium lines were interpolated using a third order Legendre polynomial.  GJ~2060~A and B spectra could then be divided by the telluric standard star spectrum to correct for atmospheric absorptions. 
 	
 	We computed 2MASS J, H and K-band contrasts as well as the K1 and K2 SPHERE contrasts from GJ~2060~A and B spectra  prior to the telluric line correction (Table \ref{table:photGJ2060}). The H, K1 and K2 synthetic contrasts match those derived from the SPHERE and NaCo data within error bars. We therefore used the synthetic 2MASS H and K-band contrasts and the 2MASS magnitude of the system \citep{2003yCat.2246....0C} to retrieve the individual magnitudes of GJ~2060~A and B.  J-band contrasts could be extracted from the SPHERE IFS data. They agree with the one derived with SINFONI. We used the contrast value of the 2015 SPHERE data to derive the J-band magnitude of the system components. 
 	
 	The 2MASS J magnitudes could then be used to flux-calibrate the J-band spectra using the 2MASS filter response curves and tabulated zero points\footnote{https://www.ipac.caltech.edu/2mass/releases/allsky/doc/sec6\_4a.html}.  We used the K1 magnitude measured with VLT/SPHERE and a spectrum of Vega \citep{1985A&A...151..399M, 1985IAUS..111..225H}  to flux-calibrate the H+K spectra. 
 
\subsubsection{HARPS data}

High S/N spectra have been acquired with HARPS \citep{mayor2003}: 1 night in April 2014  (JDB = 2456774.493808)  and 5 nights in October 2016 (between JDB = 2457666.881702 and 2457671.850830). Each spectrum contains 72 spectral orders, covering the spectral window [3800 Å, 6900Å]. The spectral resolution is approximately 100 000. The S/N of the spectra is  $\approx 100$ at 550 nm. The number of spectra per night is 2 (consecutive), except for the first night, for which only one has been taken. The data as provided by HARPS’s Data Reduction Software 3.5 (DRS) were first processed with SAFIR, a home-built tool that uses the Fourier interspectrum method described in \cite{chelli2000} and in \cite{galland2005} to measure radial velocities of stars with high $v\sin i$. SAFIR also estimates other observables, such as the cross-correlation functions, as defined in \cite{queloz2001}, and the bisector velocity spans (BVS), R0HK indexes, etc. For a detailed description of SAFIR, see \cite{galland2005}.

 		\begin{table}[h]
		\caption{Summary of HARPS radial velocity measurements of the SB1 GJ~2060.}\label{table:HARPS}
		\centering
		\begin{tabular}{ccc}
			\hline\\[-0.8em]
			Obs. JD & Radial velocity \\
			-2454000 & (km/s)\\
			\hline
			\hline\\[-0.8em]
			2774.49 & 28.99 $\pm$ 0.01\\
			3666.88 & 28.34 $\pm$ 0.02\\
			3666.89 & 28.27 $\pm$ 0.02\\
			3668.86 & 27.86 $\pm$ 0.02\\
			3668.88 & 27.89 $\pm$ 0.02\\
			3669.88 & 27.91 $\pm$ 0.01\\
			3669.89 & 27.93 $\pm$ 0.01\\
			3670.89 & 27.91 $\pm$ 0.02\\
			3670.90 & 27.79 $\pm$ 0.02\\
			3671.84 & 28.18 $\pm$ 0.02\\
			3671.85 & 28.15 $\pm$ 0.02\\	
			\hline
		\end{tabular}
	\end{table}
	
The values obtained in October 2016 show a very strong dispersion, probably due to the high magnetic activity of the star. Indeed, the orbit of the binary is $\sim 8$ years long, so  that we do not expect the radial velocities to vary more than $\sim 0.01$ km/s within a few consecutive days, very different from the 0.40 km/s variation we observed. Moreover, we note a strong correlation between the star bisector and the radial velocity measurements. We will therefore add this noise to the instrument uncertainty. 

\subsubsection{FEROS data}

Ten radial velocity measurements  have been obtained using the Fiberfed Extended Range Optical Spectrograph \citep[FEROS;][]{kaufer1999} mounted at the ESO-MPG 2.2 m telescope at La Silla Observatory. The data reduction process is described therein. FEROS is an echelle spectrograph covering the wavelength range 3500 -- 9200 \AA~across 39 orders with R $\approx$ 48000. The measurements are reported in \cite{durkan2018}, as part of a radial velocity monitoring survey of young, low-mass binaries. They cover a 12 year span, from 2005 to 2017.

The jitter evidenced in the HARPS data (section 3.2.6) must be taken into account in the FEROS set. Thus, we combined quadratically this estimated activity-related noise (0.40 km/s) to each FEROS uncertainty. 	
	
\section{Spectrophotometric analysis}
\label{section:Spectrophotometry}

\begin{figure*}[h]
	\centering
   	\includegraphics[width=\linewidth]{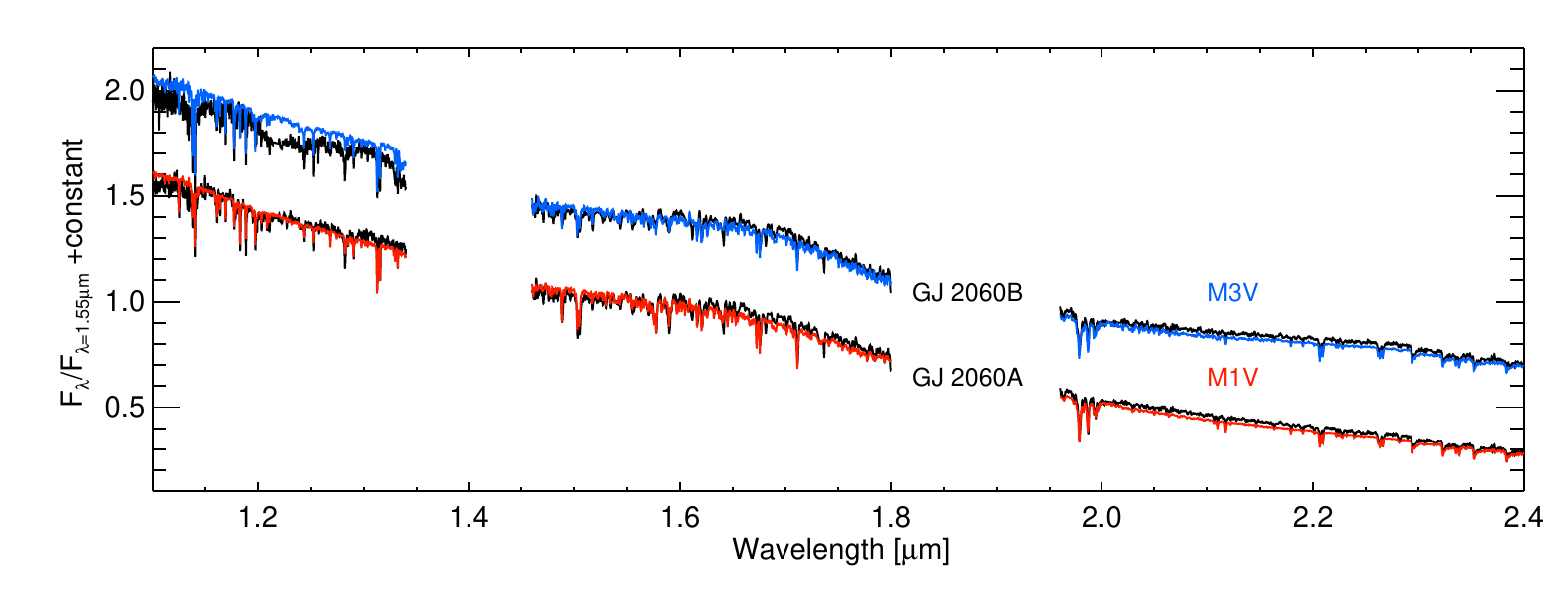}
    \caption{1.1-2.45 $\mu$m SINFONI spectra of GJ~2060~A and B renormalized at 1.55 $\mu$m.}
    \label{fig:Spectra}
\end{figure*}

We compared the SINFONI spectra of GJ2060~A and B to the medium-resolution (R$\sim$2000) spectra of K and M-dwarfs from the IRTF library \citep{2005ApJ...623.1115C, 2009ApJS..185..289R}. The 1.1-2.5 \textmu m spectral slopes of GJ~2060~A is best reproduced by the  Gl 229 A spectrum (M1V; Fig. \ref{fig:Spectra}). The detailed absorptions and slopes of the J-band and K-band spectra are also reproduced by that template (Fig. \ref{fig:SpectraJK}). The lack of water band absorption from 1.3 to 1.4 $\mu$m in the spectrum of GJ~2060~A confirms that the object has a spectral type earlier than M2.  The M0.5 and M1.5 dwarfs Gl 846 and Gl 205 fit equally well the K and H band spectrum of GJ~2060~A, respectively. Therefore, we estimate that GJ~2060~A is a M1$\pm$0.5 dwarf. 

\begin{figure*}[h]
	\centering
   	\includegraphics[width=0.8\linewidth]{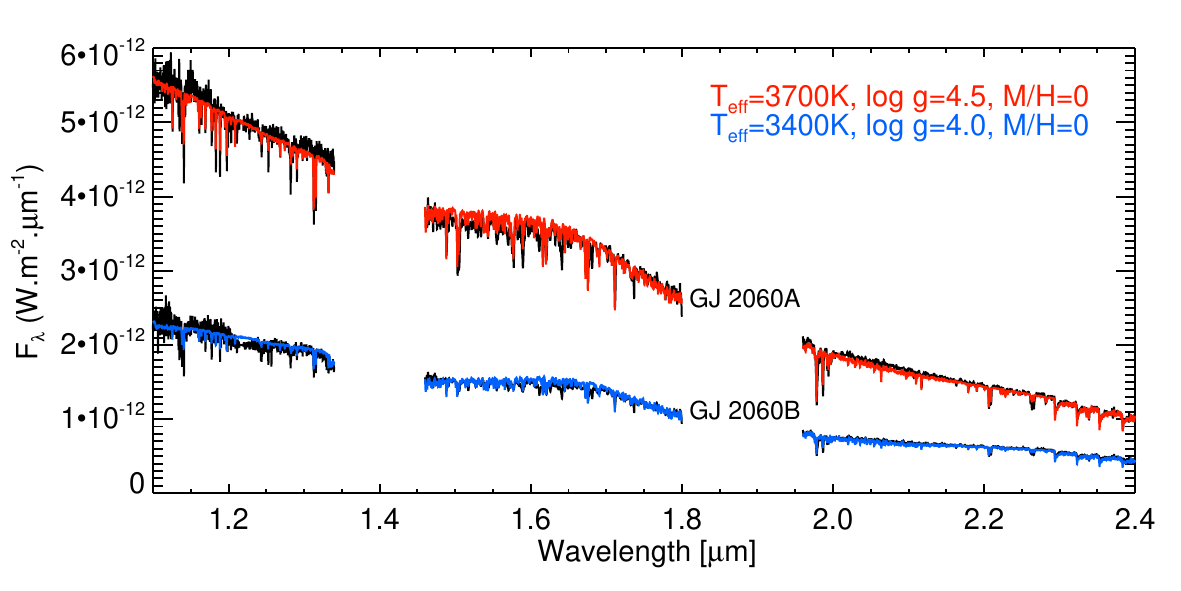}
    \caption{Spectra (apparent flux) of GJ~2060~A and B compared to the best fitting BT-SETTL synthetic spectra.}
    \label{fig:SynSpectra}
\end{figure*}

The spectral slope of GJ~2060~B is reproduced by the spectrum of the M3 dwarf Gl 388.  The  comparison at J-band  evidences departures from 1.1 to 1.2 \textmu m and 1.24 to 1.33 \textmu m between our object spectrum and the templates (Fig. \ref{fig:SpectraJK}). These departures are also evidenced in the SINFONI spectra of GJ~3305A and B obtained as part of our observation program \citep{durkan2018}. It likely arises from the SINFONI instrument. The multiple atomic lines (K I, Na I, Fe I, Al I) and  the water band absorption from 1.3 to 1.35 \textmu m indicate that the object has a spectral type later than M2. The H band spectrum is best represented by the one of the M3.5 dwarf Gl 273 while the K-band is perfectly reproduced by the spectrum of the M3 template (Fig. \ref{fig:SpectraJK}). We conclude that GJ~2060~B is a M3$\pm$0.5 dwarf.  

Those spectral types confirm the estimates made in \cite{bergfors2010} from the optical colors. We used them together with the bolometric corrections of  \cite{2013ApJS..208....9P} and the J-band magnitude (Tab. \ref{table:photGJ2060}) of each component to infer a $\mathrm{log(L/L_{\odot})=-1.20\pm0.05~dex}$ and $\mathrm{log(L/L_{\odot})=-1.63\pm0.05~dex}$ for GJ~2060~A and B, respectively.  

We performed a $\chi^{2}$ comparison of GJ~2060~A and B spectra to a grid of BT-SETTL atmosphere models \citep{baraffe2015} and show the best fitting solutions in Fig. \ref{fig:SynSpectra}. The grid covers $\mathrm{1500  \leq T_{eff} (K) \leq 5500}$ (in steps of 100 K), $\mathrm{2.5  \leq log \:g (dex) \leq 5.5}$ (in steps of 0.5 dex) and considers solar abundances. We  find $\mathrm{T_{eff}=3700\pm100}$ K and log g$>$4.0 dex for GJ~2060~A. Similarly, we find $\mathrm{T_{eff}=3400 \pm 100}$ K and log g$\geq$3.5 dex for GJ~2060~B. The $\mathrm{T_{eff}}$ are in good agreement with the estimates ( $\mathrm{T_{eff}=3615-3775}$ K for A and  $\mathrm{T_{eff}=3300-3475}$ K for B) derived from Table 5 of \cite{2013ApJS..208....9P} for the estimated spectral type of the binary components. As these results rely on atmosphere models, they do not depend on the system age. Both the $\mathrm{T_{eff}}$ and bolometric luminosities are used as input of evolutionary models for the calibration of their mass predictions in Section \ref{section:Models}. 

\section{Orbital fit and dynamical mass}
\label{section:MCMC}

Both systems orbits have been observed on several occasions covering a timespan longer than their periods, so that we are now able to derive precise estimates of their orbital elements. In both cases, we fit the relative orbit of the B component with respect to the A component, assuming a Keplerian orbit projected on the plane of the sky. In this formalism, the astrometric position of the companion can be written as:

\begin{align}
	x &= \Delta Dec = r\left(\cos(\omega  + \theta)\cos\Omega - \sin(\omega + \theta)\cos i \sin\Omega \right)\label{xMCMC}\\
	y &= \Delta Ra = r\left(\cos(\omega  + \theta)\sin\Omega + \sin(\omega + \theta)\cos i \cos\Omega \right)\label{yMCMC}
\end{align} 
	
\noindent where $\Omega$ is the longitude of the ascending node (measured counterclockwise from North), $\omega$ is the argument of the periastron, $i$ is the inclination, $\theta$ is the true anomaly, and $r = a(1-e^2)/(1+e\cos\theta)$ is the radius, where $a$ stands for the semi-major axis and $e$ for the eccentricity.

The orbital fit we performed uses the observed astrometries depicted in Table \ref{table:astromTWA22} and \ref{table:astromGJ2060} to derive probability distributions for elements $a$, $P$ (period), $e$, $i$, $\Omega$, $\omega$, and time for periastron passage $t_p$. Elements $a$ and $P$ are probed separately, so that we can deduce the probability distribution of the total mass as a by-product, as a function of the distance of the star $d$.
 
We used two complementary fitting methods, as described in \cite{chauvin2012}, (i) a least squares Levenberg-Marquardt (LSLM) algorithm to search for the model with the minimal reduced $\chi^2$, and (ii) a more robust statistical approach using the Markov-Chain Monte Carlo (MCMC) Bayesian analysis technique \citep{ford2005,ford2006} to probe the distribution of the orbital elements. Ten chains of orbital solutions were conducted in parallel, and we used the Gelman-Rubin statistics as convergence criterion \citep[see the details in][]{ford2006}. We picked randomly a sample of 500,000 orbits into those chains following the convergence. This sample is assumed to be representative of the probability (posterior) distribution of the orbital elements, for the given priors. We chose the priors to be uniform in x = $(\ln a,\ln P,e,\cos i,\Omega+\omega,\omega-\Omega,t_p)$ following \cite{ford2006}. For any orbital solution, the couples ($\omega$,$\Omega$) and ($\omega + \pi$,$\Omega + \pi$) yield the same astrometric data, this is why the algorithm fits $\Omega+\omega$ and $\omega-\Omega$, that are not affected by this degeneracy. The system distance has to be given to the algorithm. No input on the mass is needed, as it can be derived directly from $a$ and $P$ by Kepler's third law. The resulting MCMC distributions are well peaked when the data sample adequately the orbits, as is the case in this study. The complete set of posterior distributions and correlations are given in the appendix. 

\subsection{TWA~22}

\cite{bonnefoy2009} already performed an orbital fit of TWA~22 based on astrometric data from 2004 to 2007. The data covered at that time about three-quarters of a period. The authors used a pure Levenberg-Marquardt algorithm, that finds local minima and estimates the uncertainties from the resulting covariance matrix. We intend here to improve the orbital fit by using the new astrometric data (two periods are now covered) and a refined algorithm, described above, that allow a fine sampling of the phase parameters and a robust determination of the probability distributions.

The astrometric measurements gathered with NaCo on the system are particularly homogeneous and sample well the orbit. Therefore, we excluded the SPHERE point from the fit at first in order to avoid the possible bias associated with the change of instrument. We then checked the agreement between the results and the SPHERE point afterwards.

\begin{figure}[h]
	\centering
   	\includegraphics[width=\linewidth]{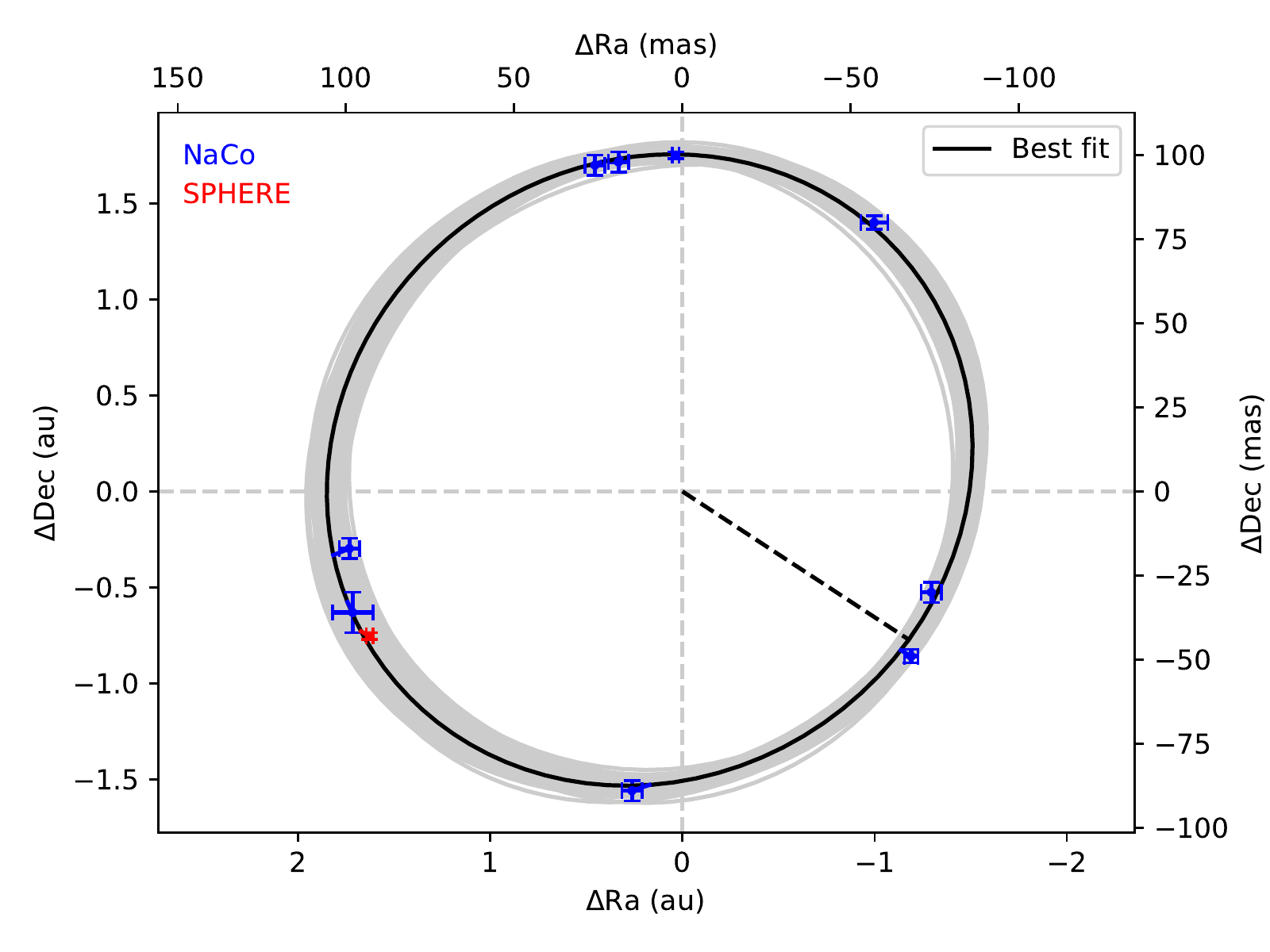}
    \caption{Plots of a hundred orbits obtained with the MCMC algorithm for system TWA~22. Astrometric measurements are displayed in their instrument color, along with their position on the fit. Only NaCo data are used in the orbital fit. The orbit in black, obtained with the LSLM algorithm, corresponds to the lower $\chi^2$.}
    \label{fig:FitTWA}
\end{figure}

The MCMC algorithm gives an estimate of the orbital elements (see Table \ref{StatisticsTWA}), with a precision of 0.02 on the eccentricity, 0.05 au on the semi-major axis, 0.04 yr for the period or 6\degree~on the inclination (see appendix A). The portrayed orbit has a low eccentricity ($\sim 0.1$) and inclination ($\sim 22 \degree$), as can be hinted from its on-sky representation in figure \ref{fig:FitTWA}. This figure shows the best fit obtained with the LSLM algorithm together with a hundred orbits picked up randomly within the 500,000 total sample used to derive the posteriors. The orbital elements derived by \cite{bonnefoy2009} are all retrieved within 1\textsigma.

\begin{table}[h]
	\centering
	\caption{Orbital elements from the MCMC fit of TWA~22 relative orbit, compared to the last orbit determination by 	\cite{bonnefoy2009}. The uncertainties on the fitted parameters  correspond to the 68\% interval of confidence of the distribution probabilities (see appendix A). The astrometric data  only allow determination of the couple $(\Omega,\omega)$  modulo \textpi.}\label{StatisticsTWA}
	\begin{tabular}{ccc}
		\hline\\[-0.8em]
		Parameter & This work & \cite{bonnefoy2009}\\
		\hline
		\hline\\[-0.8em]
		$a$ (au) & $1.72 \pm 0.05 \left(\frac{d}{17.5 \text{ pc}}\right)$ & $1.77 \pm 0.04$ \\
		$P$ (yr) & $5.35 \pm 0.04$ & $5.15 \pm 0.09$ \\
		$e$ & $0.13 \pm 0.02$ & $0.10 \pm 0.04$ \\
		$i$ (\degree) & $22 \pm 6$ & $27 \pm 5$\\
		$\Omega$ (\degree) & $129$ or $-51$  $\pm 18$ & $135 \pm 1$ \\
		$\omega$ (\degree) & $106$ or $-74$ $\pm 17$ & $100 \pm 10$\\
		$t_p$ (yr,AD) & $2006.04 \pm 0.07$ & $2006.04 \pm 0.01$ \\
		\hline
	\end{tabular}
\end{table}

The total system mass was computed from the semi-major axis and period corresponding to each orbit explored by the MCMC chains. For any distance $d$, we find a resulting total mass of $\mathrm{m_\text{tot} = {a^3}/{T^2} = 0.179 \pm 0.018~M_\odot} \left(\frac{d}{17.5 \text{ pc}}\right)^3$. Using the parallax distance and propagating its uncertainty, we finally obtain a dynamical mass of $\mathrm{m_\text{tot} = 0.18 \pm 0.02~M_\odot}$ for the pair.

We checked the consistency between the fitted orbit and the SPHERE point that we did not consider: the astrometry falls within the 68\% of confidence interval of the orbital fit, between 0.4 and 0.9 \textsigma~ from the probability peak (2-3\degree~in position angle, 2-3 mas in radius). Running the algorithm with this extra data point give very similar orbital elements (all well inside the 68\% confidence interval). It yields to the same dynamical mass, with smaller error-bars ($\mathrm{0.18 \pm 0.01~M_\odot}$).

A dynamical mass of $\mathrm{m_\text{tot} = 0.21 \pm 0.02~M_\odot}$ was obtained in \cite{bonnefoy2009} with less than 4 years coverage via a LSLM algorithm. This value is close to the one we obtain, but our peak value is outside the 1 \textsigma~confidence interval. However, the error bars on the orbital elements in \cite{bonnefoy2009} may be slightly underestimated, as they are roughly estimated from the covariance matrix. The present determination should therefore be more robust.

\subsection{GJ~2060}

Radial velocity measurements (RVs) from HARPS and FEROS \citep{durkan2018} help here refining the orbital fit. The binary is not resolved by the spectrometers (SB1). We only considered the FEROS data to get homogeneous measurements. This is legitimate, as taking into account HARPS data would not bring significant constraints. Indeed, HARPS data come down to two epochs (April 2014 and October 2016) that are close to FEROS epochs (see Fig. \ref{fig:RV}), and we have to fit an additional RV offset if we want to include data from another instrument.

The code we use is a slightly modified version of the code used for TWA~22, similar to the code used in \cite{bonnefoy2014b} for \textbeta~Pic b. In addition to the orbital elements, it evaluates the probability distributions of the offset velocity $v_0$ and amplitude $K$ of the radial velocity, with a prior uniform in $(v_0,\ln K)$ assumed for these extra variables \citep{ford2006}. In the formalism described previously, assuming a Keplerian orbit, the radial velocity is

\begin{equation}
	v_\text{rad} = K \frac{\cos{\omega}(\cos{\theta}+e)-\sin{\omega}\sin{\theta}}{\sqrt{1-e^2}} + v_0.\label{vrad}
\end{equation}

\noindent If the binary is a pure SB1, the amplitude derives from the fractional secondary mass $\mathbf{m_B/m_\text{tot}}$ as

\begin{equation}
K = \frac{2\pi}{P}\frac{m_B}{m_\text{tot}} a\sin i.\label{Amplitude}
\end{equation}

\noindent The introduction of the radial velocity breaks the degeneracy of the couple $(\Omega,\omega)$ and unique values can thus be derived for these two variables.

\begin{figure}[h]
	\centering
   	\includegraphics[width=\linewidth]{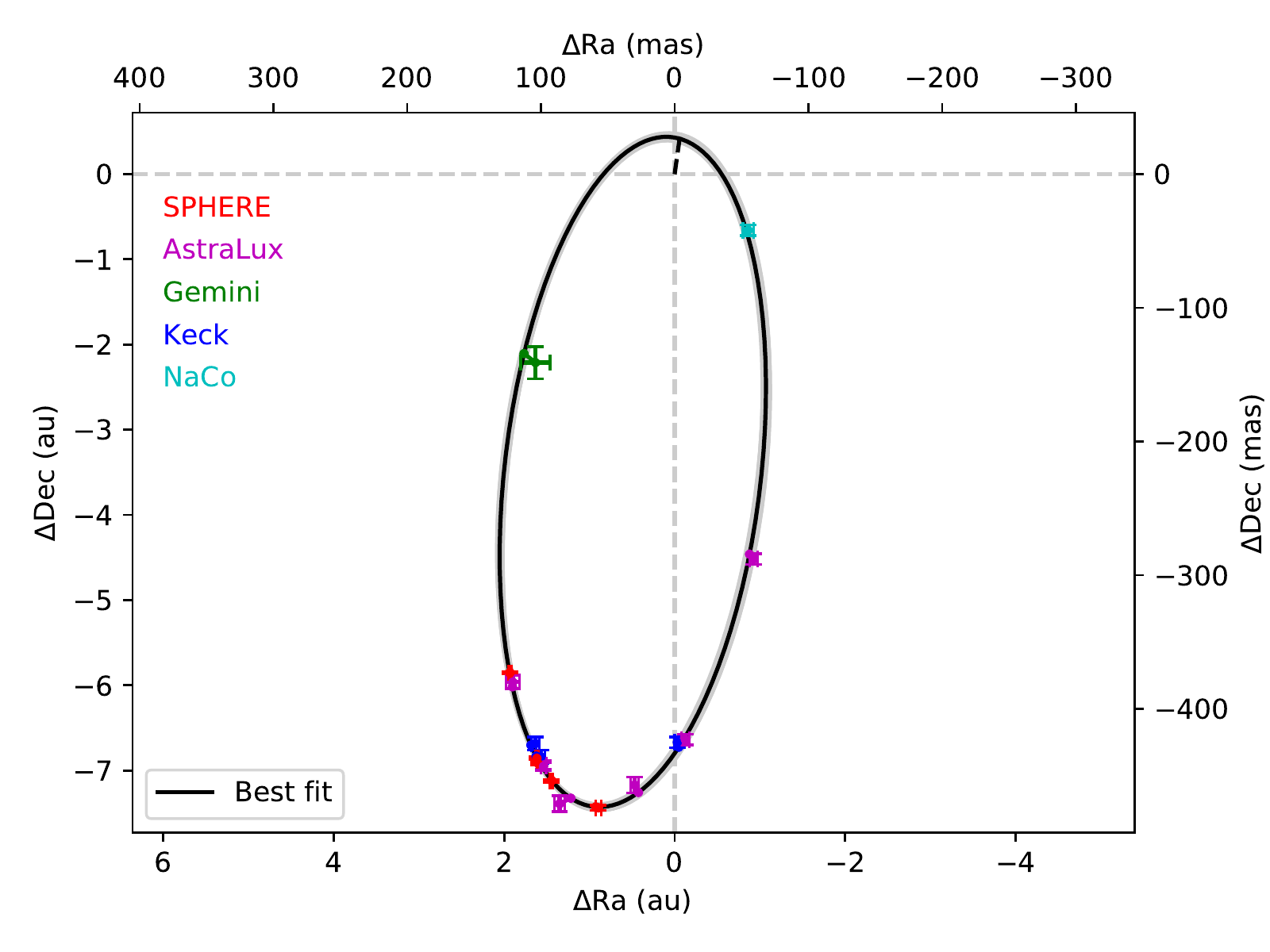}
    \caption{Plots of a hundred orbits obtained with the MCMC algorithm for system GJ~2060. Astrometric measurements are displayed in their instrument color, along with their position on the fit. The orbit in black, obtained with the LSLM algorithm, corresponds to the lower $\chi^2$.}
    \label{fig:Fit}
\end{figure}

\begin{figure}[h]
	\centering
   	\includegraphics[trim = {2cm 15cm 1.5cm 2cm},width=\linewidth]{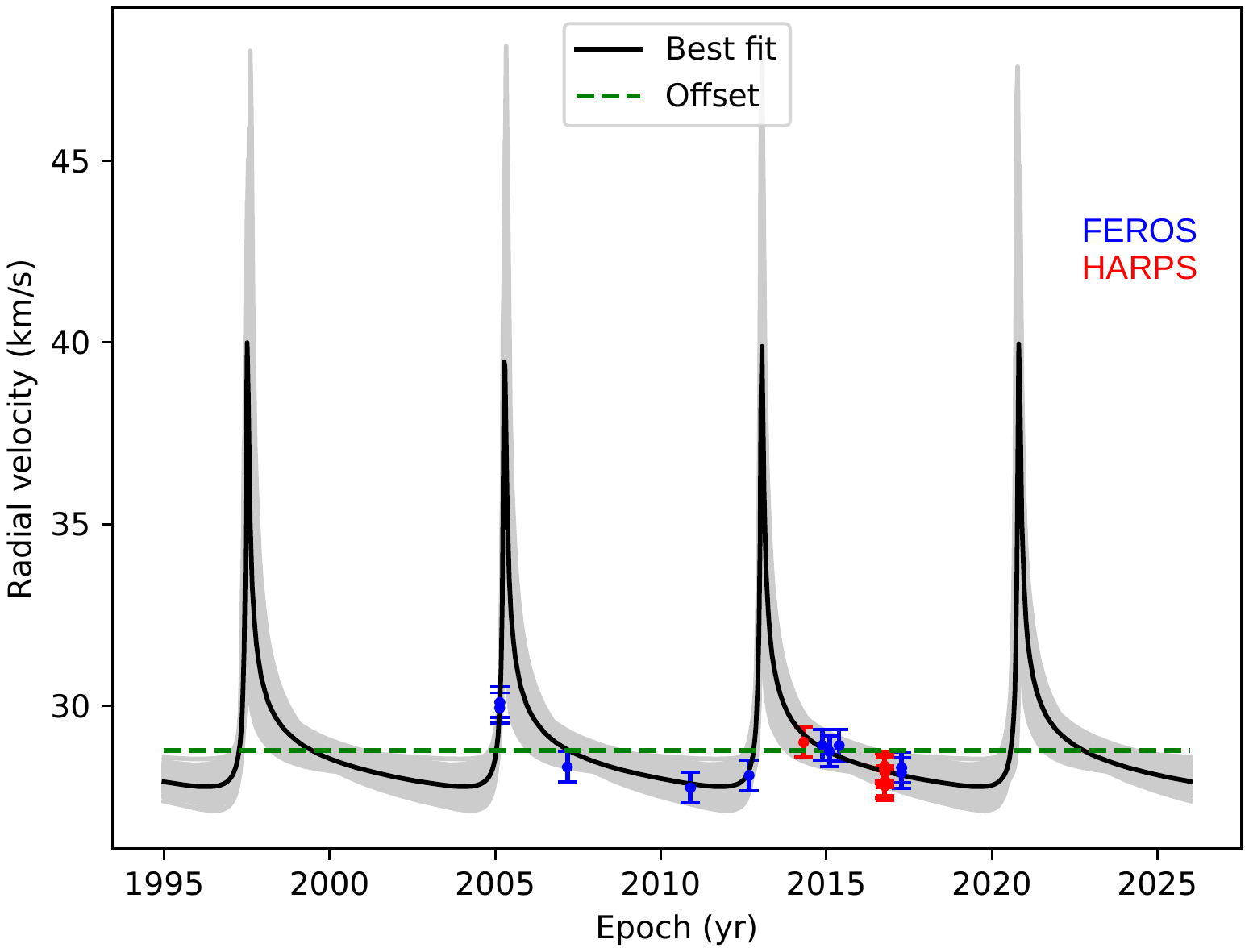}
    \caption{Plots of a hundred radial velocity evolution obtained with the MCMC algorithm for system GJ~2060. Radial velocity measurements are displayed in their instrument color, along with their position on the fit. Only FEROS data are used in the orbital fit. The orbit in black, obtained with the LSLM algorithm, corresponds to the lower $\chi^2$.}
    \label{fig:RV}
\end{figure}

The astrometric data are more numerous than in the case of TWA~22, but less homogeneous. Therefore, small systematic errors may bias the orbital fit (Table \ref{table:astromGJ2060}). Such errors are discussed in section 6.1. These 15 years of data cover approximately twice the relative orbit, but the passages near the periastron are not very well constrained and suggest a very quick displacement in that zone, hinting for a high eccentricity. The results of the MCMC algorithm are displayed in Table \ref{Statistics}. The distribution of orbital elements are very peaked, especially the one on the eccentricity (see appendix B). Indeed, we obtain a precision of 0.01 on the eccentricity, 0.04 au on the semi-major axis, 0.04 yr on the period and 3\degree~on the inclination. Noticeably, the eccentric distribution  peaks at $e=0.89$, but does not extend up to $e=1$: the components are bound. This orbital elements, and in particular the eccentricity, are very robust, and we obtain the same constraint when we fit only the astrometry. A hundred orbits, selected randomly within the 500,000 orbits used in to derive the posteriors, are plotted on Fig. \ref{fig:Fit}. Fig. \ref{fig:RV} displays the radial velocity data. The portrayed orbit confirms the very high $<1$ eccentricity.

\begin{table}[h]
	\centering
	\caption{Orbital elements from the MCMC fit of GJ~2060~AB relative orbit. The uncertainties on the fitted parameters  correspond to the 68\% interval of confidence of the distribution probabilities (see appendix B).}\label{Statistics}
	\begin{tabular}{ccc}
		\hline\\[-0.8em]
		Parameter & This work \\
		\hline
		\hline\\[-0.8em]
		$a$ (au) & $4.03 \pm 0.03 \left(\frac{d}{15.69 \text{ pc}}\right)$\\
		$P$ (yr) & $7.77 \pm 0.03$\\
		$e$ & $0.89 \pm 0.01$\\
		$i$ (\degree) & $36 \pm 3$\\
		$\Omega$ (\degree) & $8 \pm 4$\\
		$\omega$ (\degree) & $-20 \pm 5$\\
		$t_p$ (yr,AD) & $2005.27 \pm 0.03$\\
		$v_0$ (km/s) & $28.8 \pm 0.2$\\
		$K$ (km/s) & $2.3 \pm 0.9$\\
		\hline
	\end{tabular}
\end{table}

For any distance $d$, we infer a dynamical mass of $m_\text{tot} = 1.09 \pm 0.03~\mathrm{M_\odot}  \left(\frac{d}{15.69 \text{ pc}}\right)^3$ for the pair. The fractional mass could be computed for each orbit thanks to the fit of the radial velocity amplitude (Eq. \ref{vrad}). Considering our system as a pure SB1 (Eq. \ref{Amplitude}), we obtain a fractional mass of $m_\text{B}/m_\text{tot}  = 0.26 \pm 0.10  \left(\frac{15.69 \text{ pc}}{d}\right)$. However, this naive approach is questionable giving the flux ratio of the two components at FEROS wavelengths ($\sim$ 0.25). Thus, we used the method proposed by \cite{montet2015} and considered our RVs to be the flux-weighted sum of the two individual RVs. The amplitude $K$ fitted by the orbital fit could then be written as

\begin{align}
	K &= (1-F)K_A - FK_B\\
	&= \frac{2\pi}{P} a \sin{i}\left((1-F)\frac{m_B}{m_\text{tot}}-F\frac{m_A}{m_\text{tot}}\right)
\end{align}

\noindent where $F = L^V_B/(L^V_A+L^V_B) = 0.2$, $L^V_A$ and $L^V_B$  are the components luminosities in the visible spectrum and $K_A$ and $K_B$ are respectively the amplitudes from A and B. From this relation,  for any distance $d$, we obtain a fractional mass of  $m_\text{B}/m_\text{tot}  = 0.46 \pm 0.10  \left(\frac{15.69 \text{ pc}}{d}\right)$. Using the parallax distance and propagating its uncertainty, we finally obtain $m_\text{tot} = 1.09 \pm 0.10~\mathrm{M_\odot}$ and $\frac{m_\text{B}}{m_\text{tot}} = 0.46 \pm 0.10$.

The uncertainty on the total mass mainly comes from the uncertainty on the distance $d$, as $\Delta m_\text{tot}/m_\text{tot} = 3 \Delta d/d$. $d$ and $\Delta d$ derive from the parallax released within the new reduction of Hipparcos data \citep{vanleeuwen2007}. The binarity of the system was taken into account in Hipparcos reduction through two additional variables in the proper motion fit. Moreover, the high eccentricity of the orbit prevents a good sampling of the radial velocity, in particular close to the periastron passage. That leads to high error bars in the inclination and velocity amplitude $K$. These errors propagate on the fractional secondary mass (see equation \ref{Amplitude}). A higher accuracy on the orbital elements determination would certainly be achieved if the periastron passage was sampled in the available astrometric and spectroscopic data. This is unfortunately not the case yet.

On the other hand, the uncertainty on the fractional mass mainly comes from the very low constraints provided by the RVs. During most of the orbital revolution, the RV variation has a similar magnitude than the noise evidenced by the HARPS measurements. Only the sampling of the periastron passage could provide meaningful points that can refine the fractional mass. The next passage corresponds to October 2020. We can also notice that the difference is very significant between the naive (SB1) and corrected (flux-weighted) approach: the fractional mass nearly doubles. Averaging the flux-weighted RVs is a first order method, and is probably not precise enough to disentangle the two lines in our case, where the luminosity of the secondary is not negligible compared to the primary. A more refined method \citep[e.g.,][]{czekala2017} would be necessary to trace back the individual RVs from our measurements and compute a robust fractional mass. Thus, we use only the total mass in the next sections.

\section{Comparison to the models}
\label{section:Models}

Both our systems now have a dynamical mass and an estimated age given by their membership to moving groups, as well as a robust estimate of their bolometric luminosities $L$ and effective temperatures $T_\text{eff}$. Thus, we are able to probe the accuracy of the PMS evolutionary models at these mass ranges.

There exist several evolutionary models for PMS stars, relying on different physics (e.g., atmospheric models, convection efficiency). Two of them are suitable for 0.1 M$_\odot$ objects, the DM97 model \citep{dantona1997} and the BHAC15 model \citep{baraffe2015}. Four more models becomes suitable for higher-mass PMS stars: the SDF00 model \citep{siess2000}, the PISA model \citep{tognelli2011,tognelli2012}, the PARSEC model \citep{bressan2012} with \cite{chen2014}'s corrections for low-mass stars, and the Darmouth model \citep{dotter2008} with \cite{feiden2016}'s integrations of magnetic field. Testing the predictions of different models enables to compare the relevance of their approach, and thus to achieve a better understanding of the underlying physics.
 
\subsection{TWA~22}

According to the previous sections, TWA~22 has a total dynamical mass of $0.18 \pm 0.02~\mathrm{M_\odot}$ and an age of $\sim$25 Myr. We first considered the isochrones and iso-masses predicted by evolutionary models in a $(T_\text{eff},L)$ plane. We used the bolometric luminosities and effective temperatures derived by \cite{bonnefoy2014a}. Fig. \ref{fig:Isochrones} compares our observed $T_\text{eff}$ and $L$ to the BHAC15 tracks. The two components, A and B, are not located on the same isochrone, the primary on 10 Myr and the secondary 20 Myr, but their positions are consistent with coevality between 10 and 25 Myr within 1 sigma. On the other hand, the predicted masses are respectively around 0.06 and 0.07 $\mathrm{M_\odot}$ for A and B, at the lower end of the stellar regime, which gives a total mass of 0.13 $\mathrm{M_\odot}$. Nevertheless, when we impose coevality at the moving group age and allow for a shift of $T_\text{eff}$ within the 1\textsigma~interval, we retrieve the total dynamical mass, with masses of about 0.08 and 0.1 $\mathrm{M_\odot}$ for A and B. The corresponding diagram is shown in the appendix for the DM97 model. Underprediction of the total mass and non coevality are again retrieved, but once again the discrepancy disappears when we impose coevality at the moving group's age and allow for a shift of $T_\text{eff}$.

\begin{figure}[h]
	\centering
   	\includegraphics[width=\linewidth]{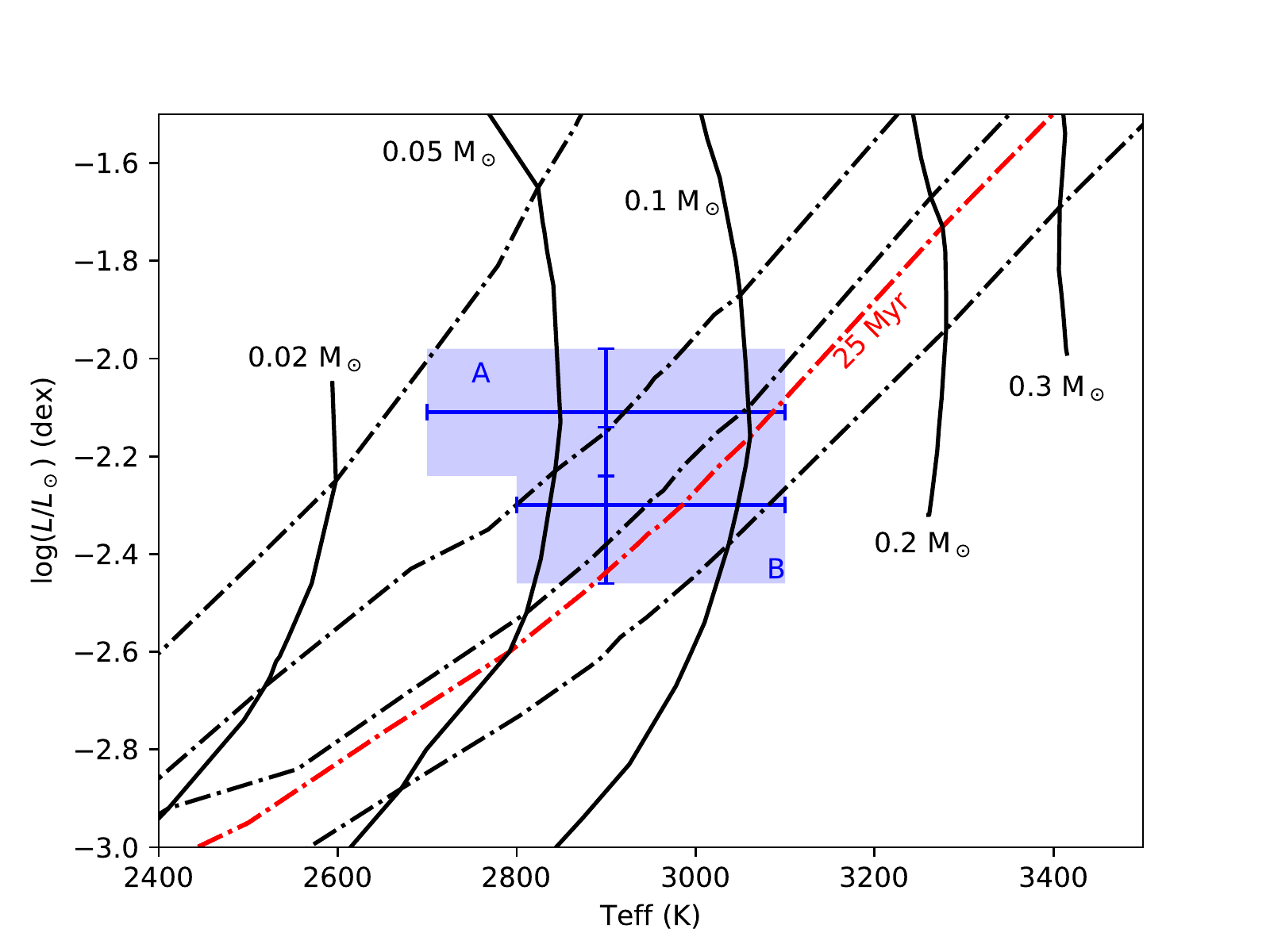}
    \caption{Isochrones and iso-mass curves predicted by BHAC15. The 1, 10, 20, 25 and 50 Myr isochrones have been drawn in dotted lines (1 Myr is up), while the iso-mass curves are in solid line . The 25 Myr isochrone correspond to the age of the $\beta$Pic-MG, and is drawn in red. The blue patterns correspond to the observed values and their error bars for each component of system TWA 22, A and B.}
    \label{fig:IsochronesTWA}
\end{figure}

\begin{figure}[h]
	\centering
   	\includegraphics[width=\linewidth]{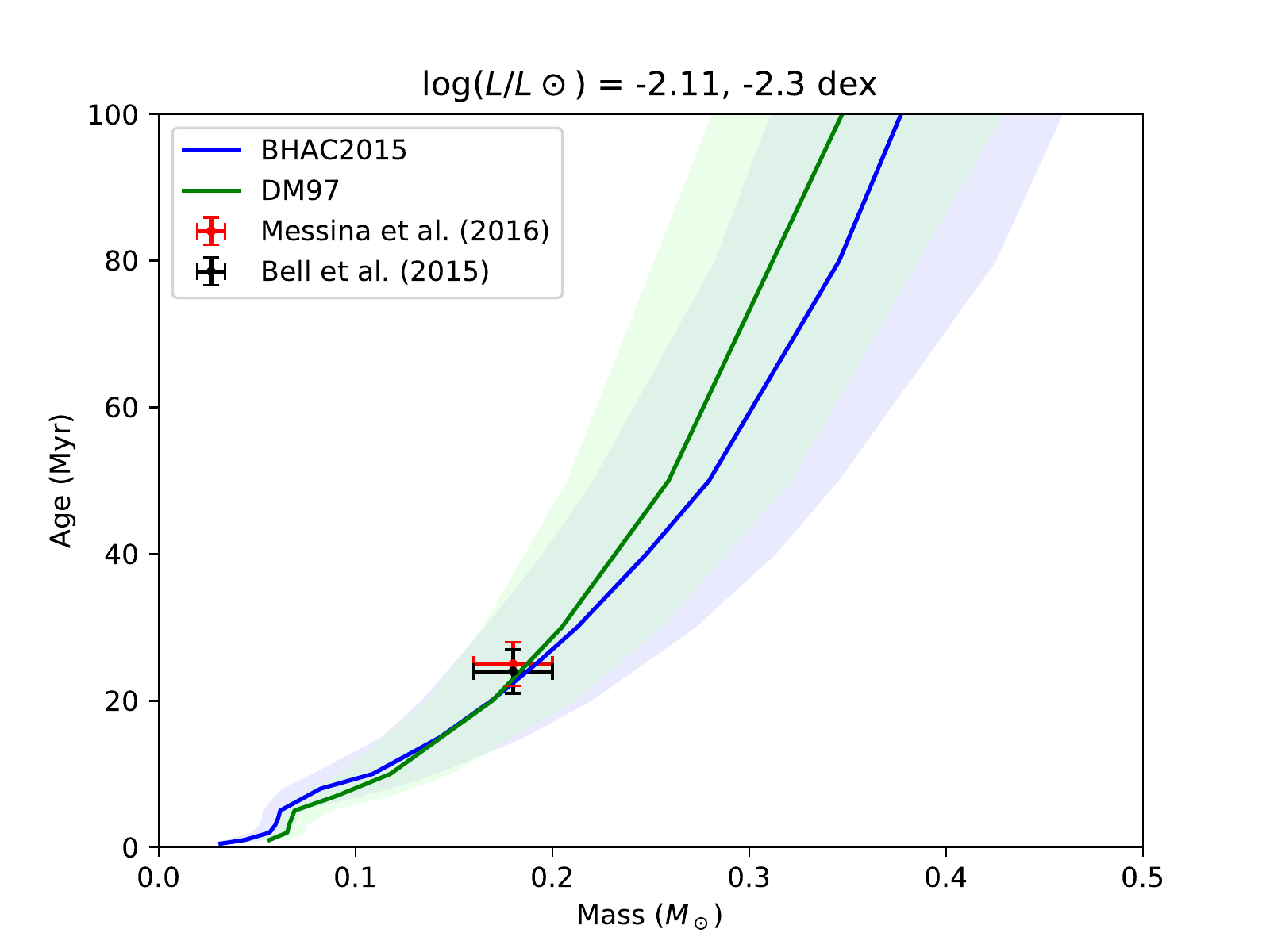}
    \caption{Comparison of TWA 22 direct mass measurement
for two different \textbeta Pic-MG age estimations with the predicted
masses of the BHAC15 and DM97 tracks derived from the bolometric luminosity. Errors on the photometry are propagated on predictions (colored zone). The error on the distance is taken here into account.}
    \label{fig:ModeleTWA}
\end{figure}

As opposed to the bolometric luminosity and dynamical mass, the effective temperature predicted by the models is not robust, as it depends strongly on the atmosphere model. For each component, we thus used the measured luminosity to compute the predicted mass for a range of ages with the BHAC15 and DM97 models. The corresponding plot is displayed in Fig. \ref{fig:ModeleTWA}, data have been linearly interpolated where necessary to provide predictions at the required ages. The prediction at the moving group age is consistent with the dynamical mass.

\subsection{GJ~2060}

According to the previous sections, GJ~2060 has a total dynamical mass of $m_\text{tot} = 1.09 \pm 0.10~\mathrm{M_\odot}$, and its age estimate can go from 30 to 200 Myr.

\begin{figure}[h]
	\centering
   	\includegraphics[width=\linewidth]{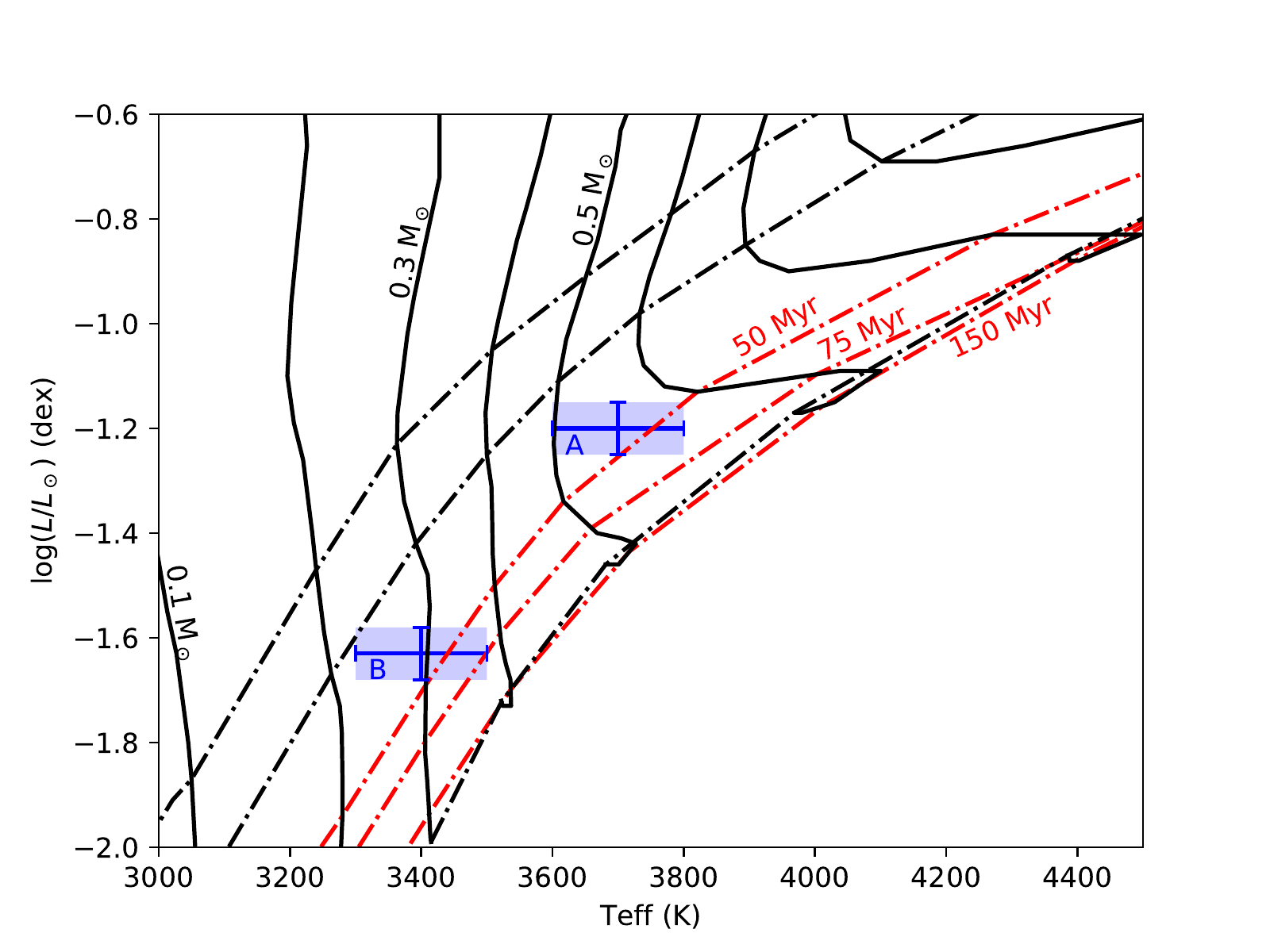}
    \caption{Isochrones and iso-mass curves predicted by BHAC15. The 10, 20, 50, 75, 150 and 600 Myr isochrones have been drawn with dashed lines (1 Myr is up), while one iso-mass is drawn in solid line every 0.1 $\mathrm{M_\odot}$ from 0.1 (left) to 1 $\mathrm{M_\odot}$ (right). The 50, 75 and 150 Myr isochrones correspond to possible ages for ABDor-MG, and are drawn in red. The blue patterns correspond to the observed values and their error bars for each component of system GJ~2060, A and B.}
    \label{fig:Isochrones}
\end{figure}

We first considered the isochrones and iso-masses predicted by evolutionary models in a $(T_\text{eff},L)$ plane. Fig. \ref{fig:Isochrones} compares our observed $T_\text{eff}$ and $L$ to the BHAC15 tracks. The two components, A and B, are located on the same isochrone, around 40 Myr, which is consistent with the younger estimations of the ABDor-MG age. On the other hand, the predicted masses are respectively around 0.55 and 0.3 $\mathrm{M_\odot}$ for A and B, which gives a total mass of 0.85 $\mathrm{M_\odot}$, far (2\textsigma) from the 1.09 $\mathrm{M_\odot}$ obtained by the orbital fit. We tried to impose a total mass of 0.85 $\mathrm{M_\odot}$ in the orbital fit in order to evaluate how this would change the distribution of $\chi^{2}$. In that case, this leads to orbits with $\chi^2_\text{red}>8.5$ for a distance of 15.69 pc, and $\chi^2_\text{red}>2.5$ for the 1 \textsigma~distance 15.24 pc, compared to 0.5 when the mass is set free. Therefore, the predicted mass does not account for the astrometry of the system.

The corresponding diagrams are shown in the appendix for all the other models. Underprediction of the total mass are retrieved in each case. Coevality is sometimes only marginally achieved (PARSEC), and a very young age can be predicted (20 Myr, DM97).

\begin{figure}[h]
	\centering
   	\includegraphics[width=\linewidth]{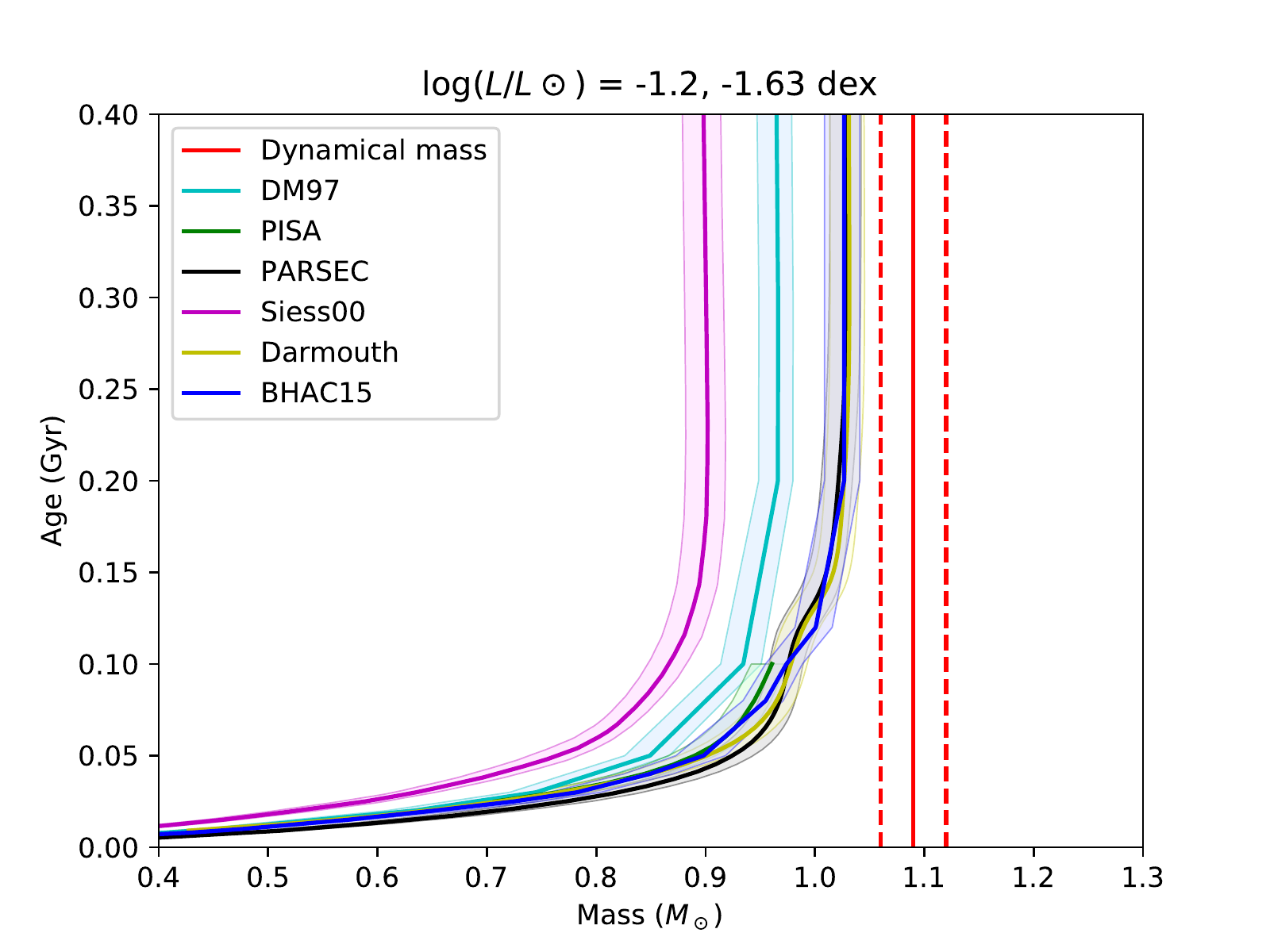}
    \caption{Mass age relations according according to the six different evolutionary models, for GJ~2060 observed luminosities. A distance of 15.69 pc is assumed. The dynamical mass and its uncertainty are depicted in red.}
    \label{fig:M1M3}
\end{figure}

As in the TWA 22 case, we then used each component measured luminosity to compute the predicted mass for a range of ages with the six models (BHAC15, DM97, PARSEC, PISA, Darmouth and SDF00) and we deduce a plot linking the mass and age for the observed luminosity.  These plots are displayed in Fig. \ref{fig:M1M3}. We retrieve the $\sim$ 20\% underestimation of the total mass (2\textsigma~deviation) if a young age is assumed. Conversely, an old age (> 150 $\mathrm{M_\odot}$) gives a mass marginally compatible (1\textsigma) with the orbital fit. 

From the plots, we computed the predicted mass for each model and different ages of the AB Dor moving group. The results are displayed in Table \ref{Ages}. In order to avoid summing correlated errors, we drew the mass-age relation for several distances, and determined the system mass in each case. We computed the spread and deduced the uncertainty due to the distance \textsigma$_d$. For the most probable distance, we then derived the age uncertainty due to the luminosities \textsigma$_L$. The final ages uncertainties are then the quadratic sum of the independent errors \textsigma$_L$ and \textsigma$_d$. Only the >100 Myr case fits marginally within the 68 \% interval of confidence of the MCMC probability distribution of the dynamical mass. That age is inconsistent with the positions of both stars on the temperature-luminosity diagram for all models, except for PARSEC. 

\begin{table}[h]
	\centering
	\small
	\caption{Predicted mass (in solar mass) for system GJ~2060 depending on the evolutionary model, from its luminosity and for several assumed ages. The error on the distance is taken here into account. The dynamical mass is $1.09 \pm 0.10~\mathrm{M_\odot}$.}\label{Ages}
	\renewcommand{\arraystretch}{1.5}
	\begin{tabular}{lccccc}
		\hline
		Model &  50 Myr & 75 Myr & 100 Myr & 150 Myr & 200 Myr\\
		\hline
		\hline
		BHAC15 & 0.90$^{+0.03}_{-0.04}$ & 0.95$^{+0.03}_{-0.04}$ & 0.97$^{+0.03}_{-0.03}$ & 1.01$^{+0.03}_{-0.03}$ & 1.03$^{+0.02}_{-0.04}$\\
		PISA &  0.89$^{+0.03}_{-0.04}$ & 0.94$^{+0.03}_{-0.04}$ & 0.96$^{+0.03}_{-0.03}$ & x & x\\
		PARSEC & 0.92$^{+0.05}_{-0.03}$ & 0.97$^{+0.02}_{-0.04}$ & 0.98$^{+0.02}_{-0.04}$ & 1.01$^{+0.03}_{-0.03}$ & 1.02$^{+0.03}_{-0.02}$\\
		SDF00 & 0.76$^{+0.05}_{-0.04}$ & 0.83$^{+0.03}_{-0.03}$ & 0.87$^{+0.03}_{-0.04}$ & 0.90$^{+0.03}_{-0.04}$ & 0.90$^{+0.03}_{-0.03}$\\
		DM97 & 0.85$^{+0.03}_{-0.04}$ & 0.89$^{+0.03}_{-0.03}$ & 0.93$^{+0.05}_{-0.03}$ & 0.95$^{+0.03}_{-0.03}$ & 0.97$^{+0.01}_{-0.04}$\\
		Darmouth & 0.90$^{+0.05}_{-0.04}$ & 0.96$^{+0.03}_{-0.04}$ & 0.98$^{+0.03}_{-0.04}$ & 1.02$^{+0.03}_{-0.04}$ & 1.03$^{+0.03}_{-0.03}$\\
		\hline
	\end{tabular}
\end{table}

\begin{table}[h]
	\centering
 	\small
	\caption{Mean predicted mass (in solar mass) for GJ~2060~A and B, from their luminosities.}\label{Masses}
	\begin{tabular}{lccccc}
		\hline\\[-0.8em]
		 Component &  50 Myr & 75 Myr & 100 Myr & 150 Myr & 200 Myr\\
		\hline
		\hline\\[-0.8em]
		GJ~2060~A & 0.55 & 0.55 & 0.55 & 0.57 & 0.57\\
		GJ~2060~B &  0.32 & 0.37 & 0.40 & 0.41 & 0.42\\
		\hline
	\end{tabular}
\end{table}

\section{Discussion}
\label{section:Discussion}

The discrepancy evidenced on the dynamical mass of GJ~2060~AB ranges from 1 to 2 \textsigma, depending on the system age. Such a disparity is not statistically impossible, as it represents respectively the edge of the 68\% and 95\% confidence interval. As an example, a 1 to 2 \textsigma~overestimation of the parallax could solve the issue while being a legit statistical realization. We present below some alternative hypotheses.

\subsection{Data derivation and interpretation}

The SINFONI spectra of GJ~2060~A and B  are well fitted by the BT-SETTL model (Fig. \ref{fig:SynSpectra}). Consistent estimates of the bolometric luminosities can be inferred from the multi-epoch observations  of the pair (Table \ref{table:photGJ2060}). Thus, we can focus on vetting the dynamical mass estimate.

GJ~2060 astrometry has been measured with many different instruments, so that systematic errors can lead to important biases if they are not accounted for in the error bars. However, we performed the MCMC fit to each instrument astrometry, with and without the radial velocity measurements, and found very close values to the one given in section \ref{section:MCMC}, for both orbital elements and total mass peak values, that can definitely not account for the 0.1 or 0.2 $\mathrm{M_\odot}$ difference. We find a total mass of $m_\text{tot} = 1.10 \pm 0.14~\mathrm{M_\odot}$ when we only consider the largest homogeneous sample of astrometric epochs (AstraLux), and $m_\text{tot} = 1.05 \pm 0.12~\mathrm{M_\odot}$ when we consider the three main sets of astrometric epochs (AstraLux, SPHERE, Keck). We estimate $1.08 \pm 0.12~\mathrm{M_\odot}$ when we consider all the astrometry and exclude the radial velocity measurements. In all these cases, as the orbit is less constrained, the mass can agree within 1\textsigma~with the model predictions for the old age ranges of the AB Dor moving group.

The absolute orientation of the field are usually inferred from the observations of different reference astrometric fields (clusters). This orientation could not be derived in a homogeneous way for all our astrometric epochs and instruments. Therefore, it may introduce a bias on the orbital parameter determination of GJ~2060~AB. A systematic on the pixel scale of the instruments is less likely to change our results given the short separation of GJ~2060~AB. Therefore, we added parameters to the MCMC algorithm in order to estimate and account for systematic angular offsets in the astrometry. Our astrometry consists of 5 different samples, but only three of them contain more than 2 data points, AstraLux, SPHERE and Keck. We performed an orbital fit with only these samples along with the RVs, with AstraLux data (more numerous) taken as reference. Two offset parameters $\alpha_1$ and $\alpha_2$ had thus to be added to the original MCMC algorithm : the AstraLux data are fitted as they are, with a model corresponding to Eqs. (\ref{xMCMC}) and (\ref{yMCMC}), while the SPHERE data are first rotated through the angle $\alpha_1$ and the Keck data through the angle $\alpha_2$. We modified the algorithm to allow any number N of samples (that is N different instruments) as astrometry input. One sample has to be designated as the reference and the algorithm fits N-1 angular offsets, assuming a flat prior \citep[e.g.,][]{montet2015}. The results are then displayed with posterior distributions and correlations to the other parameters. We found a distribution centered around -0.18\degree for the offset between AstraLux and SPHERE data, with standard deviation 0.2\degree. For the offset between AstraLux and Keck, the distribution is centered around 0.10\degree, with standard deviation 0.3\degree. Near apoastron, an angular offset of 0.3\degree~corresponds to 2-3 mas of offset on the right ascension. However, the total and fractional mass and remain unaffected. We are then confident that our dynamical mass estimate is not strongly affected by these angular offsets.

\subsection{Models imprecision at the moving group ages}

Pre-main sequence models have a well known tendency to underestimate significantly the mass of low-mass stars \citep{hillenbrand2004,mathieu2007}. \cite{mathieu2007} studied the 23 PMS stars for which a dynamical mass had been derived and compared these masses to the predictions of the evolutionary models, given the bolometric luminosity and effective temperature of the stars. They highlighted a mean underestimation of 20-30\% for low-mass stars ($\mathrm{< 1~M_{\odot}}$), similar to the underestimation of GJ~2060 mass. Since then, new evolutionary models have been designed. Moreover, dozens of new dynamical masses have been obtained for PMS stars in the mean time (most of them for stars younger than 10 Myr). These studies often confirm the previously noticed mass discrepancy \citep[e.g., ][]{simon2017,mizuki2018}.

Among these systems, some are comparable to our objects. In the ABDor moving group, the systems ABDor~Bab \citep{azulay2015,janson2018} and C \citep{close2005,luhman2006,close2007,boccaletti2008,azulay2017} have been deeply analyzed in relation to the discussion about the age of the moving group. ABDor Ba and Bb's dynamical masses and luminosities are not consistent with the PMS isochrones \citep[][Paper II]{janson2018}. At the given luminosities and for a moving group age of 150 Myr, the predicted masses are $\sim$ 25\% below the dynamical masses, which are similar to GJ 2060 B. The study of AB Dor C is consistent with any age between 20 and 120 Myr, and the mass derived from the models is slightly underestimated (10\%) but still consistent with the dynamical mass, which is similar to that of TWA~22 components \citep{azulay2017}.

On the other hand, NTT~33370~AB is a 80 Myr low-mass binary very similar to TWA~22 in terms of masses \citep{schlieder2014,dupuy2016}. Both individual masses are strongly underpredicted by the BHAC15 model (2$\sigma$, $46^{+16}_{+19}$\%), which contrasts with the perfect agreement we found for TWA~22.

This issue does not disappear for older ages of the PMS regime. System 2M1036 is a triple M-dwarf stellar system in the Ursa Major moving group whose age is estimated at 400-500 Myr \citep{brandt2015,jones2015}. \cite{calissendorff2017} evidenced a 1\textsigma~underestimation on each component mass. Conversely, in the same moving group, the K binary system NO UMa dynamical masses are in good agreement with the model predictions \citep{schlieder2016}.

The evolutionary models have not yet entirely mastered the physics of PMS stars, as can be seen from the frequent mass underestimation. It is in particular surprising that two very similar systems can encounter very different prediction agreements.  Facing up to the mass over-estimation of system GJ 1108 A, \cite{mizuki2018} compared a dozen PMS stars dynamical masses with the predictions of the BHAC15 model, and report a $\leq$ 10\% offset towards underestimation, and $\sim$ 20\% scatter. Their results also confirm that the tendency to underpredict the mass is neither associated with a mass range nor with an age.  

Magnetic activity is also often brought up as a cause of discrepancy in low-mass stars, as it affects greatly the convection and induces large spot coverage fractions, that may lead to displacements on the HR diagram \citep[eg.,][]{feiden2015,somers2015}.  The high jitter in HARPS RV measurements ($\sim$ 400 m/s) indicates that GJ 2060 has a strong magnetic activity. \cite{somers2015} studied the influence of spots on PMS stars and showed that it could lead to non negligible radius inflation, that would then lower the stars effective temperatures and luminosities. The gap bewteen the normal and spotted case depends a little on the star's age and strongly on the star's mass. Following Fig. 1 B, we assumed a $\Delta$L of -10\% and a $\Delta$T$_\text{eff}$ of -5\% for the primary, and $\Delta$L of -20\% and a $\Delta$T$_\text{eff}$ of -8\% for the secondary. We then plot the new position of GJ 2060 A and B on the BHAC15 isochrones on Fig. 18 (f) in the appendix. The positions are shifted of 0.04 dex and 0.1 dex towards the brighter luminosities, and 200 K and 300 K towards the hotter temperatures. The diagram is now consistent with coevality at 150 Myr, and the total mass that is derived matches the dynamical estimate. These corrections are computed with a spot surface coverage of 50\%. The intense activity of the stars could thus account for the disagreement with the models. A shift in temperature could also resolve the slight mismatch of TWA 22  that appears in the HR diagram. However, a higher luminosity would lead to an excessive mass. This hints for a lesser activity-induced effect for TWA 22 components. In order to test this hypothesis, it would be worth comparing the activity indicators of different PMS binaries with their predicted mass discrepancy. Such study was done recently by \cite{stassun2014} for eclipsing binaries, where they evidenced that activity was not the only cause of the disparity. Finally, in our case where the two components are regularly (at each periastron passage) very close to each other (< 1 au), tidal interactions may also affect the evolution of the stars, although the effects are expected to be weaker than in the eclipsing binary cases, that are constantly undergoing strong interactions. An in-depth study would be needed however to determine the effect of tidal forces on the interiors of eccentric binaries.

On the other hand, \cite{simon2017} suggested that all underestimations come from hidden components within the systems. If it is unlikely that such explanation accounts for all the observed discrepancies, in particular within tight binaries, it is nevertheless a suggestion worth studying for GJ~2060, especially given its unusually high eccentricity ($e \sim 0.9$).

\subsection{Missing mass: existence of GJ~2060 Ab or Bb}

Hidden mass close to the primary could explain the strong disagreement between models and data for GJ~2060. Indeed, a 0.1 to 0.2 $\mathrm{M_\odot}$ (depending on the system age) additional companion could account for the mass underestimation (see table \ref{MissingMass}) and could have been missed in the SPHERE datasets if close enough from one of the two components.

\begin{table}[h]
	\centering
	\small
	\caption{GJ~2060 missing mass (in solar mass) depending on models and age. Errors propagation have been obtained from the MCMC posteriors dispersion, the luminosities uncertainty at given distances and the errors on the distance, assuming independancy. }\label{MissingMass}
	\renewcommand{\arraystretch}{1.5}
	\begin{tabular}{lccccc}
		\hline
		Model &  50 Myr & 75 Myr & 100 Myr & 150 Myr & 200 Myr\\
		\hline
		\hline
		BHAC15 & 0.18$^{+0.09}_{-0.08}$ & 0.13$^{+0.09}_{-0.08}$ & 0.11$^{+0.09}_{-0.09}$ & 0.07$^{+0.09}_{-0.09}$ & 0.05$^{+0.11}_{-0.08}$\\
		PISA &  0.19$^{+0.09}_{-0.08}$ & 0.14$^{+0.09}_{-0.08}$ & 0.12$^{+0.09}_{-0.09}$ & x &  x\\
		PARSEC & 0.16$^{+0.09}_{-0.10}$ & 0.11$^{+0.10}_{-0.08}$ & 0.10$^{+0.10}_{-0.08}$ & 0.07$^{+0.09}_{-0.08}$ & 0.06$^{+0.10}_{-0.09}$\\
		SDF00 & 0.32$^{+0.09}_{-0.09}$ & 0.25$^{+0.09}_{-0.09}$ & 0.21$^{+0.09}_{-0.09}$ & 0.18$^{+0.09}_{-0.09}$ & 0.18$^{+0.09}_{-0.08}$\\
		DM97 & 0.23$^{+0.09}_{-0.08}$ & 0.19$^{+0.09}_{-0.08}$ & 0.15$^{+0.09}_{-0.09}$ & 0.13$^{+0.09}_{-0.08}$ & 0.11$^{+0.11}_{-0.08}$\\
		Darmouth & 0.18$^{+0.09}_{-0.09}$ & 0.12$^{+0.09}_{-0.08}$ & 0.10$^{+0.09}_{-0.08}$ & 0.06$^{+0.09}_{-0.08}$ & 0.05$^{+0.10}_{-0.08}$\\
		Mean & 0.21 & 0.16 & 0.13 & 0.10 & 0.09\\
		\hline
	\end{tabular}
\end{table}

Quick dynamical simulations with a symplectic integrator (SWIFT\_ HJS \citeauthor{beust2003}\citeyear{beust2003}, see appendix) evidenced that a additional companion should orbit closer than 0.1 au (6 mas) from either of the companion to remain bound for the system's lifetime.

The PSFs of GJ~2060~A and B are not elongated, even in the SPHERE images (FWHM$\sim$30 mas). In the latter, we injected models of putative companions  with different fluxes and separations and checked whether they would induce a PSF lengthening  that could be noticed by eye. The models of the putative companion were built using a flux-normalized PSF. The PSF is the other component of the system (e.g., B if the binarity of A is investigated, and vice versa). Using the BHAC15 models, we estimate that we would have just missed a 0.25 $\mathrm{M_\odot}$  companion at a projected separation $0.45$ au. Thus, 0.2 solar mass object at 0.1 au would have gone unnoticed from the imagers. 

\begin{figure}[h]
	\centering
   	\subfloat[Mass, coplanar case]{\includegraphics[width=\linewidth]{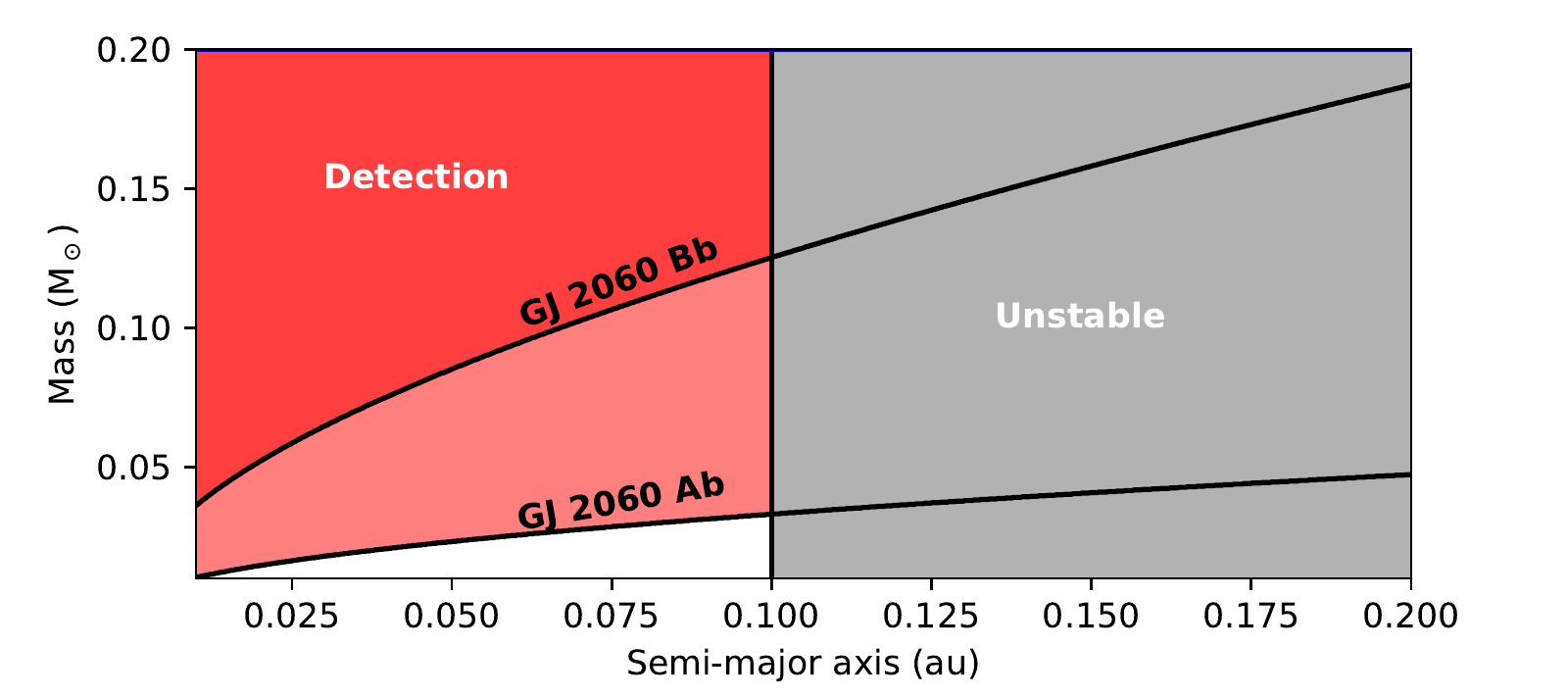}}\\
   	\subfloat[Inclination, m$_\text{inner}$ = 0.1 $\mathrm{M_\odot}$]{\includegraphics[width=\linewidth]{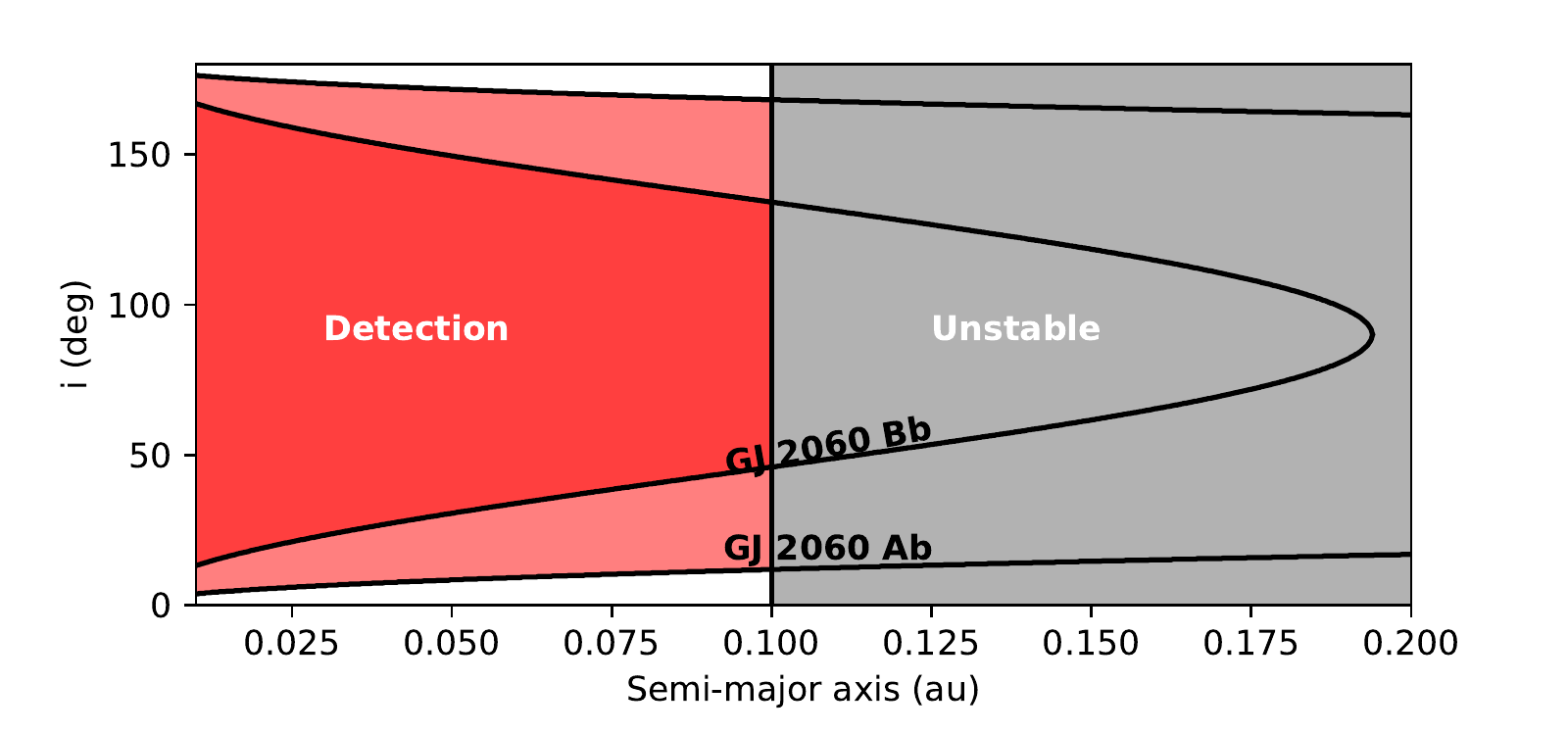}}
    \caption{Radial velocity detection limits around GJ~2060~A and B as a function of the inner orbit radius (coplanar case), in terms of the inner companion (a) mass (b) inclination. For (a), the orbits have been chosen coplanar (i = 36\degree) and for (b), the inner mass has been set to 0.1 $\mathrm{M_\odot}$.}
    \label{fig:LimDet}
\end{figure}

As for the spectrograph, the available RV data are too sparse to resolve in frequency an additional orbit, and the flux ratio in the optical prevents the detection of any spectral signature. However, the closer the object, the stronger the radial velocity perturbation amplitude. A simple comparison between the perturbation amplitude on GJ~2060~A and B radial velocities and our measurements standard deviation \textsigma~is summarized in Fig. \ref{fig:LimDet} for the circular case, in a (semi-major axis, mass) and a (semi-major axis, inclination) diagrams. We used the predicted mass of GJ~2060~A and B from Table \ref{Masses}  at an age of 100 Myr for that purpose. We chose 3\textsigma~dispersion of the RVs as a detection threshold, and represented the corresponding frontier on the plots. The limit of the dynamical stability has been set to 0.1 au; the accurate stability limit depends on the third companion mass and inclination, but is in all cases $\lesssim 0.1$ au. In the coplanar case, a mass higher than 0.1 $\mathrm{M_\odot}$ could have been unnoticed around the secondary. That is not the case for a putative companion around the primary, because the light we observe comes mostly from the primary, so that most signals would be easily spotted. However, a 0.1 or even 0.2 $\mathrm{M_\odot}$ at 0.1 au could be compatible with our deviation in both situations, primary or secondary, for small enough inclinations, respectively 10 and 5\degree~around the primary, and 45 and 25\degree~around the secondary. 

If there is indeed a hidden companion, its luminosity would add to the luminosity of the nearest component, so that the latter measured flux would be biased. According to the BHAC15, a 50-75 Myr 0.2 $\mathrm{M_\odot}$ companion have $\mathrm{log(L/L_{\odot})=-1.9, -2}$ dex and a 0.1 $\mathrm{M_\odot}$ companion have $\mathrm{log(L/L_{\odot})\sim -2.3}$ dex. The component hosting a hidden companion would appear overluminous for its temperature (slightly for the primary, significantly for the secondary), shifting its position on Fig. \ref{fig:Isochrones} towards the younger isochrones and straining coevality. In the PARSEC isochrones in the appendix, a significant luminosity shift (corresponding to a 0.2 $\mathrm{M_\odot}$ companion) of the primary could achieve coevality. Conversely, the same companion around the secondary would induce a luminosity shift that would break coevality in all models.

\begin{figure*}[h]
	\centering
   	\subfloat[ORB6]{\includegraphics[width=0.5\linewidth]{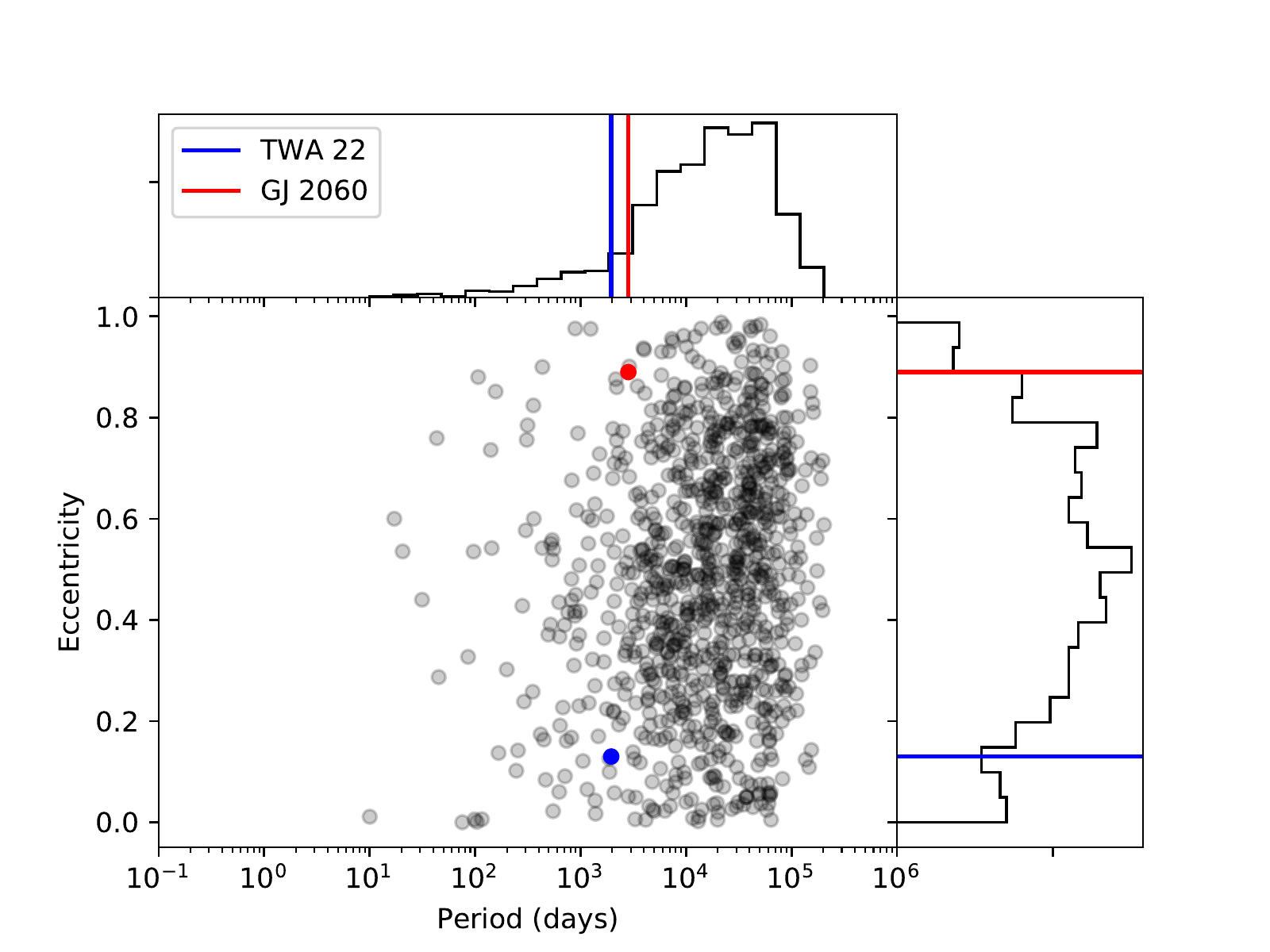}}
   	\subfloat[SB9]{\includegraphics[width=0.5\linewidth]{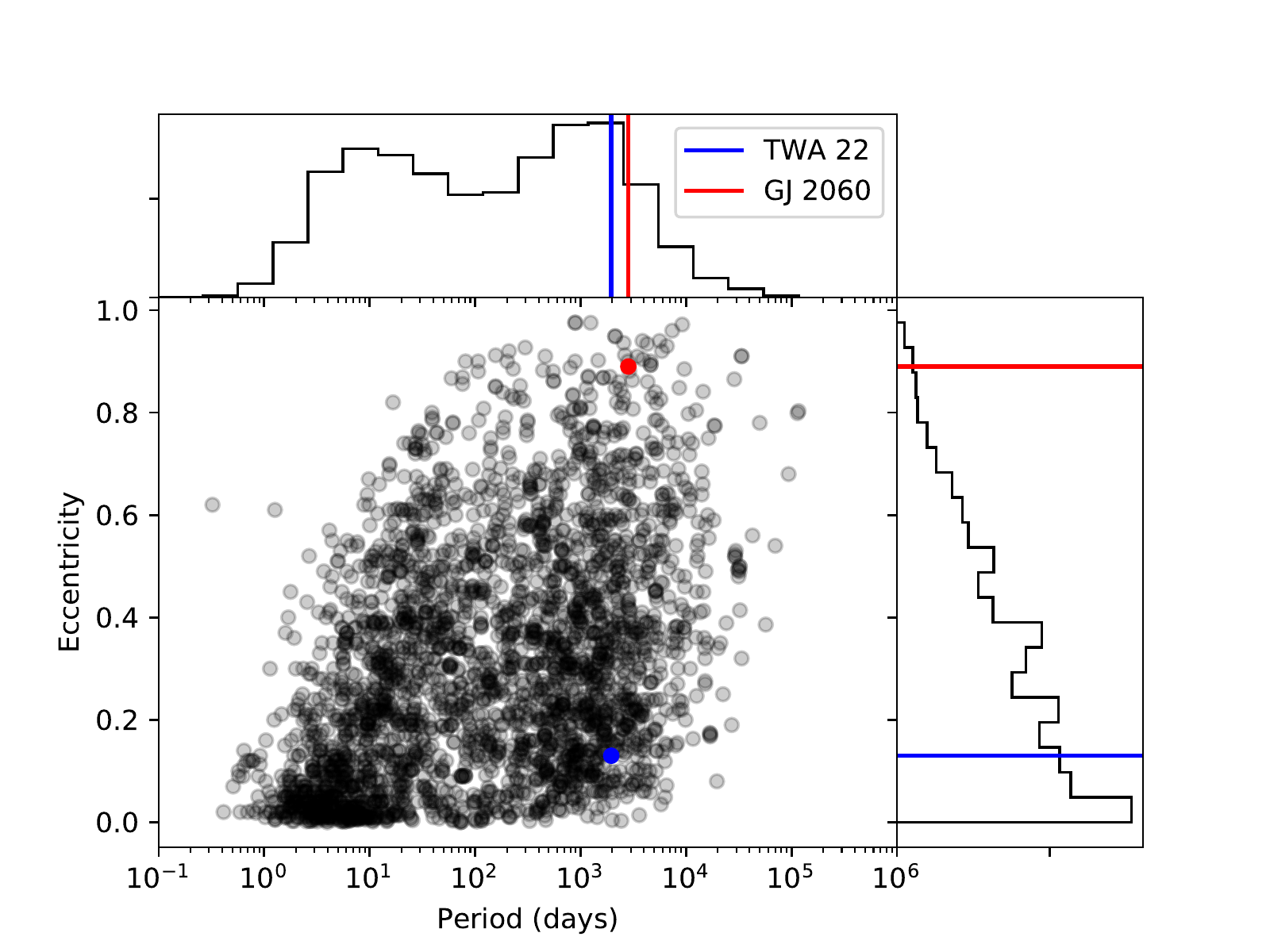}}
    \caption{Period-eccentricity diagrams from two catalogs of binary stars: (a) Visual binaries, from ORB6 and (b) spectroscopic binaries, from SB9.}
    \label{fig:eT}
\end{figure*}

Finally, the high eccentricity ($0.90 \pm 0.01$) of the visual orbit is noticeable, and one can wonder if it could indicate strong dynamical interactions. From the ORB6 catalog\footnote{http://www.usno.navy.mil/USNO/astrometry/optical-IR-prod/wds/orb6}, we computed the periods and eccentricities of visual binary stars with reliable orbital elements (according to the grades given in the catalog).  From the SB9 catalog \citep{pourbaix2004}, we computed the periods and eccentricities of spectroscopic binary stars. Our two binaries fall near the limit of each catalog's period coverage, so that while it gives an interesting overview, more binaries would be needed to draw robust statistical conclusions. Gaia next data releases will significantly contribute to overcome this lack. The resulting diagrams are shown in Fig. \ref{fig:eT}.  While not so common, GJ 2060's eccentricity seems not so rare at this range of periods and age (before circularization). Moreover, no mechanism are known to enhance the eccentricity of an outer companion at such period ratio (greater than 200, that excludes any meaningful mean-motion resonance). Only close encounters could dynamically raise the eccentricity, but the configuration would then not be stable. All in all, the high eccentricity is likely uncorrelated to the potential existence of a third companion.

\section{Conclusion}
\label{section:Conclusion}

We considered two systems of young astrometric M-dwarf binaries, TWA~22 and GJ~2060, and used existing astrometric and spectroscopic data along with new Keck, SPHERE, NaCo, HARPS and FEROS data to derive the total mass of these systems. We consolidated the total dynamical mass estimate of TWA~22: $0.18 \pm 0.02$ M$_\odot$. We derived the first estimate of the total mass of GJ~2060: $1.08 \pm 0.10$ M$_\odot$. The orbits of the two systems are well constrained thanks to the use of our additional data and the errors are carefully estimated through the MCMC approach. The orbit of GJ~2060 has an unusually high eccentricity, around 0.9. The cross-contamination of GJ 2060 primary and secondary spectra into the FEROS and HARPS data prevents us from deriving accurate dynamical masses of the individual components.

The study of the photometry and spectroscopy of the two systems, along with their membership to moving groups and accurate distances, allow to test the PMS evolutionary models predictions. The dynamical mass of TWA22~AB is  correctly predicted by the models at the age of the \textbeta~Pictoris moving group. The placement of GJ~2060 A and B on evolutionary tracks confirms the system coevality at an age compatible with the AB Doradus moving group ($\sim$ 50 Myr). However, all models underpredict the total mass of GJ~2060 AB, by 10 to 20\% (0.1 to 0.2 $\mathrm{M_\odot}$, 1-2 \textsigma). A new precise parallax (likely to come in Gaia DR3 release) would strongly decrease the uncertainty on the dynamical mass and could improve the statistical relevance of the discrepancy. 

GJ~2060 AB's underpredicted mass is consistent with a trend found on other systems in the same mass range. It could be explained by luminosity and temperature drop caused by high starspots coverage. In that case, we would retrieve coevality at 150 Myr. We  also discussed the potential existence of a third companion close to one component of GJ~2060 that could account for this disagreement.  Dynamical modeling shows that such a companion would have to be very close to one of the stars, less than 0.1 au (6 mas), in order to remain stable for millions of years. Such a close companion could have gone unnoticed, although the RVs are putting some constraints on its mass and inclination. Astrometric and spectroscopic data at periastron as well as the use of RV disentanglement techniques could help clarifying the origin of the discrepancy, and in particular if only one of GJ~2060~AB's component has an underpredicted mass.

A dozen of new PMS stellar mass measurements have become available in the last decade. A  complete reassessment of the dynamical mass determinations of sub-solar mass stars and a homogeneous comparison of theses measurements to the latest PMS models would help to conclude on the models reliability,  On the other hand, the upcoming data releases of the Gaia mission should yield a statistical sample of dynamical mass determination of low-mass stars \citep{pourbaix2011}. Additional studies should in any case be needed to infer the luminosity and temperatures of theses many systems and allow for a detailed comparison of the masses to evolutionary models predictions.

\begin{acknowledgements}
We thank Carine Babusiaux, Benjamin Montet and Thierry Forveille for fruitful discussions, as well as the anonymous referee for reviewing our work and for insightful and constructive comments which improved the manuscript. The   project   is   supported   by   CNRS,   by   the   Agence Nationale de la Recherche (ANR-14-CE33-0018, GIPSE), the OSUG@2020 labex and the Programme National de  Planétologie  (PNP,  INSU)  and  Programme  National  de  Physique  Stellaire (PNPS,  INSU). Most of the computations presented in this paper were performed using the Froggy platform of the CIMENT infrastructure (https://ciment.ujf-grenoble.fr), which is supported by the Rh\^one-Alpes region (GRANT CPER07\_13 CIRA), the OSUG@2020 labex (reference ANR10 LABX56) and the Equip@Meso project (reference ANR-10-EQPX-29-01) of the programme Investissements d'Avenir, supervised by the Agence Nationale pour la Recherche. This research has made use of the SIMBAD database operated at the CDS (Strasbourg, France). Research on moving groups by Thomas Henning is supported by the DFG-SFB 881 "The Milky Way System".

SPHERE is an instrument designed and built by a consortium consisting of IPAG (Grenoble, France), MPIA (Heidelberg, Germany), LAM (Marseille, France), LESIA (Paris, France), Laboratoire Lagrange (Nice, France), INAF–Osservatorio di Padova (Italy), Observatoire de Genève (Switzerland), ETH Zurich (Switzerland), NOVA (Netherlands), ONERA (France) and ASTRON (Netherlands) in collaboration with ESO. SPHERE was funded by ESO, with additional contributions from CNRS (France), MPIA (Germany), INAF (Italy), FINES (Switzerland) and NOVA (Netherlands).  SPHERE also received funding from the European Commission Sixth and Seventh Framework Programmes as part of the Optical Infrared Coordination Network for Astronomy (OPTICON) under grant number RII3-Ct-2004-001566 for FP6 (2004–2008), grant number 226604 for FP7 (2009–2012) and grant number 312430 for FP7 (2013–2016). We also acknowledge financial support from the Programme National de Planétologie (PNP) and the Programme National de Physique Stellaire (PNPS) of CNRS-INSU in France. This work has also been supported by a grant from the French Labex OSUG@2020 (Investissements d’avenir – ANR10 LABX56). The project is supported by CNRS, by the Agence Nationale de la Recherche (ANR-14-CE33-0018). It has also been carried out within the frame of the National Centre for Competence in  Research PlanetS supported by the Swiss National Science Foundation (SNSF). MRM, HMS, and SD are pleased to acknowledge this financial support of the SNSF. Finally, this work has made use of the the SPHERE Data Centre, jointly operated by OSUG/IPAG (Grenoble), PYTHEAS/LAM/CESAM (Marseille), OCA/Lagrange (Nice) and Observatoire de Paris/LESIA (Paris). We thank P. Delorme and E. Lagadec (SPHERE Data Centre) for their efficient help during the data reduction process. 
\end{acknowledgements}

\bibliographystyle{aa}
\bibliography{Biblio}

\onecolumn
\clearpage
\section*{Appendix A: Spectrophotometry}
\label{section:Appendix}

\begin{figure*}[h]
	\centering
	\begin{tabular}{cc}
   	\includegraphics[width=8cm]{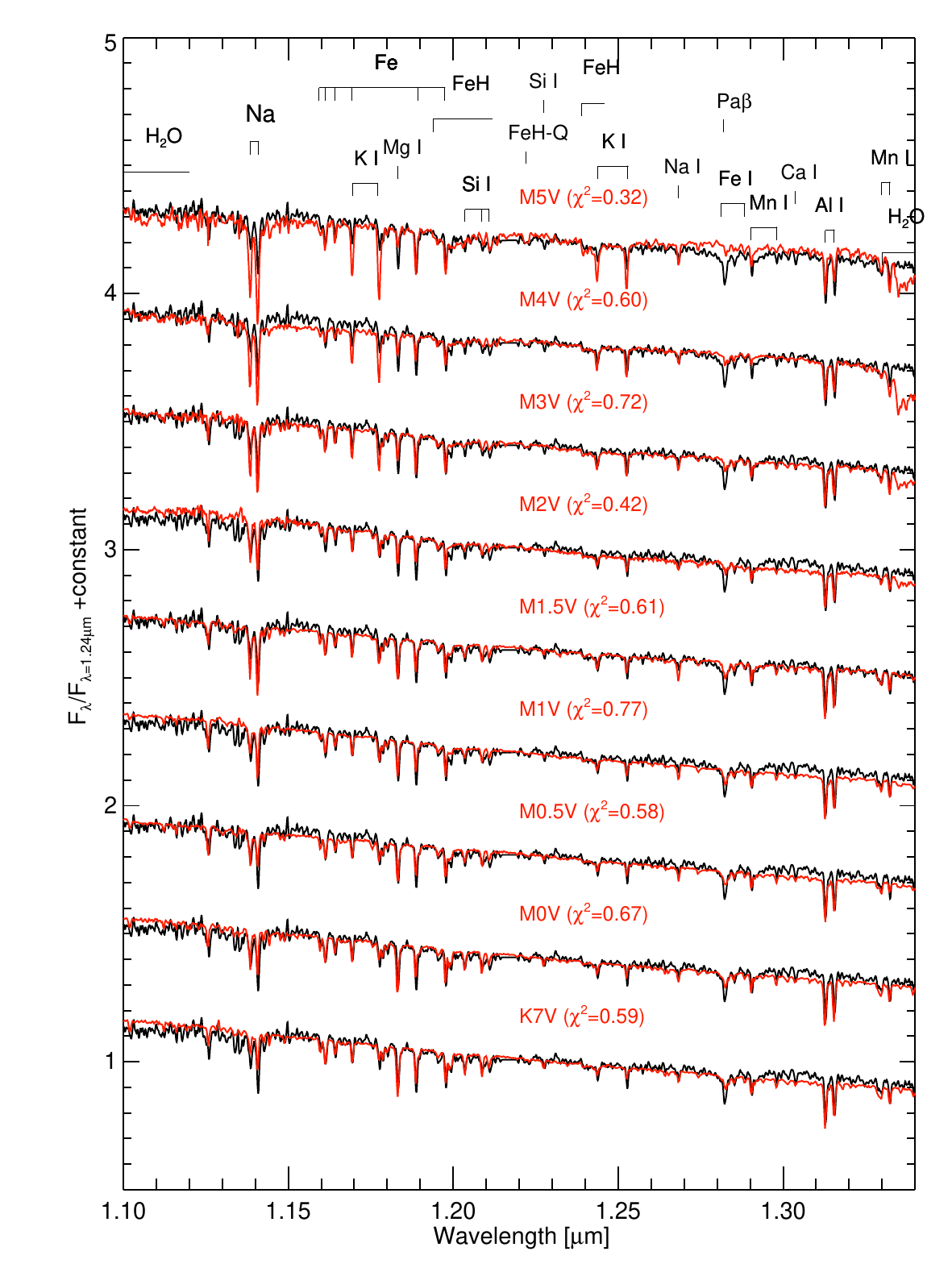}	 &   	\includegraphics[width=8cm]{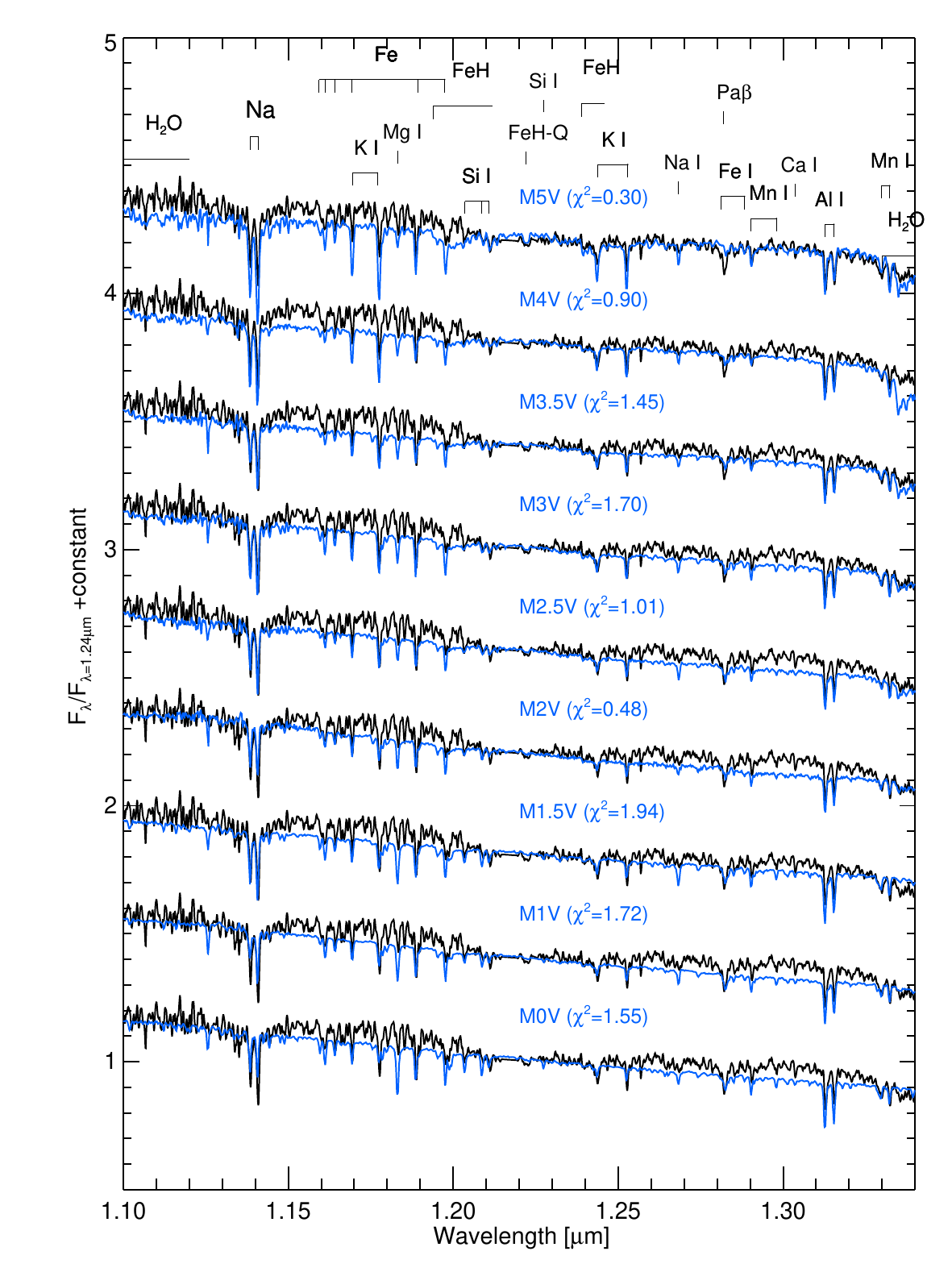} \\
   	\includegraphics[width=8cm]{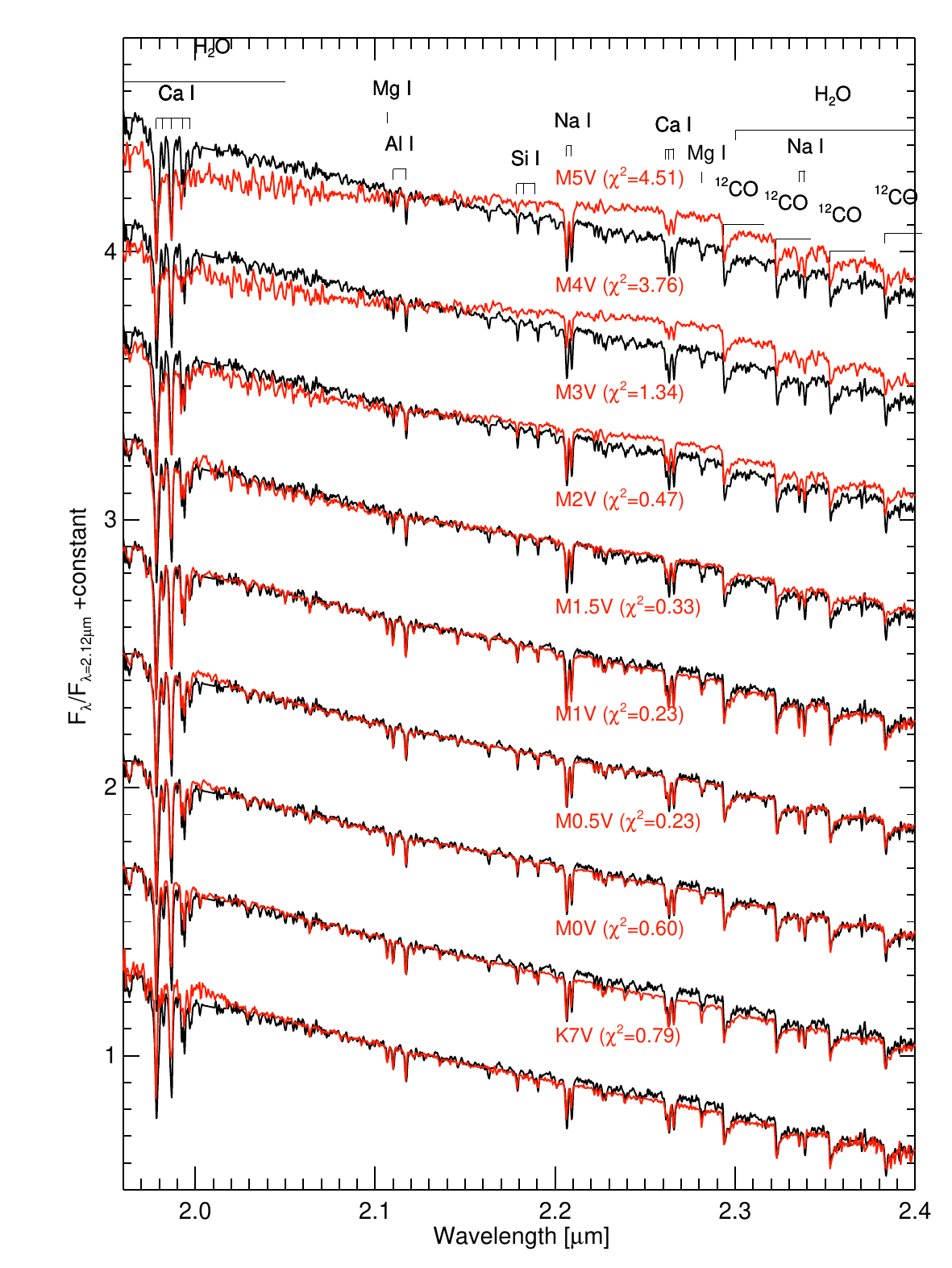}	&   	\includegraphics[width=8cm]{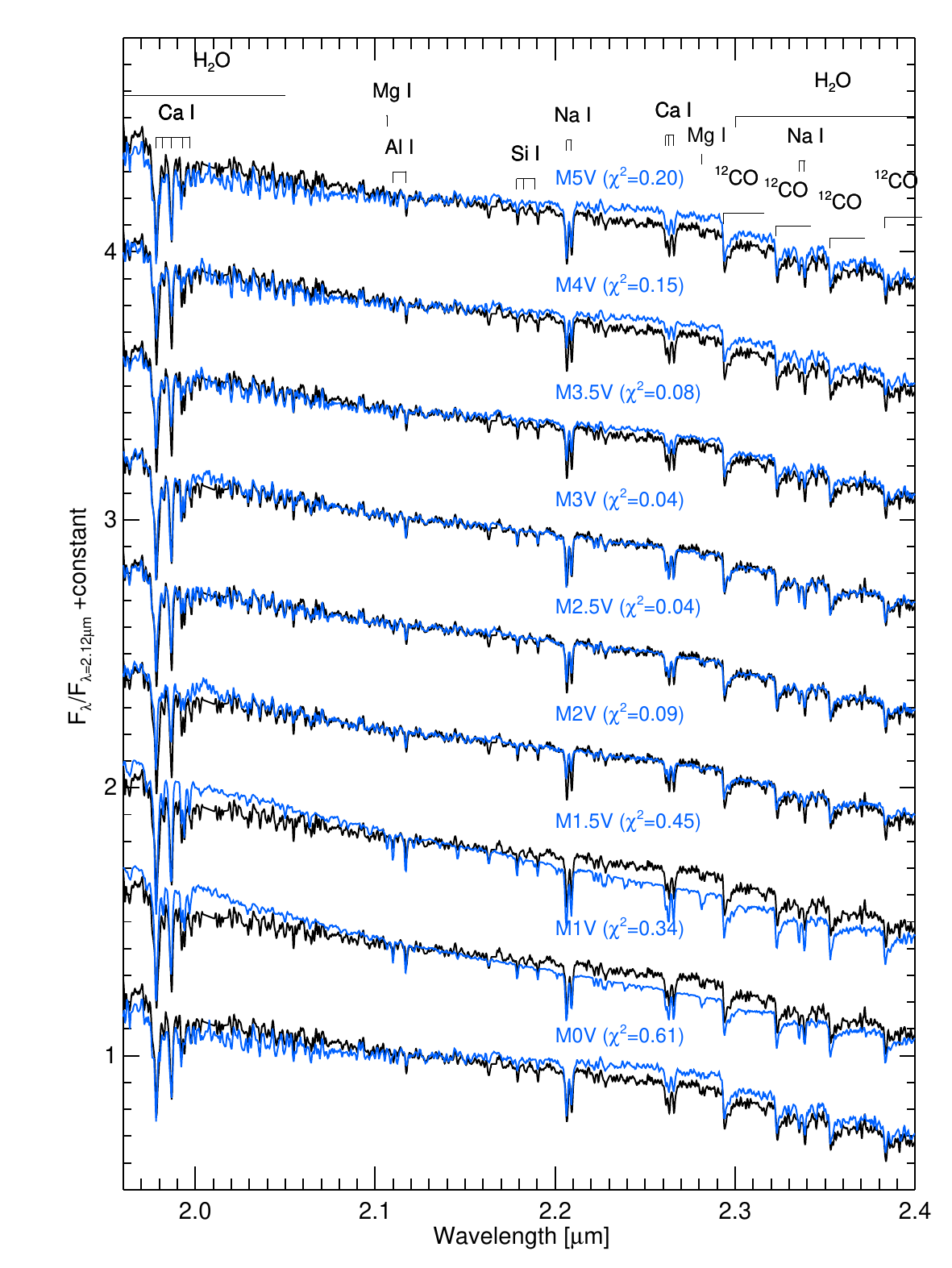} \\
	\end{tabular}
    \caption{Comparison of the J and K band spectra of GJ~2060~A (red) and GJ~2060~B (blue) to M-dwarf spectra.}
    \label{fig:SpectraJK}
\end{figure*}

\begin{figure*}[h!]
	\centering
	\begin{tabular}{cc}
   	\includegraphics[width=8cm]{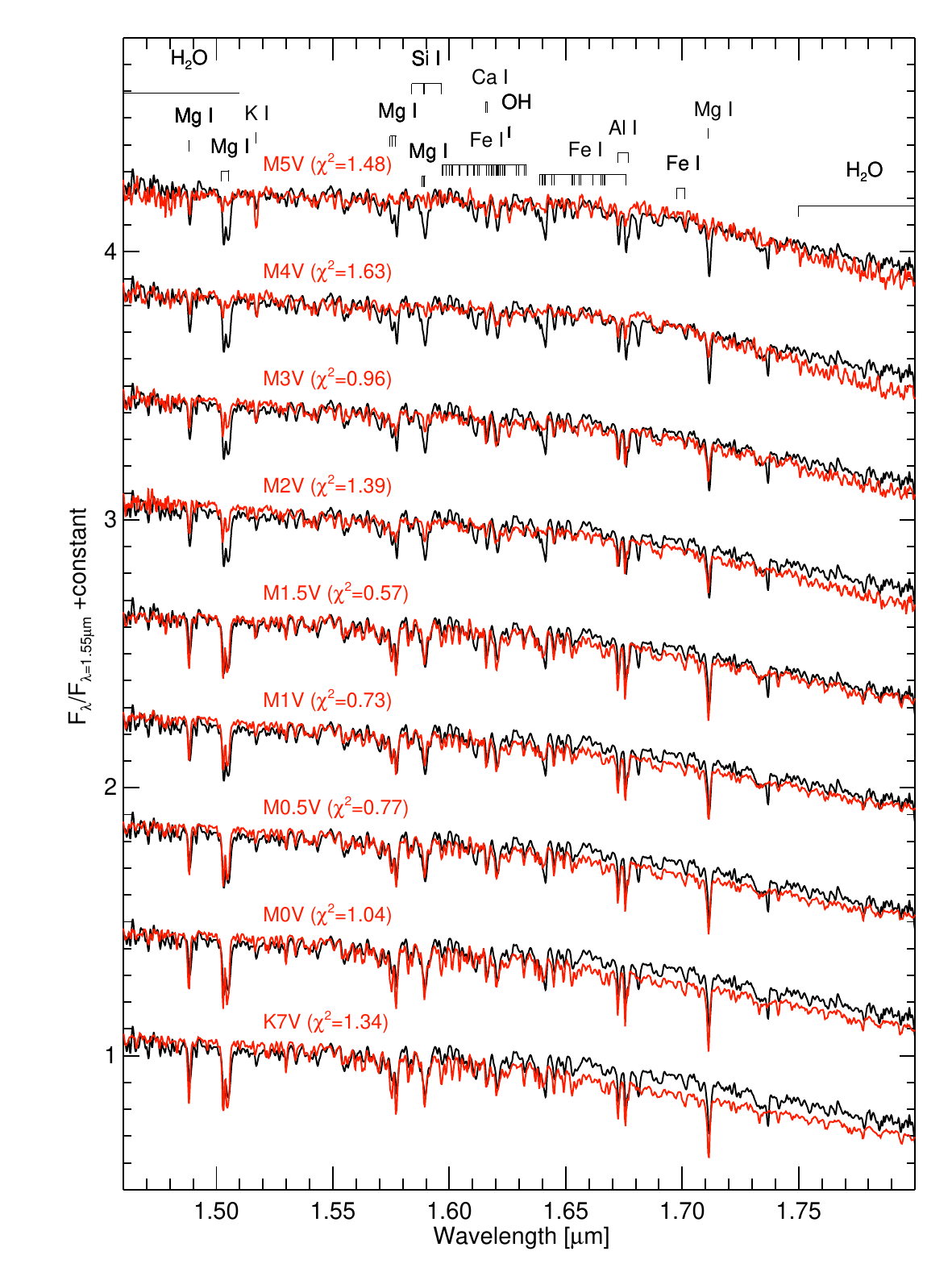}	 &   	\includegraphics[width=8cm]{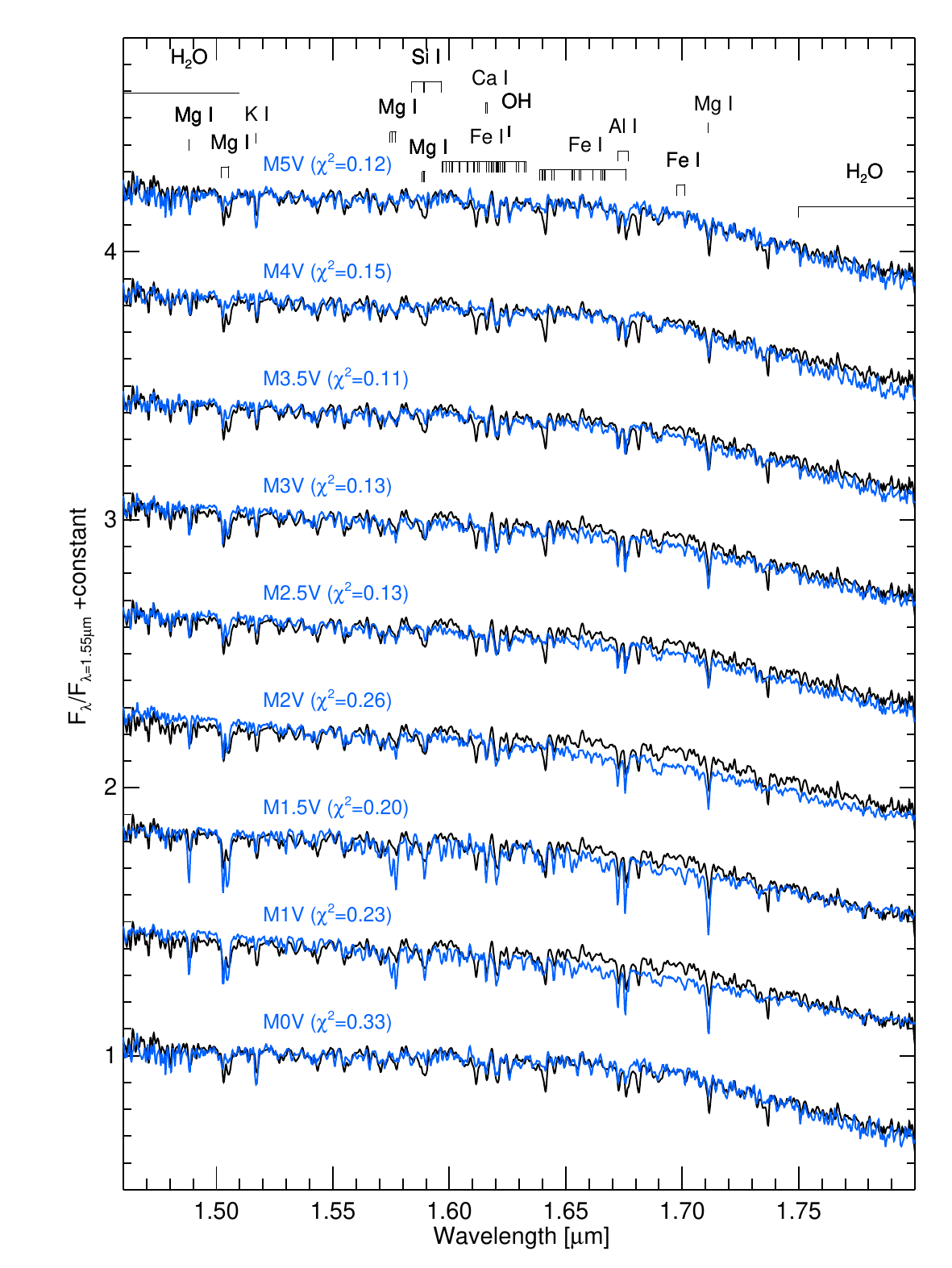} \\
	\end{tabular}
    \caption{Comparison of the H band spectra of GJ~2060~A (red) and GJ~2060~B (blue) to M-dwarf spectra.}
    \label{fig:SpectraH}
\end{figure*}\clearpage

\section*{Appendix B: Orbital fit}

\begin{figure}[h!]
	\centering
   	\includegraphics[trim = {1cm 1cm 1cm 1cm},width=\linewidth]{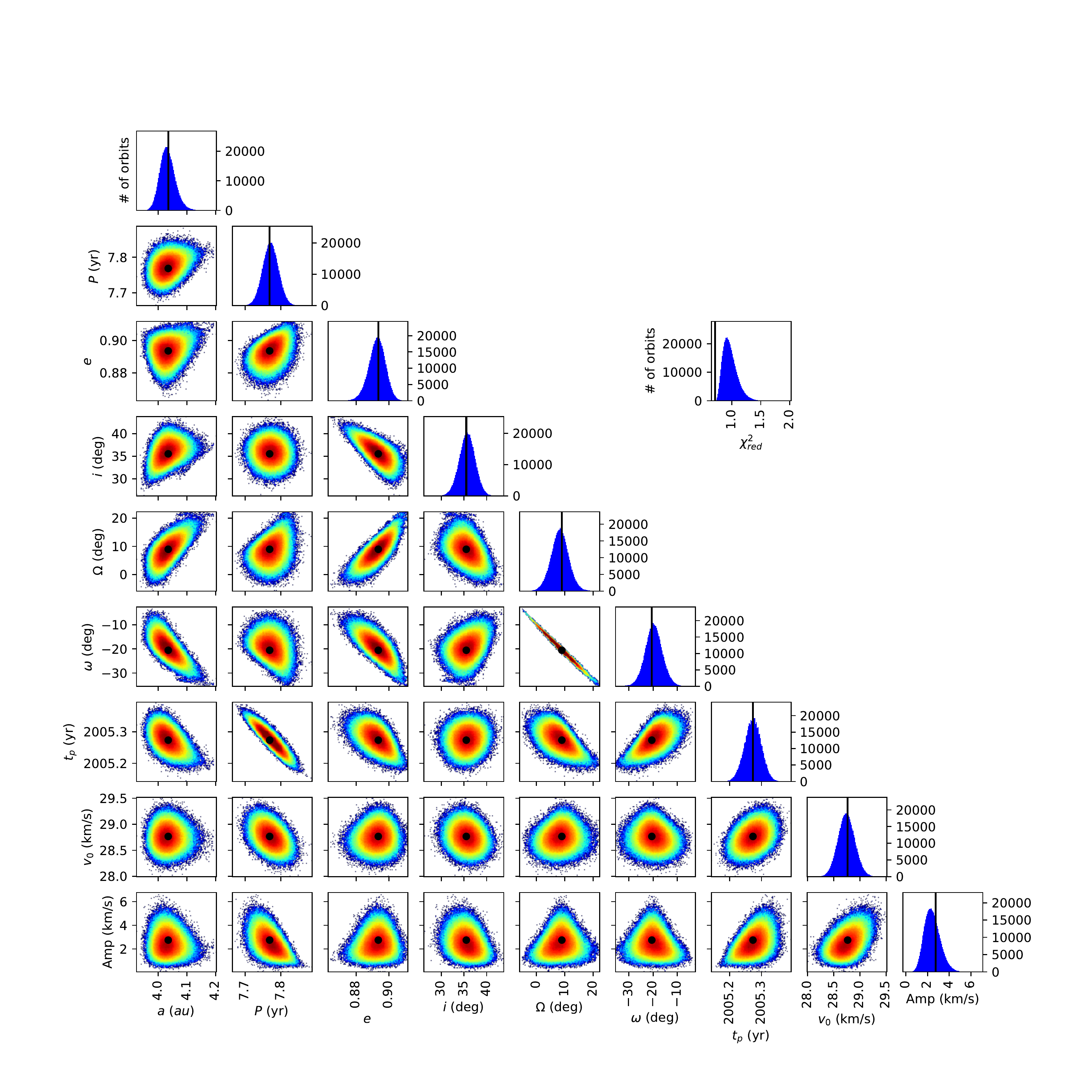}
    \caption{Distribution and correlations of each of the orbital element fitted by the MCMC algorithm for system GJ~2060. The black lines and points depict the best fitting orbit (better $\chi^2$), obtained with the LSLM algorithm. The color scale is logarithmic, blue corresponds to 1 orbit and red to 1000.}
    \label{fig:Posterieures}
\end{figure}

\begin{figure}[h!]
	\centering
   	\includegraphics[trim={1cm 0 0 0},width=\linewidth]{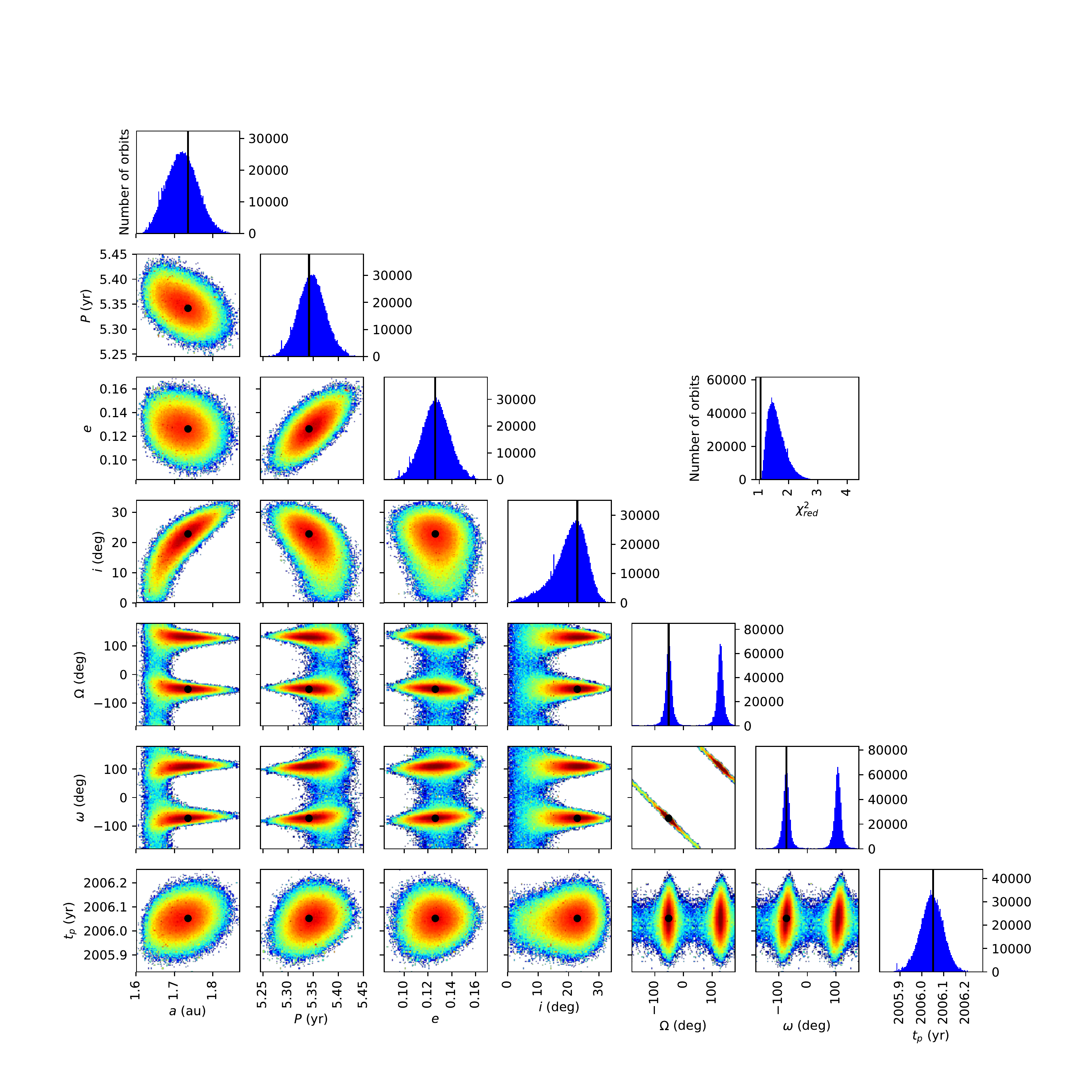}
    \caption{Distribution and correlations of each of the orbital element fitted by the MCMC algorithm for system TWA~22. The black lines and points depict the best fitting orbit (better $\chi^2$), obtained with the LSLM algorithm. The color scale is logarithmic, blue corresponds to 1 orbit and red to 1000.}
    \label{fig:PosterieuresTWA}
\end{figure}\clearpage

\section*{Appendix C: Model comparison}

\begin{figure}[h]
	\centering
   	\subfloat[d = 16.14 pc]{\includegraphics[width=0.5\linewidth]{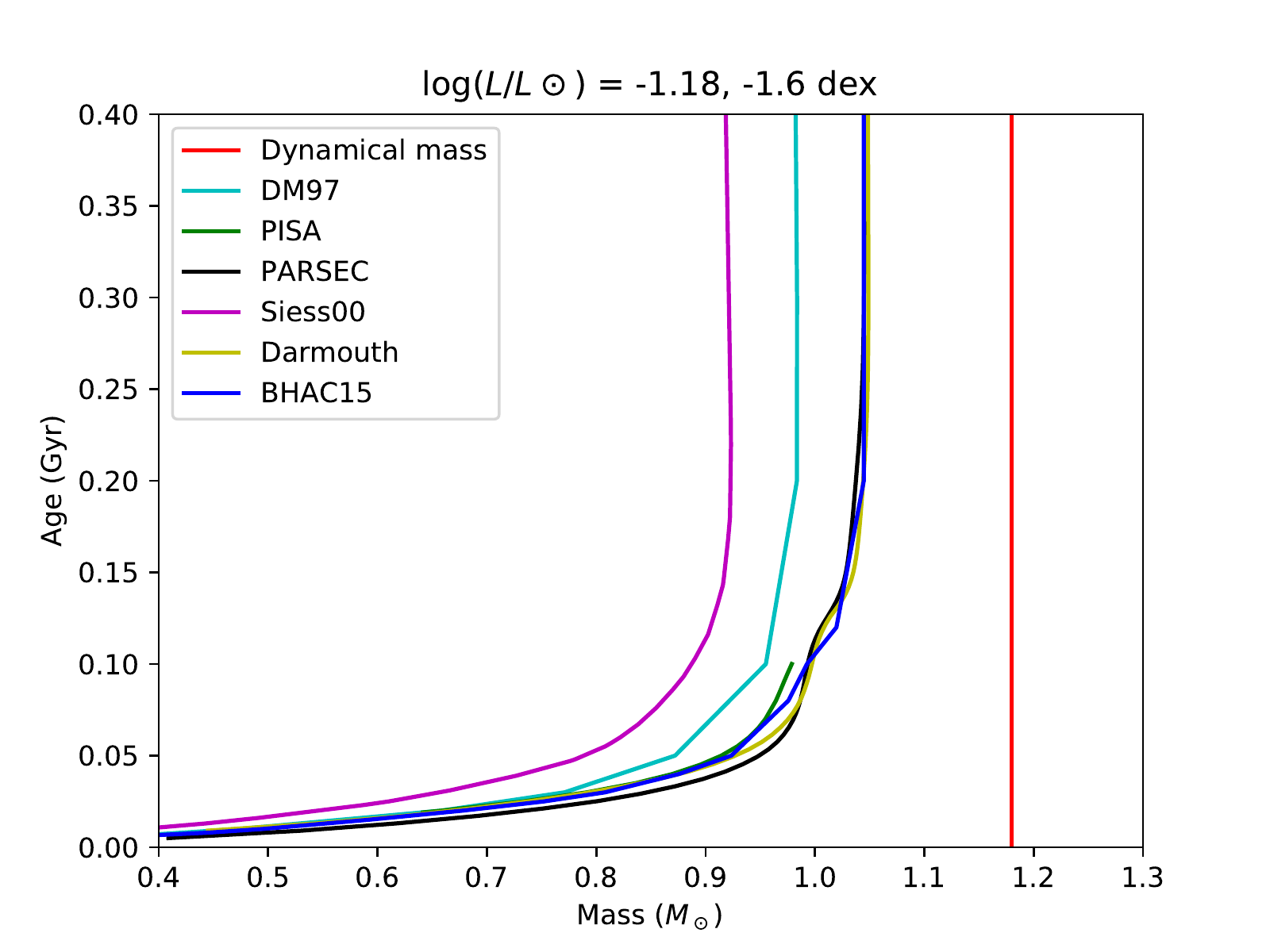}}
   	\subfloat[d = 15.24 pc]{\includegraphics[width=0.5\linewidth]{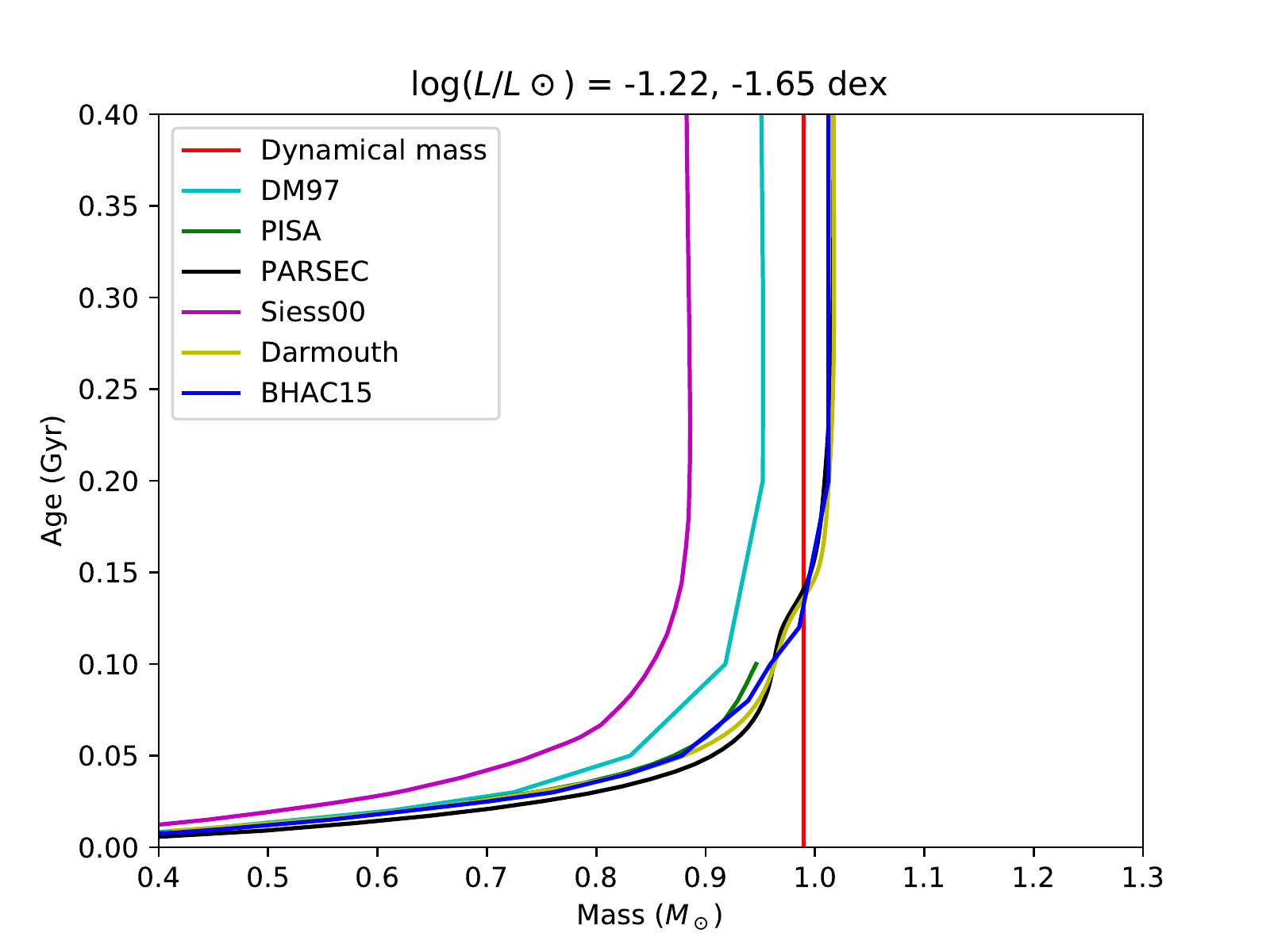}}
    \caption{Mass age relations according to the six different evolutionary models, for GJ~2060 observed luminosities. The dynamical mass is depicted in red. Panel (a) corresponds to the higher boundary of the distance (16.14 pc), panel (b) to the lower boundary (15.24 pc).}
\end{figure}

\begin{figure}[h]
	\centering
   	\subfloat[GJ~2060~A]{\includegraphics[width=0.5\linewidth]{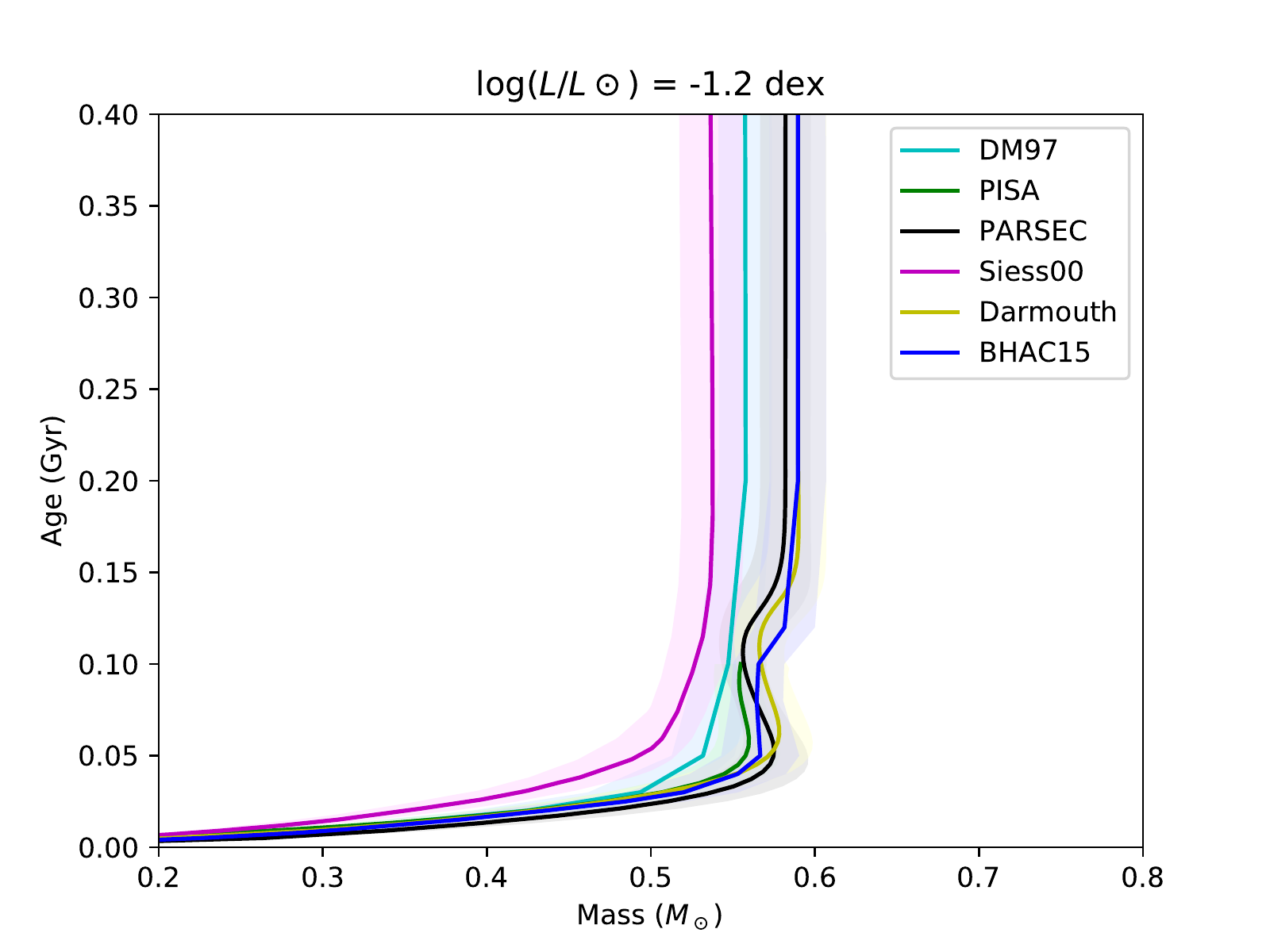}}
   	\subfloat[GJ~2060~B]{\includegraphics[width=0.5\linewidth]{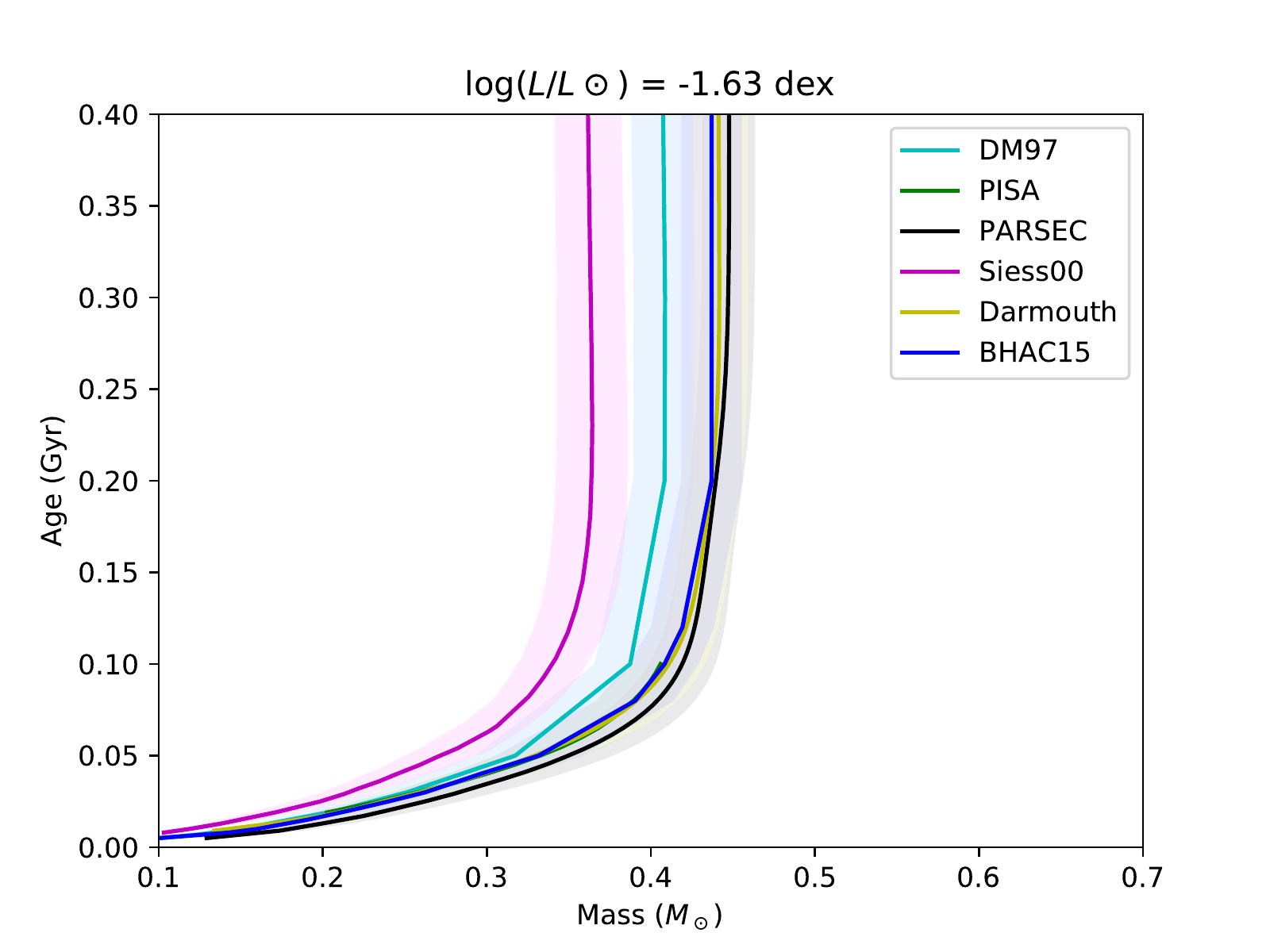}}
    \caption{Mass age relations according to the six different evolutionary models, for GJ~2060 observed luminosities. Panel (a) corresponds to the primary, panel (b) to the secondary. The error on the distance is here taken into account.}
\end{figure}

\newpage

\begin{figure}[h]
	\centering
	\subfloat[Darmouth]{\includegraphics[width=0.45\linewidth]{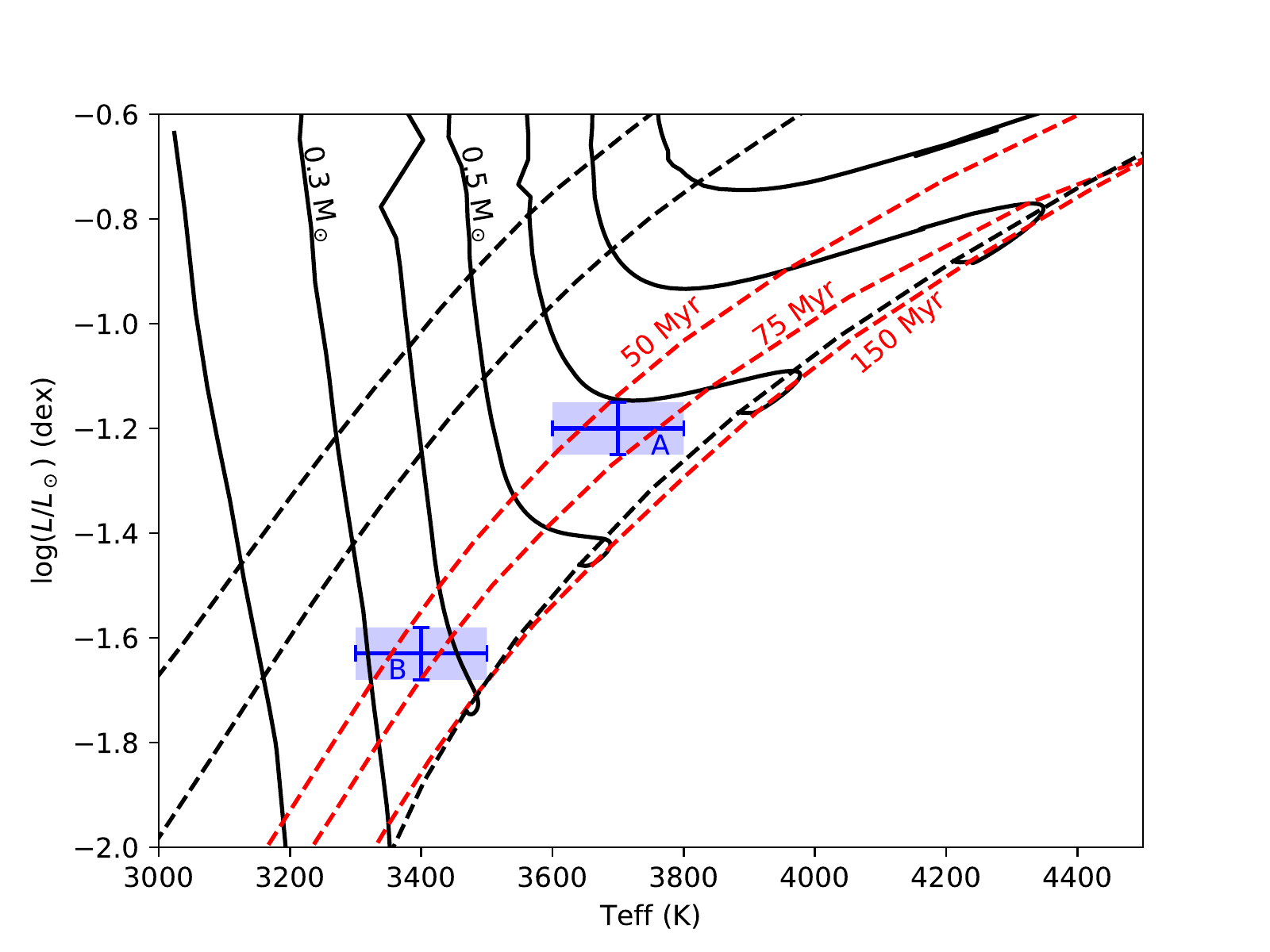}}
	\subfloat[DM97]{\includegraphics[width=0.45\linewidth]{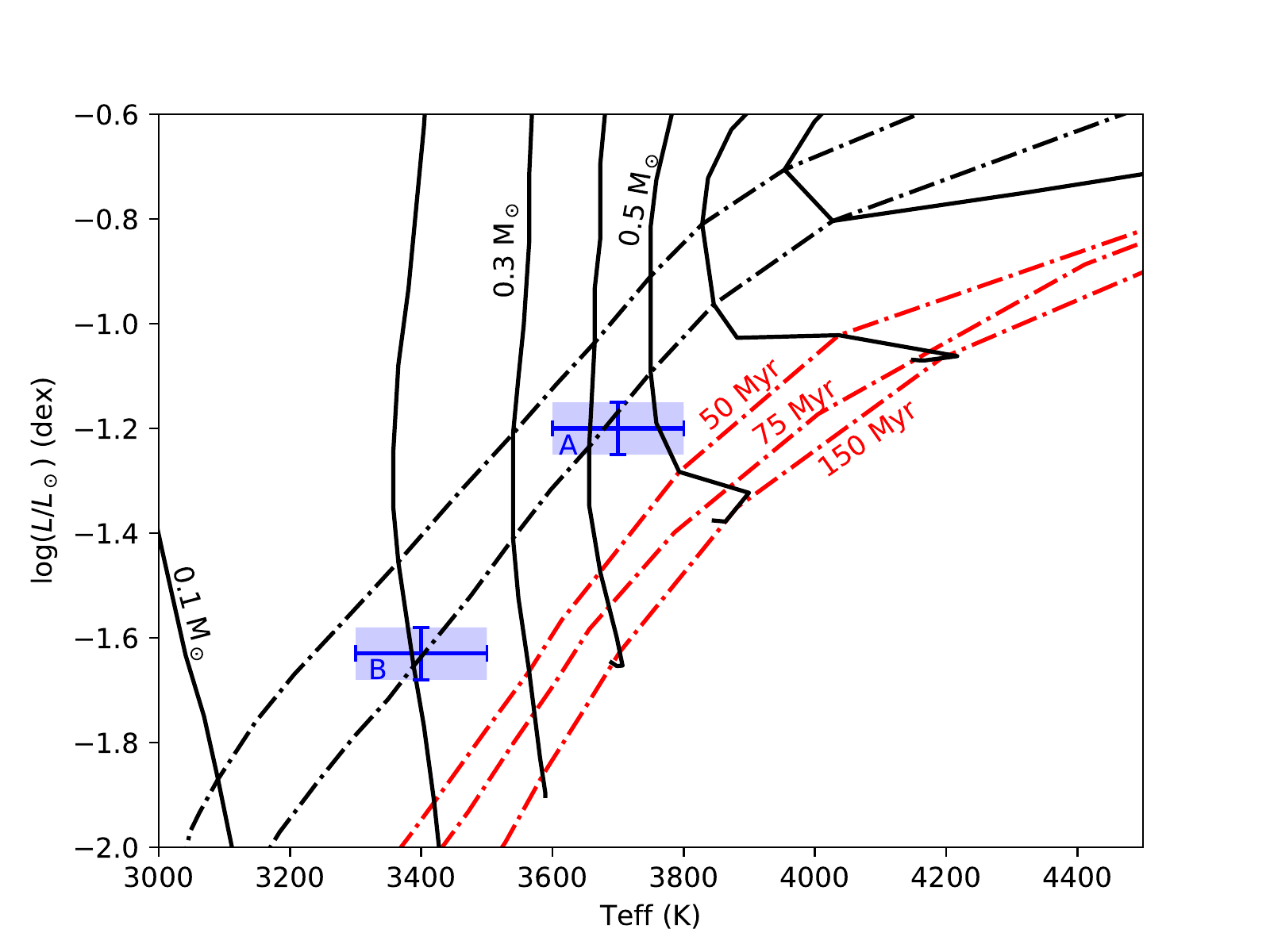}}\\
   	\subfloat[PARSEC]{\includegraphics[width=0.45\linewidth]{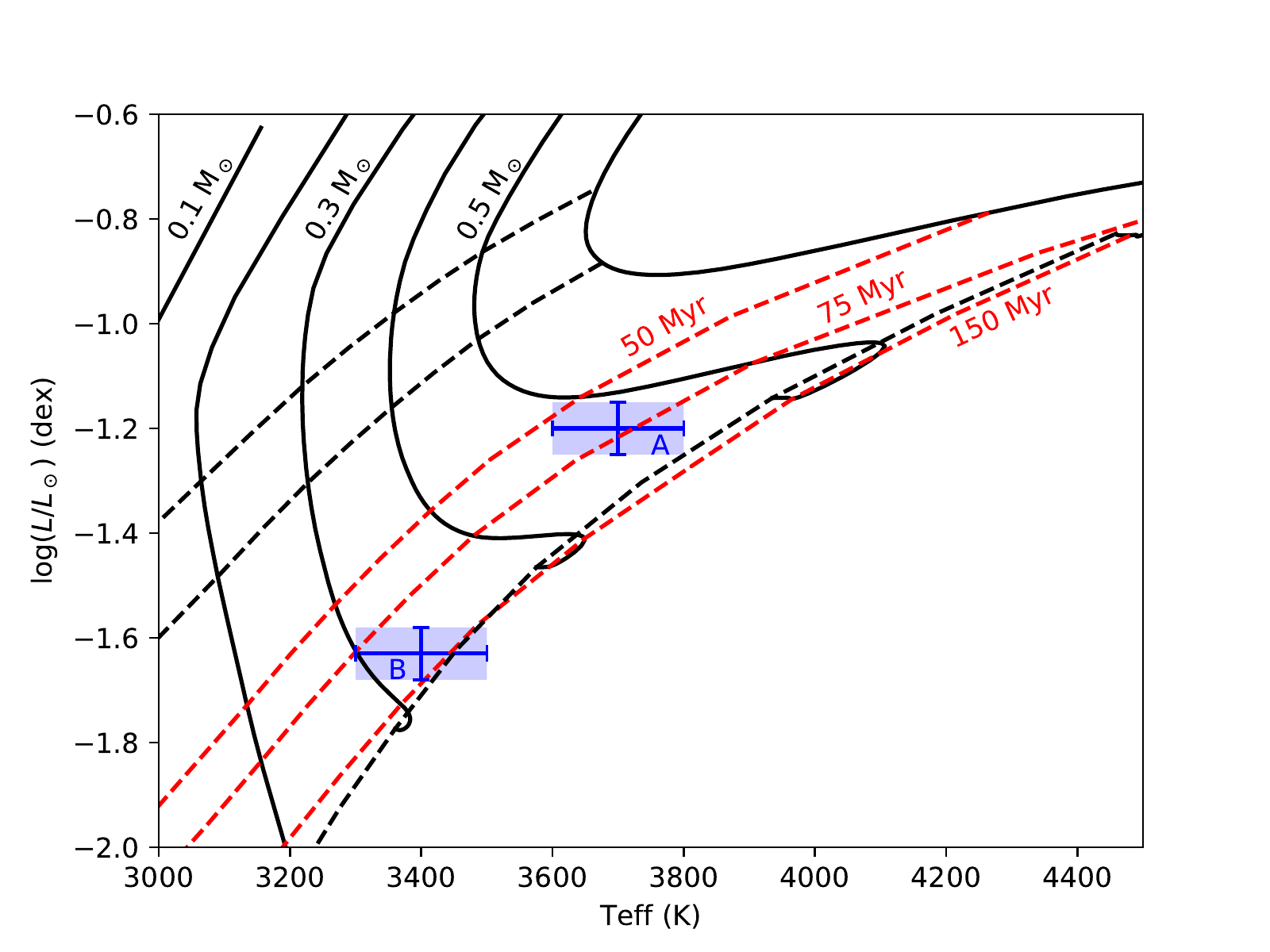}}
   	\subfloat[PISA]{\includegraphics[width=0.45\linewidth]{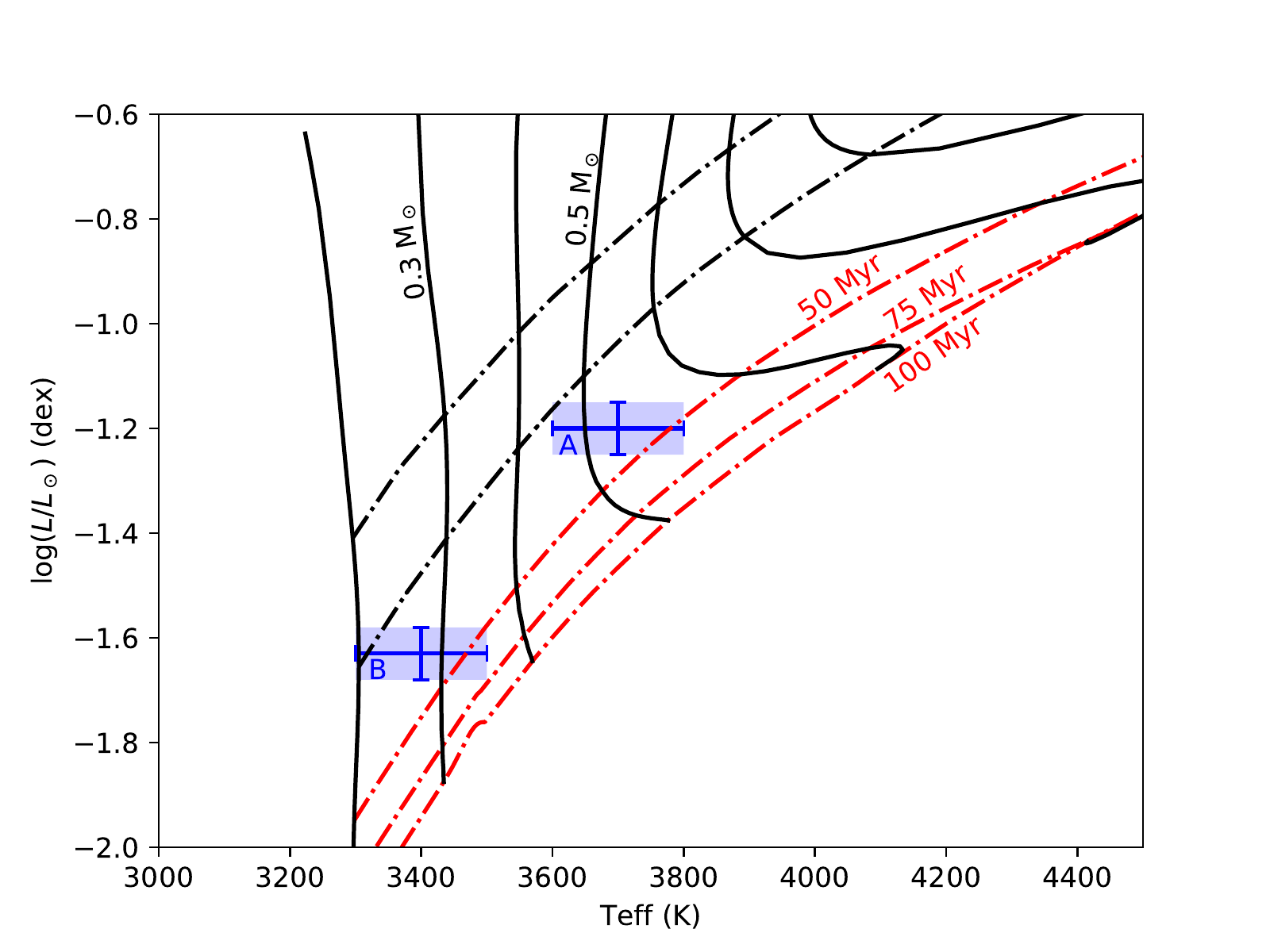}}\\
   	\subfloat[SDF00]{\includegraphics[width=0.45\linewidth]{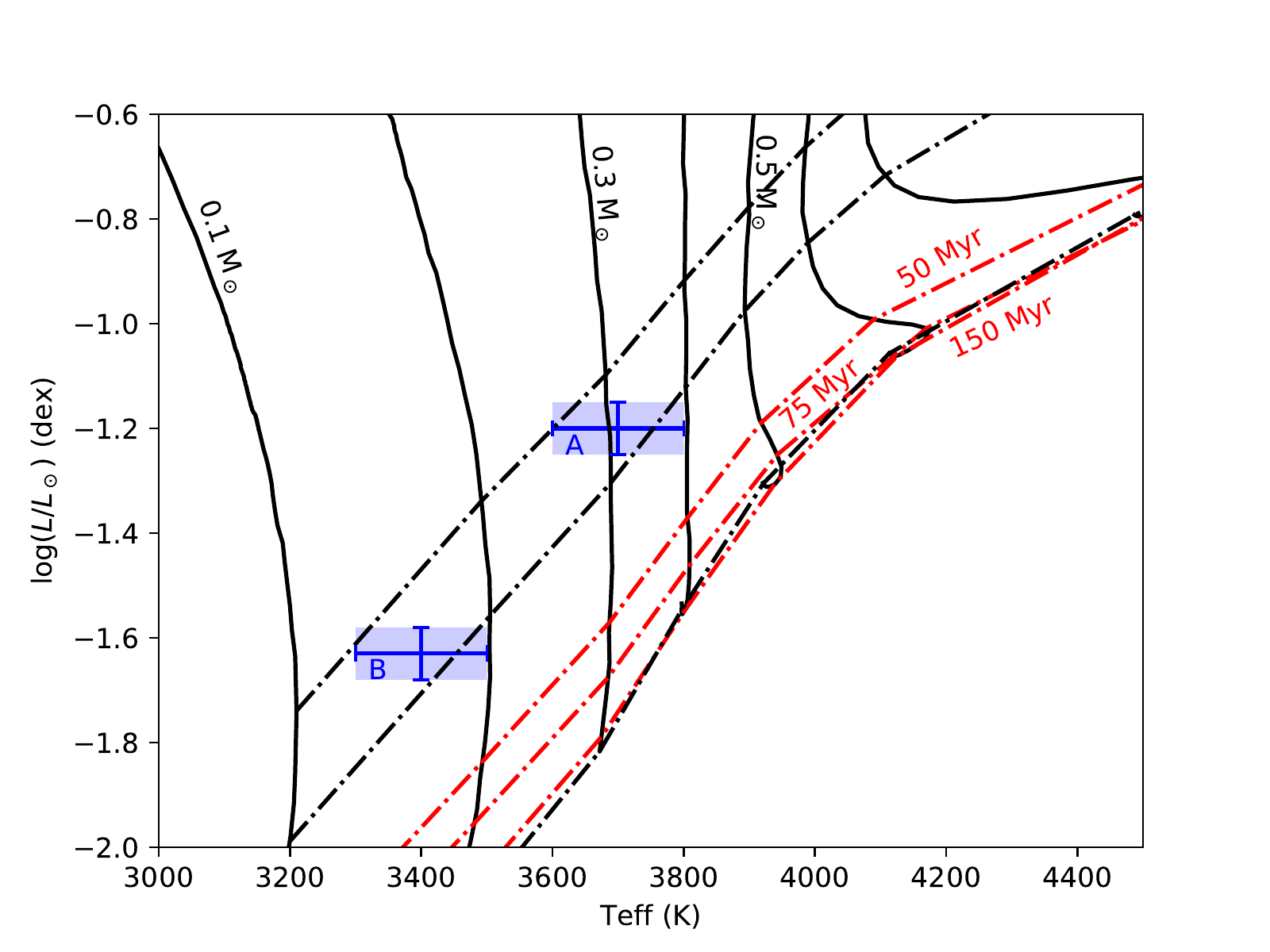}}   	
   	\subfloat[BHAC15+spots, $f_\text{spot} = 0.5$]{\includegraphics[width=0.45\linewidth]{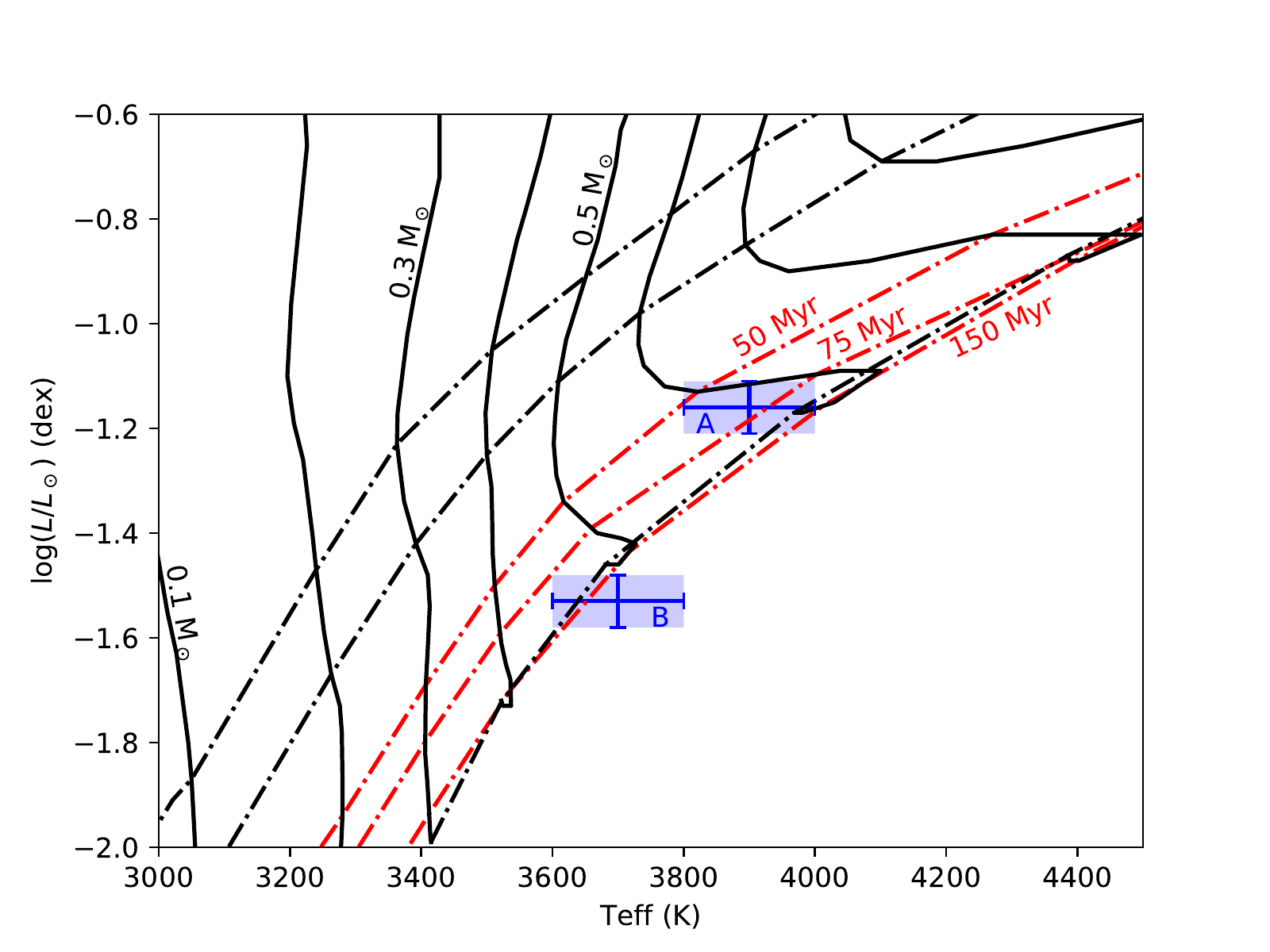}}   	
    \caption{Isochrones and iso-mass curves predicted by different evolutionary models. The 10, 20, 50, 75, 150 and 600 Myr isochrones have been drawn (1 Myr is up) (except for the Pisa and DM97 models, that stop respectively at 100 and 500 Myr), while one iso-mass is drawn every 0.1 $\mathrm{M_\odot}$ from 0.1 (left) to 1 $\mathrm{M_\odot}$ (right). The 50, 75 and 150 Myr isochrones correspond to possible ages for ABDor-MG, and are drawn in red. The blue points correspond to the observed values and their error bars for each component of system GJ~2060, A and B.}
\end{figure}

\begin{figure}[h]
	\centering
   	\subfloat[DM97]{\includegraphics[width=0.45\linewidth]{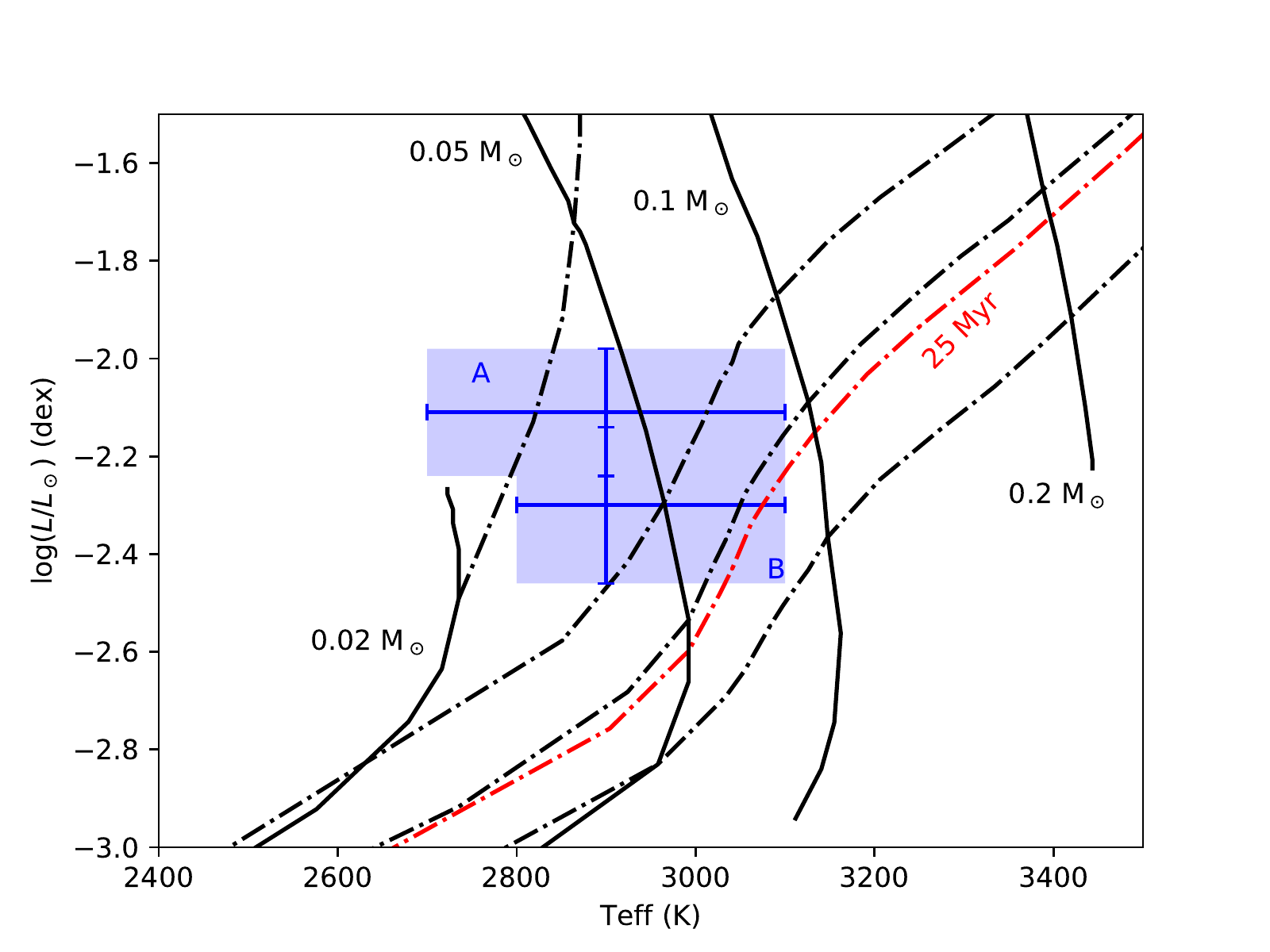}}
   	\subfloat[BHAC15+spots, $f_\text{spot} = 0.5$]{\includegraphics[width=0.45\linewidth]{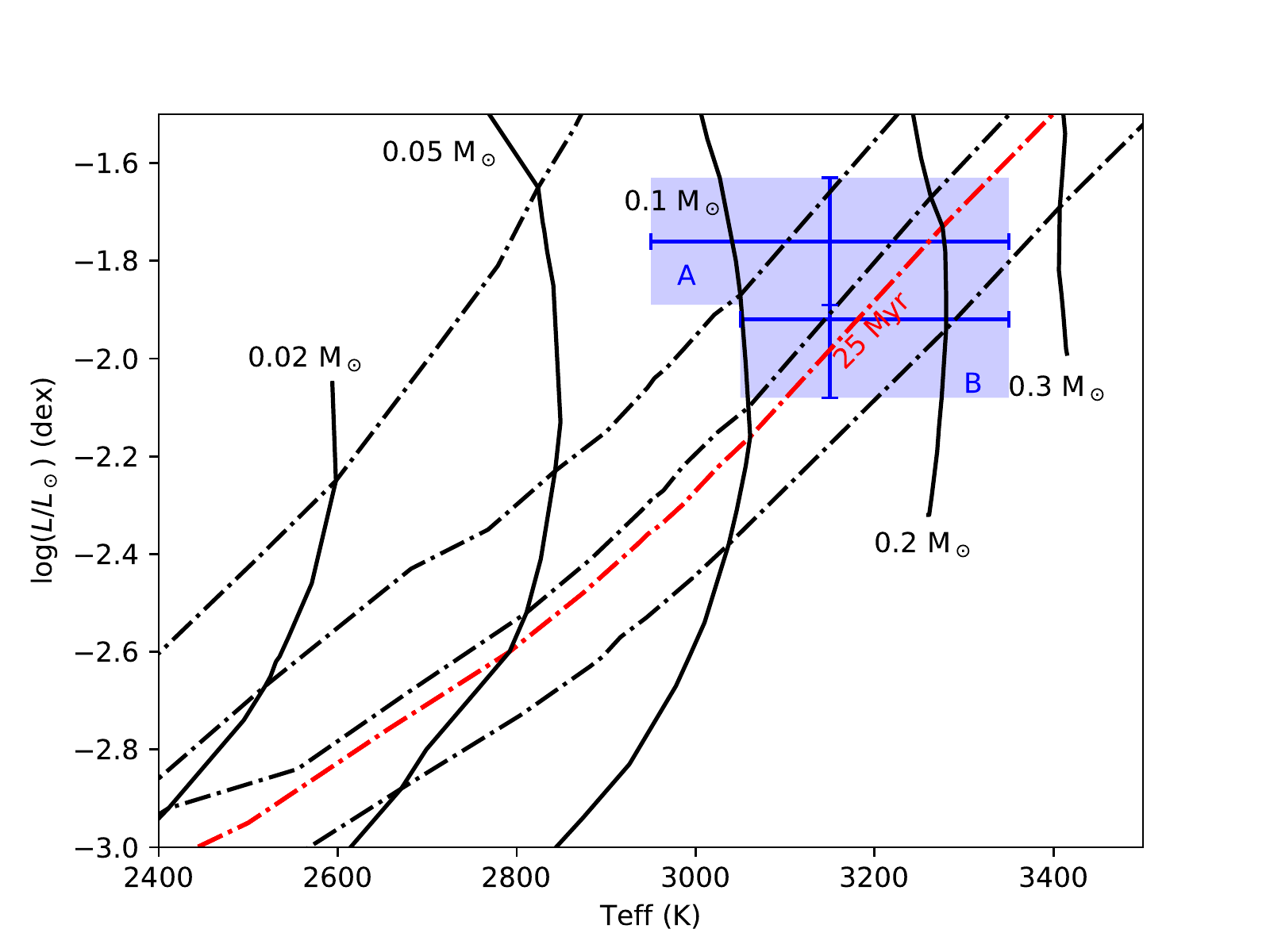}}   
    \caption{Isochrones and iso-mass curves predicted by different evolutionary models. The 1, 10, 20, 25 and 50 Myr isochrones have been drawn in dotted lines (1 Myr is up), while the iso-mass curves are in solid line . The 25 Myr isochrone correspond to the age of the $\beta$Pic-MG, and is drawn in red. The blue patterns correspond to the observed values and their error bars for each component of system TWA 22, A and B.}
\end{figure}\clearpage

\section*{Appendix D: Dynamical analysis}

\begin{figure}[h]
	\centering
   	\subfloat[]{\includegraphics[width=0.5\linewidth]{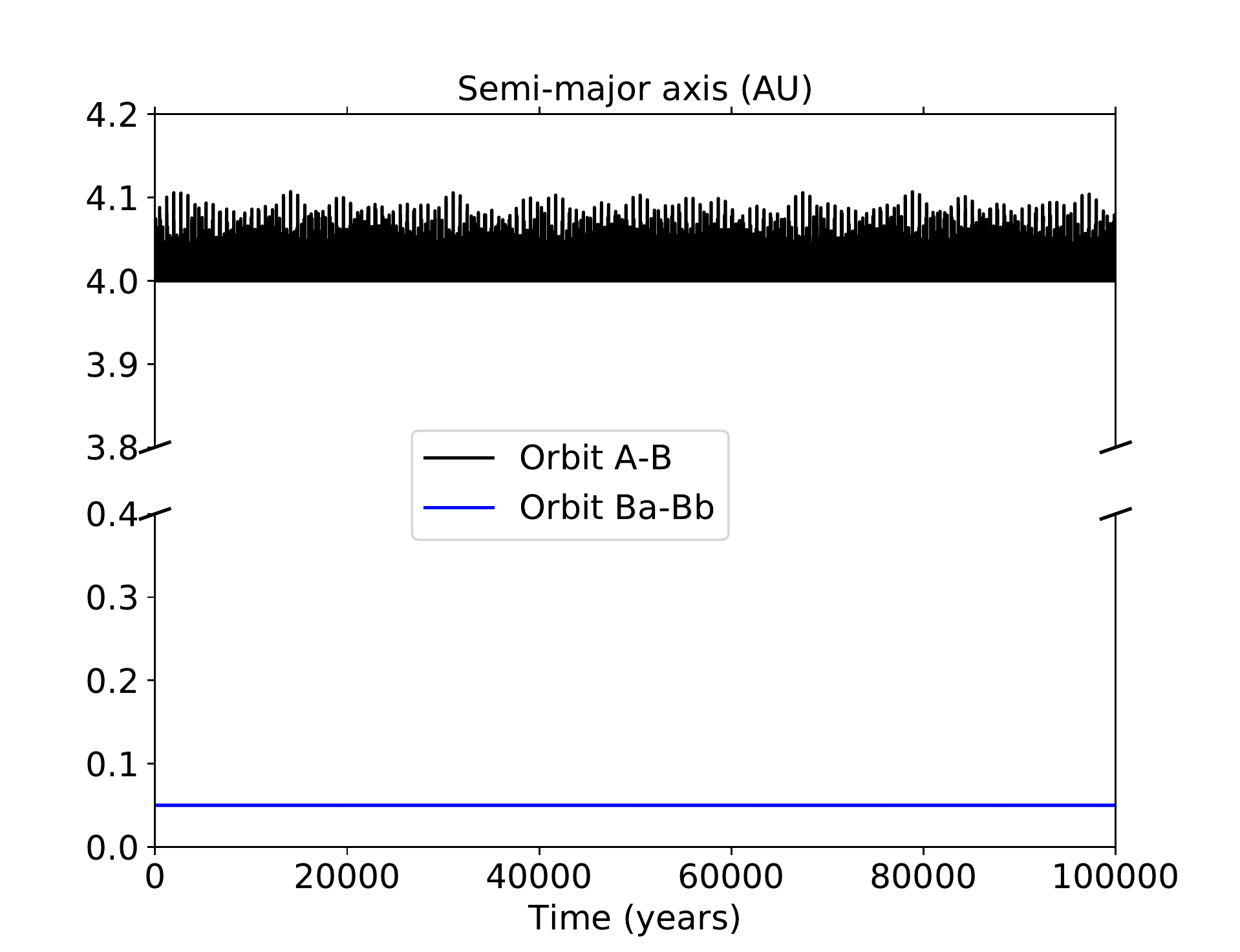}}
   	\subfloat[]{\includegraphics[width=0.5\linewidth]{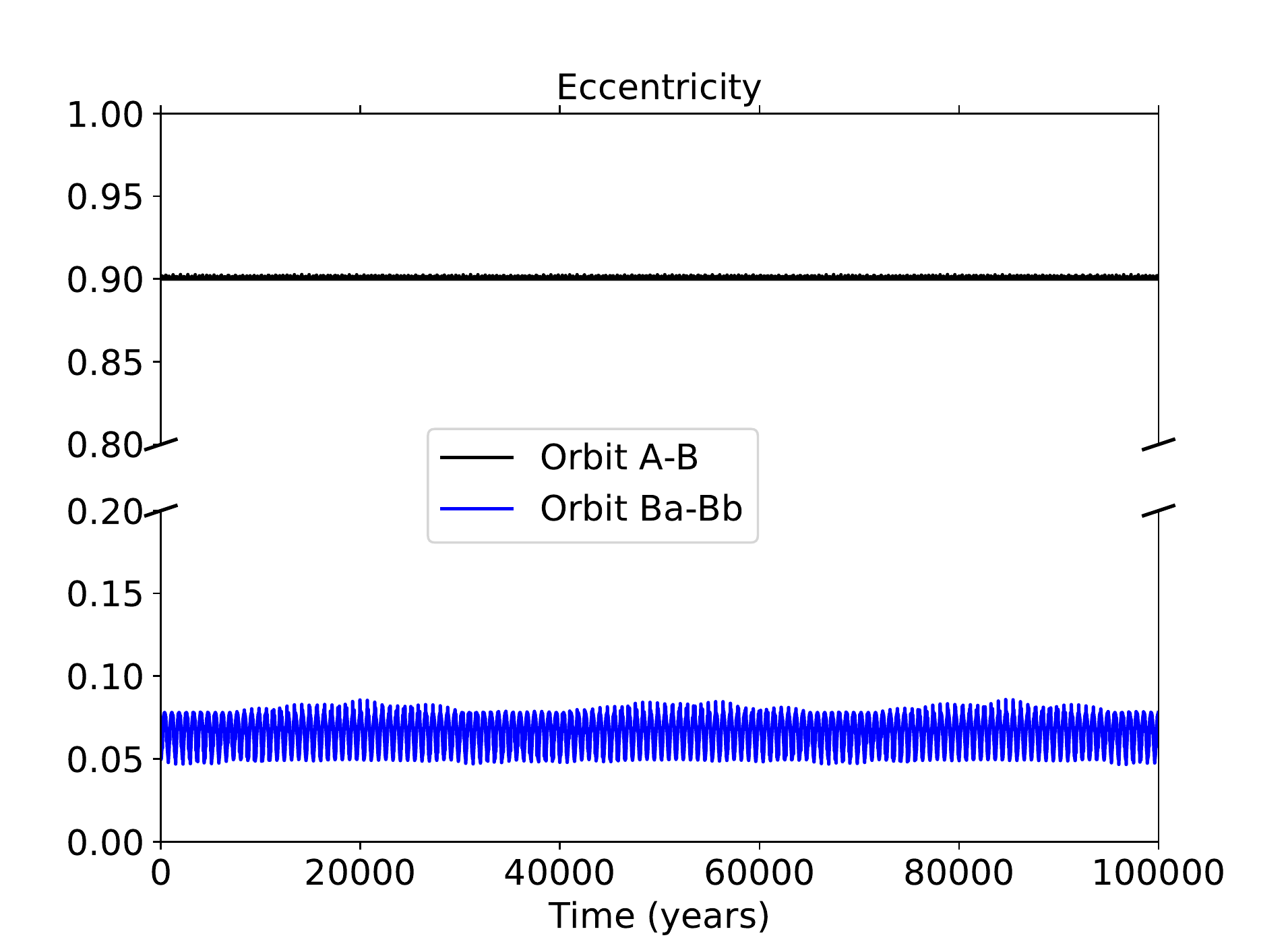}}
    \caption{100,000 yr evolution of the semi-major axis and eccentricity of the two orbits represented in Fig. \ref{fig:Orbite}.}
    \label{fig:ae}
\end{figure}

\begin{figure}[h]
	\centering
   	\includegraphics[width=0.5\linewidth]{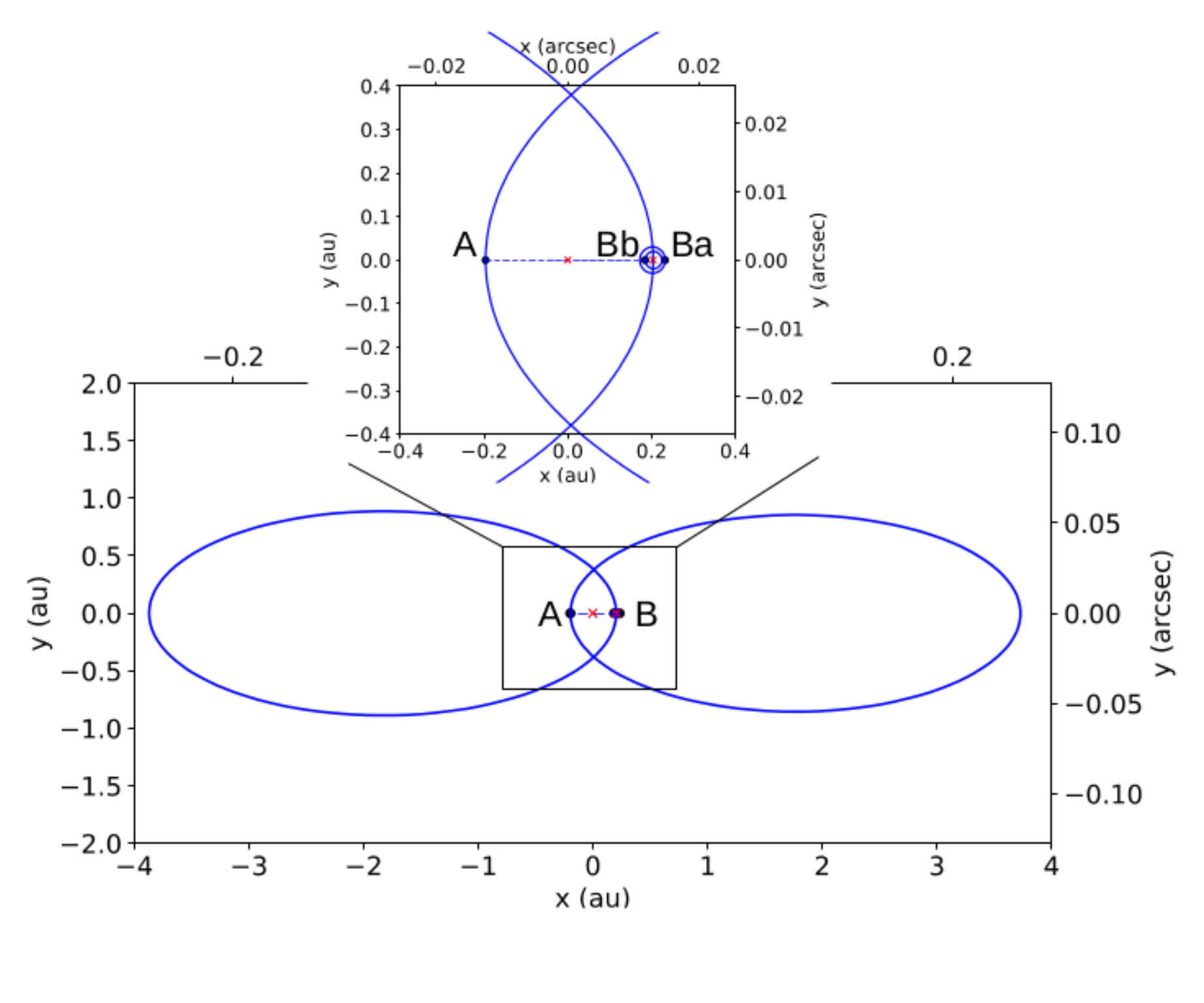}
    \caption{Face-on example of a hypothetical stable three-body configuration consistent with the data. Only components A and Ba are currently detected.}
    \label{fig:Orbite}
\end{figure}

Dynamical simulations were performed with SWIFT{\_}HJS, a symplectic N-body code designed for multiple systems \citep{beust2003}, to test the stability of a three-body evolution. Indeed, no stability criterion can be easily derived for three close bodies with similar masses, especially in the case of highly eccentric perturbers such as here. Some configurations were tested, around both components, assuming null eccentricity for the internal orbit (most stable case) and a coplanar situation. An example of stable configuration around the primary is depicted on figure \ref{fig:Orbite}. The corresponding semi-major axis and eccentricity evolution for 100,000 yr, more than 10,000 times the larger period, is depicted in Fig.\ref{fig:ae}. The parameters of the simulations are described below. Within the constraints we imposed ourselves (circular coplanar orbit), our dynamical simulations show that the high eccentricity of A-B relative orbit would force the putative component to be closer than 0.1 au from the primary. The same criterion holds for an orbit around the secondary. 

A 100,000 years dynamical simulation has been performed with the configuration of Fig. \ref{fig:Orbite}, with SWIFT\_HJS. A time step of 0.001 years has been chosen. The inner orbit has initially a semi-major axis of 0.05 au and eccentricity 0.05, while the outer orbit is set with semi-major axis 4 au and eccentricity 0.9. The masses are respectively 0.55, 0.21 and 0.32 $\mathrm{M_\odot}$ for the primary, the putative companion and the secondary. The orbits are taken coplanar, with an initial mean anomaly difference of 45\degree . The evolutions of the semi-major axis and eccentricity show a strong stability of the orbits. 

\end{document}